\documentclass[%
 reprint,
superscriptaddress,
 amsmath,amssymb,
 aps,
]{revtex4-1}

\usepackage{graphicx}
\usepackage{dcolumn}
\usepackage{bm}
\usepackage{hyperref}
\setcitestyle{super}

\begin{document}


\title{Model for quantitative tip-enhanced spectroscopy and the extraction of nanoscale-resolved optical constants}


\author{Alexander S. McLeod}
\email[]{asmcleod@physics.ucsd.edu}
\affiliation{University of California, San Diego}


\author{P. Kelly}
\affiliation{University of California, San Diego}

\author{M. D. Goldflam}
\affiliation{University of California, San Diego}

\author{Z. Gainsforth}
\affiliation{Space Sciences Laboratory, University of California at Berkeley}

\author{A. J. Westphal}
\affiliation{Space Sciences Laboratory, University of California at Berkeley}

\author{Gerardo Dominguez}
\affiliation{California State University San Marcos}
\affiliation{University of California, San Diego}

\author{Mark H. Thiemens}
\affiliation{University of California, San Diego}

\author{Michael M. Fogler}
\affiliation{University of California, San Diego}

\author{D. N. Basov}
\affiliation{University of California, San Diego}

\date{\today}

\begin{abstract}
Near-field infrared spectroscopy by elastic scattering of light from a probe tip resolves optical contrasts in materials at dramatically sub-wavelength scales across a broad energy range, with the demonstrated capacity for chemical identification at the nanoscale.  However, current models of probe-sample near-field interactions still cannot provide a sufficiently quantitatively interpretation of measured near-field contrasts, especially in the case of materials supporting strong surface phonons.  We present a model of near-field spectroscopy derived from basic principles and verified by finite-element simulations, demonstrating superb predictive agreement both with tunable quantum cascade laser near-field spectroscopy of SiO$_2$ thin films and with newly presented nanoscale Fourier transform infrared (nanoFTIR) spectroscopy of crystalline SiC.  We discuss the role of probe geometry, field retardation, and surface mode dispersion in shaping the measured near-field response.  This treatment enables a route to quantitatively determine nano-resolved optical constants, as we demonstrate by inverting newly presented nanoFTIR spectra of an SiO$_2$ thin film into the frequency dependent dielectric function of its mid-infrared optical phonon.  Our formalism further enables tip-enhanced spectroscopy as a potent diagnostic tool for quantitative nano-scale spectroscopy.
\end{abstract}


\maketitle

\section{\label{sec:intro} Introduction}
Since Synge's 1928 letter to Einstein proposing a bold method for optical imaging beyond the diffraction limit \cite{NovotnyHistory}, sub-wavelength optical characterization techniques have remained subjects of intensive interest and fierce debate owing to their transformative potential.  Among such techniques, apertureless near-field scanning optical microscopy (ANSOM)\cite{Pohl,Inouye} has shattered the diffraction limit, achieving optical resolutions better than $\lambda/1000$ at infrared and THz frequencies \cite{Demodulation,ElasticScattering,THzNearField}.

Recent coupling of ANSOM to a broadband coherent infrared light source and asymmetric Michelson interferometer has enabled Fourier transform infrared spectroscopy at the nanometer length scales (nanoFTIR)\cite{AmarieNanoFTIR,BrehmNanoFTIR,HuthNanoFTIR}, in switchable combination with single-frequency imaging by the pseudo-heterodyne (PSHet) detection scheme \cite{GomezPSHet,PSHet}.  These novel interferometric techniques detect both amplitude and phase\cite{ComplexOpticalConstants,PlasmonsPhase,NearFieldPhase} of the probe-scattered ``near-field signal", which encodes nano-scale near-field optical contrasts from the sample and transmits them to the far-field. While applications to nanoscale chemical sensing at vibrational ``fingerprint" energies are obvious \cite{RaschkeVibration,HuthNanoFTIR,AmarieBones}, the utility of this instrument for fundamental nano-scale studies of correlated electron systems are equally compelling \cite{RaschkeReview,Qazilbash,RaschkeVO2,PhononEnhancement,FeiGraphenePlasmons,ChenGraphenePlasmons,Charnukha,MartinPumpProbe}.

ANSOM employs a conductive or dielectric AFM probe as both an intense near-field source and scatterer of light into the far-field.  The mechanism of optical contrast has long been understood intuitively via the simple point dipole model \cite{FordWeber,PureOpticalContrast}, in which radiation scattered from a small polarizable sphere of radius $a$ illuminated by an incident field $E_\mathrm{inc}$ is modulated through electrostatic interaction with a material surface located a distance $d$ away in the $z$-direction:

\begin{gather}
\alpha_\mathrm{eff}\equiv P_z\,/\,E_\mathrm{inc}=\frac{\alpha}{1-\alpha \beta / \big(16 \pi (a+d)^3 \big)} \label{eq:PDP} \\
\text{with} \quad \alpha \equiv 4 \pi a^3 \quad \text{and} \quad \beta \equiv \frac{\epsilon-1}{\epsilon+1}. \nonumber
\end{gather}
Here $\alpha$ denotes the ``bare" polarizability of the sphere producing a vertical dipole moment $P_z$, and $\beta$ denotes the quasi-static limit of the Fresnel coefficient $r_p(q,\omega)$.  A function of both frequency $\omega$ and in-plane momentum $q$, the Fresnel coefficient describes the relative magnitude and phase of $p$-polarized light reflected from the surface of material with frequency-dependent dielectric function $\epsilon(\omega)$.

While important theoretical advances have brought near-field spectroscopy beyond qualitative descriptions \cite{AizpuruaTaubner,Porto,Moon,Cvitkovic,Hauer,FDPInverse}, available models describing the probe-sample near-field interaction remain beset by critical limitations:
\begin{enumerate}
\item Many general formulations, although formally exact, prove cumbersome to implement for practical calculation beyond reduction to the simple point dipole model \cite{Carminati,Esslinger}.   Field retardation and antenna effects of the probe are explored formally, but not quantitatively.
\item Although the near-field interaction may be described as an exact scattering problem, many solution methods rely on perturbation expansions in powers of the sample response factor $\beta$ or $r_p$. \cite{HuthNanoFTIR,Valle,FDPInverse,Esslinger}  One can show that such series are divergent beyond modest values of $r_p$ (the ``strong coupling" regime), leaving this method inapplicable for the analysis of crystalline solids and strongly resonant plasmonic systems. \cite{Hauer}
\item A number of tunable geometric parameters with \textit{ad hoc} or empirical values are introduced to quantitatively fit measured data.  These include fractional weights of relevant probe surface charge,\cite{Cvitkovic,Hauer} effective probe size and geometry,\cite{Moon,Porto,Cvitkovic} the bare tip polarizability, etc. \cite{PhononEnhancement}  The multitude of \textit{ad hoc} tunable parameters provides an unreliable recipe for predictive modeling or quantitative interpretation of data.
\end{enumerate}

To address these extant shortcomings, the aim of the present work is threefold.  We first present a new model of probe-sample near-field interaction, the \textit{lightning rod model}, whose generality allows exploring the influence of both probe geometry and electrodynamic effects, while remaining formally exact in both theory and implementation.  Mathematical reduction of this formalism back to the point dipole model will make clear that unnecessary \textit{ad hoc} assumptions underpin prevailing models \cite{Cvitkovic,Hauer}, and that geometric and electrodynamic considerations must ultimately play a role in their vindication.

Second, we demonstrate this model's capability to predict spectroscopic near-field contrast in the case of layered structures, which exhibit a strongly momentum-dependent optical response, as well as strongly resonant systems, through comparison with near-field spectra measured on thin films of silicon dioxide (SiO$_2$) and bulk silicon carbide (SiC).  Our measurement apparatus is a novel infrared near-field microscope equipped for both PSHet imaging and broadband nanoFTIR spectroscopy, described in Appendix \ref{sec:setup}.

Finally, we present a method to \textit{invert} the \textit{lightning rod model} to extract a material's complex dielectric function with nanoscale resolution, which we demonstrate explicitly for an SiO$_2$ thin film sample.  This procedure, combined with the unifying formalism of the \textit{lightning rod model}, provides a powerful diagnostic tool for quantitatively studying the nano-resolved optical properties of molecular systems, phase-separated materials, and confined nanostructures using ANSOM.\cite{RaschkeReview}

\begin{figure*}[t!]
\centering
   \includegraphics[width=\textwidth]{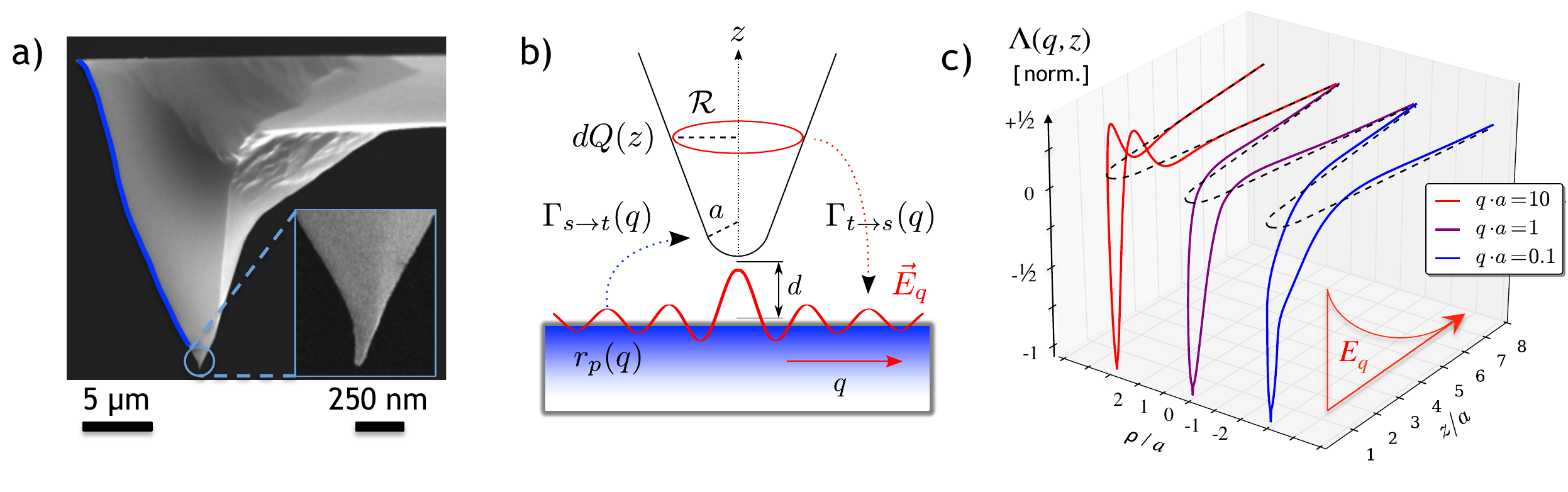}
   \vspace{-15pt}
   \caption{\textbf{a)} Scanning electron micrograph of a typical commercial near-field probe exhibiting a conical geometric profile and characteristic length scales (probe length and tip radius) separated by nearly three orders of magnitude.  The surface profile (blue) is considered in Sec. \ref{sec:SiC}.  \textbf{b)} Schematic description of the probe-sample near-field interaction, involving emission of cylindrical evanescent fields from charge elements in the probe and their reflection by the sample.  \textbf{c)} Probe response function $\Lambda(q,z)$ (defined in the main text) computed by the boundary element method (Appendix \ref{app:QSboundaryelement}) for evanescent fields $\vec{E}_q$ of increasing momenta $q$.  Dashed curves indicate the geometric profile of the probe, and surface charge distribution profiles are normalized by their minimum values for viewing purposes.}
   \label{fig:1}
\end{figure*}

\section{The Lightning Rod Model} \label{sec:model}
Our model describes the near-field interaction between an ANSOM probe and a sample surface through a general formalism that is in principle exact, without appealing to empirical or \textit{ad hoc} parameters.  The chief observable of ANSOM is the radiation field of a polarized probe placed in proximity to a sample (experimental details described in Appendix \ref{sec:setup}).  Since the field originates from reorganization of charges developing on the probe surface in response to an incident illumination field, together with the near-field of the proximate sample, we begin by forming an expression for this instantaneous charge distribution.

Constraining our attention to nearly axisymmetric probe geometries, the charge distribution is succinctly expressed through a linear charge density $\lambda_Q(z)\equiv dQ/dz(z)$, $Q$ denoting charge and $z$ the probe's axial coordinate.  In the quasi-static approximation, the field $E_\mathrm{rad}$ re-radiated or back-scattered from the probe is proportional to its induced dipole moment
\begin{equation}\label{eq:dipolemoment}
 P_z=\int dz\,z\,\lambda_Q(z).
 \end{equation}
 Appendix \ref{app:radiation} presents how the radiated field is obtained from $\lambda_Q(z)$ when electrodynamic phenomena are of fundamental importance, \textit{i.e.} when the size of the scatterer is comparable to the light wavelength.  This relationship highlights the central role of the induced charge distribution in determining the measured observables of near-field spectroscopy.

$\lambda_Q(z)$ can be written as the sum of charges induced by the incident field and those differential contributions $d\lambda_{Q\,\mathrm{nf}}(z)$ induced by reflection of probe-generated near-fields off the proximate sample:
\begin{equation}
\lambda_Q(z)=E_\mathrm{inc}\, \Lambda_0(z)+\int d\lambda_{Q\,\mathrm{nf}}(z). \label
{eq:chargedistparts}
\end{equation}
Here $\Lambda_0(z)$ denotes the induced charge per unit field resulting from incident illumination.  Its functional form depends on the nature of the incident field and the probe geometry, but its contribution to $\lambda_Q(z)$ scales with the magnitude of the incident field $E_\mathrm{inc}$. The induced charge elements $d\lambda_{Q\,\mathrm{nf}}(z)$ take the form:
\begin{align}
d\lambda_{Q\,\mathrm{nf}}(z)&=dQ \, \iint_0^\infty dq\, dq'\;  \mathcal{G}(q)\, \Gamma_{t\rightarrow s}(q) \nonumber \\
	&\quad \quad \quad \quad \quad \quad \times R(q,q')\, \Gamma_{s\rightarrow t}(q')\, \Lambda(q',z). \label{eq:parametrization}
\end{align}
Here $q$ and $q'$ denote in-sample-plane momenta for Fourier components of the near-field reflected by the sample in response to the polarized probe.  Provided a planar sample geometry, this parameterization offers a sparse basis in which to efficiently solve the problem, in contrast with real-space treatments (\textit{e.g.} the finite element method).  Eq. \eqref{eq:parametrization} can be understood through the physical mechanism shown schematically in Fig. \ref{fig:1}b and described as follows in terms of field emission from the probe and sample reflection.

Charge elements $dQ=dz'\lambda_Q(z')$ on the probe form rings with radii $\mathcal{R}_{z'}$ along its surface.  Considered in the angular spectrum decomposition (Appendix \ref{app:fielddecomposition}), each ring emits a distribution of axisymmetric $p$-polarized evanescent fields whose Fourier components are weighted by $\mathcal{G}(q)=q\, J_0(q \mathcal{R}_{z'})$.  $J_{i}(\ldots)$ is the Bessel function of the first kind at order $i$, with cylindrical coordinates $\rho$ and $z$.  These emitted fields (so-called evanescent Bessel beams) reach the sample surface a distance $d$ below the tip apex ($z=0$) via propagator $\Gamma_{t\rightarrow s}(q)=e^{-q(z'+d)}$ and in the empty tip-sample gap $-d<z<0$ take the divergence-free form (per unit charge):
\begin{equation}
\vec{E}_q(\vec{r})=\mathcal{G}(q)\left(J_0(q\rho)\,\hat{z}+J_1(q\rho)\,\hat{\rho}\right)\,e^{q (z-z')}. \label{eq:qbasis}
\end{equation}

In general, the sample may subsequently scatter evanescent fields with momentum $q$ into distinct Fourier components $q'$ as described through a differential sample response function $R(q,q')$.  For samples with continuous in-plane translational symmetry (\textit{e.g.} flat surfaces, layered structures) this response function reduces to the Fresnel reflection coefficient for $p$-polarized light,
\begin{equation}
R(q,q')=r_p(q)\, \delta(q-q'), \label{eq:planarreflection}
\end{equation}
written here as a function of the in-plane momentum $q$ of incident light, with $\delta(\ldots)$ denoting the Dirac delta distribution.  These scattered fields extend from the sample surface via propagator $\Gamma_{s\rightarrow t}(q')=e^{-q' (z+d)}$ and re-polarize the probe, inducing a linear charge density (per unit field) described by a \textit{probe response function} $\Lambda(q',z)$.  This formalism accommodates the non-trivial influence of realistic tip geometries on the functional form of illumination- and sample-induced charge distributions $\Lambda_0(z)$ and $\Lambda(q',z)$.

Induced charge densities can be pre-computed for an axisymmetric probe of arbitrary geometry using a simple boundary element method (Appendices \ref{app:QSboundaryelement} and \ref{app:EDboundaryelement}).  Fig. \ref{fig:1}c displays $\Lambda(q,z)$ for several values of $q$, computed for the case of a hyperboloidal (conical with rounded tip) probe geometry with tip curvature radius $a$.  As shown in Fig. \ref{fig:1}c, the density of charge dramatically accumulating at the probe apex - the celebrated \textit{lightning rod effect} - increases roughly as $1/q$.  This results from the requisite screening of evanescent fields by induced charges distributing a distance $\delta z=1/q$ along the probe surface.  The momentum-dependent \textit{lightning rod effect} is critically absent from models of the probe-sample near-field interaction lacking a faithful geometric description.

Confining our attention to planar sample geometries, Eqs. \eqref{eq:chargedistparts}, \eqref{eq:parametrization}, and \eqref{eq:planarreflection} together describe a self-consistent quasi-one-dimensional scattering process:
\begin{gather}
\lambda_Q(z)=E_\mathrm{inc}\, \Lambda_0(z)- \int_0^\infty dq\, \widetilde{\lambda}_Q(q)\cdot q\, e^{-2 q d}\, r_p(q)\, \Lambda(q,z) \label{eq:SSEQ} \\
\text{with}\quad \widetilde{\lambda}_Q(q) \equiv \int_0^L dz' \lambda_Q(z')\cdot e^{-qz'}\,J_0(q\,\mathcal{R}_{z'}). \label{eq:integralxform}
\end{gather}
The integral transform in Eq. \eqref{eq:integralxform} denotes summation of near-fields emitted from charges along the entire length of the probe, $0<z<L$.  A similar integral transformation $z\rightarrow s$ applied to $\lambda(z)$ and $\Lambda(q,z)$ in Eq. \eqref{eq:SSEQ} yields an integral equation in $\widetilde{\lambda}(s)$:
\begin{equation}
\widetilde{\lambda}_Q(s)=E_\mathrm{inc}\,\widetilde{\Lambda}_0(s)-\int_0^\infty dq\,\widetilde{\lambda}_Q(q)\cdot q\,e^{-2 q d}\,r_p(q)\,\widetilde{\Lambda}(q,s). \label{eq:SSEQsimplified}
\end{equation}
Eq. \eqref{eq:SSEQsimplified} resembles the Lippman-Schwinger equation of scattering theory \cite{Scattering}. Here $\widetilde{\Lambda}_0(s)$ and $\widetilde{\Lambda}(q,s)$ play the role of in- and out-going ``scattering states."  Tractability in this scattering formalism is afforded by the axisymmetric approximation, which preserves most fundamental aspects of the system geometry.

Provided knowledge of the probe response functions $\Lambda_0(z)$ and $\Lambda(q',z)$, Eq. \eqref{eq:SSEQsimplified} is soluble by traditional methods\cite{Nystrom} after discretizing $q$ to a set of Gauss-Legendre nodes $\{q_i\}$.  We found evaluation at $N_q \approx 200$ such nodes sufficient for numerical accuracy to within 1\%.  Only a finite range of momenta $0 \le q \le q_\mathrm{max}$ need be considered in practice, since $\widetilde{\Lambda}_0(s)$ and $\widetilde{\Lambda}(q,s)$ drop precipitously in magnitude above a cutoff momentum $s\sim 1/a$, with $a$ the smallest length scale relevant to the probe geometry, in this case the radius of curvature at the probe apex, $a\approx 30$ nm for many commercial probe tips.  This reflects the inability of strongly confined fields (\textit{e.g.} $q\sim \mathrm{nm}^{-1}$) to efficiently polarize the probe. 

The solution to Eq. \eqref{eq:SSEQsimplified} is then
\begin{gather}
\vec{\lambda}_Q=\frac{\vec{\Lambda}_0}{\mathbf{I}-\mathbf{\Lambda}\mathbf{G}} E_\mathrm{inc}, \end{gather}
where the denominator is taken in the sense of matrix inversion, vectors imply functional evaluation at momenta $\{q_i\}$, $\mathbf{I}$ is the identity operator, and other matrices are defined as
\begin{equation}
\mathbf{\Lambda}_{ij}\equiv \widetilde{\Lambda}(q_j,s_i) \quad \text{and} \quad \mathbf{G}_{ij} \equiv -q_i\,e^{-2 q_i d}\, r_p(q_i)\, \delta q_i \,\delta_{ij}. \label{eq:G}
\end{equation}
Here $\delta q_i$ is the measure of $q_i$ and $\delta_{ij}$ denotes the Kronecker delta function.  Defining similarly a vector of charge distribution functions $\big[\vec{\Lambda}(z)\big]_i \equiv \Lambda(q_i,z)$ together with their associated contributions to the radiated field $\vec{e}_\mathrm{rad} \equiv E_\mathrm{rad}\big[ \vec{\Lambda} \big]$ (see Appendix \ref{app:radiation}), the total induced charge and consequent back-scattered field are provided through Eq. \eqref{eq:SSEQ} as:
\begin{eqnarray}
\lambda_Q(z)\,/\,E_\mathrm{inc}&=&\Lambda_0(z)+\vec{\Lambda}(z) \cdot \mathbf{G}\, \frac{\vec{\Lambda}_0}{\mathbf{I}-\mathbf{\Lambda} \mathbf{G}}  \label{eq:LRMcharge} \\
E_\mathrm{rad}\,/\,E_\mathrm{inc}&=&e_{\mathrm{rad},\,0}+\vec{e}_\mathrm{rad} \cdot \mathbf{G}\, \frac{\vec{\Lambda}_0}{\mathbf{I}-\mathbf{\Lambda} \mathbf{G}}. \label{eq:LRMpolarization}
\end{eqnarray}
Note that dependence on both the tip-sample distance $d$ and the local optical properties of the sample enter these expressions through $\mathbf{G}$, whereas geometric properties of the probe enter separately through $\mathbf{\Lambda}$.  When applied to a realistic probe geometry, these expressions constitute the \textit{lightning rod model} of probe-sample interaction, so named for its quantitative description of the strong electric fields localized by an elongated geometry to a pointed apex.  The product of $\mathbf{\Lambda}$ and $\mathbf{G}$ signifies that strong near-fields from the probe multiplicatively enhance optical interactions with the sample surface.  Expanding the Eq. \eqref{eq:LRMpolarization} inverse matrix as a geometric series reveals an infinite sequence of probe polarization and sample reflection events, equivalent to the perturbation expansions presented elsewhere \cite{HuthNanoFTIR,Valle,FDPInverse,Esslinger}.

Eq. \eqref{eq:LRMpolarization} can also recover the point dipole model (Eq. \eqref{eq:PDP}).  After simplifying the probe geometry to a metallic sphere of radius $a$ and assuming that all center-evaluated ($z=a$) fields polarize like the homogeneous incident field $E_\mathrm{inc}$, we have:
\begin{eqnarray}
\Lambda(q,z)&=&3/2\,(z-a)\,e^{-q a}, \\
\widetilde{\Lambda}(q,s)&=&-a^3\,s\,e^{-(s+q) a}, \label{eq:PDPLambda} \\
\text{and} \quad p_z(q)&=&\int_0^{2a} dz\,z\,\Lambda(q,z)=a^3\,e^{-q a}. \label{eq:PDPpz}
\end{eqnarray}
$\widetilde{\Lambda}(q,s)$ is obtained from semicircular $\mathcal{R}_z$ in Eq. \eqref{eq:integralxform}, and exhibits a characteristic maximum followed by a sharp decay in magnitude near $s \sim a^{-1}$.  In this case, Eq. \eqref{eq:SSEQsimplified} yields $\widetilde{\lambda}_Q(s)$ in closed form owing to the separability of $\widetilde{\Lambda}(q,s)$:
\begin{equation}
\widetilde{\lambda}_Q(s)=\frac{-a^3\,s\,e^{-s a}}{1-a^3 \int_0^\infty dq\,q^2\,e^{-2 q (d+a)}\,r_p(q)}.
\end{equation}
The sphere's polarization is obtained through Eqs. \eqref{eq:SSEQ} and \eqref{eq:dipolemoment} as
\begin{equation}
\alpha_\mathrm{eff} \equiv P_z/E_\mathrm{inc}= \frac{a^3}{1-a^3 \int_0^\infty dq\,q^2\,e^{-2 q (d+a)}\,r_p(q)}.
\end{equation}
If the sample material is weakly dispersive for $q \gg \omega/c$, $r_p(q) \approx \beta$ and Eq. \eqref{eq:PDP} is recovered.

Such simplifications are instructive, but this work makes full implementation of Eq. \eqref{eq:LRMpolarization} without recourse to approximation, thus revealing aspects of the probe-sample near-field interaction unresolved by the point dipole model.  While perturbative expansions and the point-dipole model may be attractive for their relative simplicity, they are certainly not expected to be accurate. In particular, for large $\beta$, nothing short of the full numerical solution to Eq. (11) is acceptable for predicting experimental observables with quantitative reliability. Our procedure for doing so is detailed in the following sections.  

The near-field experiments presented in this work utilize lock-in detection of the probe's back-scattered field at harmonics $n$ of the probe tapping frequency $\Omega$ to suppress noise and unwanted background.  Simulating this technique, the probe's back-scattered field $E_\mathrm{rad}$ (Eq. \eqref{eq:LRMpolarization}) is connected to experimentally observed amplitude $S_n$ and phase $\phi_n$ signals through a sine transform under sinusoidally varying tip-sample distance $d$:
\begin{gather}
s_n(\omega) = \mathcal{I}(\omega) \int_{-\pi/\Omega}^{\pi/\Omega} dt\,\sin \left(n\Omega t \right) \,E_\mathrm{rad}\big(d,r_p(q,\omega)\big) \\
\text{with} \quad d=A \left(1+\sin\left(\Omega t\right) \right), \nonumber
\end{gather}
\vspace{-20pt}
\begin{equation}
S_n(\omega) \equiv \left|s_n(\omega)\right|, \quad \text{and} \quad \phi_n \equiv \mathrm{arg}\left\{s_n(\omega)\right\}.
\end{equation}
Here $\Omega$ and $A$ are the tapping frequency and amplitude of the near-field probe, respectively, and $\mathcal{I}(\omega)$ denotes the frequency-dependent instrumental response of the collection optics, interferometer and detector used for the measurement.  This factor can be removed by normalizing experimental $s_n(\omega)$ to ``reference" near-field signal values, as collected from a uniformly reflective sample material such as gold or undoped silicon.  This normalization process is further discussed in Sec. \ref{sec:electrodynamics}.

A prediction of near-field contrast using the \textit{lightning rod model} therefore requires calculation of Eq. \eqref{eq:LRMpolarization} at several values of $d$; in practice we find 20 such values sufficient, with evaluation of Eq. \eqref{eq:LRMpolarization} for each requiring several milliseconds on a single 2.7 GHz processor.  Cumulatively, the calculation remains both realistic and fast, more so than previously reported semi-analytic solutions for realistic probe geometries.\cite{Zhang,RaschkeHyperboloid}  For example, typical calculations of a demodulated and normalized near-field spectrum across 100 distinct frequencies require less than 10 seconds of computation time.

We conclude this formal introduction with a conceptual clarification.  Although the geometry and material composition of the near-field probe implicitly determine its response function $\Lambda(q,z)$, the formalism embodied by Eq. \eqref{eq:LRMpolarization} is general and outwardly irrespective of specific properties of the probe.  Also, while plasmonic enhancement may be encompassed in $\Lambda(q,z)$, it is \textit{not} a prerequisite for effective near-field enhancement at the probe apex.  Near-field enhancement is attainable through a combination of three distinct mechanisms:\cite{HartschuhReview}
\begin{enumerate}
\item the \textit{lightning rod effect} proper, due to accumulation of charge at geometric singularities, an essentially electrostatic effect,
\item \textit{plasmonic enhancement}, due to the correlated motion of surface charges near the plasma frequency of metals,
\item and \textit{antenna resonances}, in which the size of an optical antenna correlates with the incident wavelength in a resonant fashion, a purely electrodynamic effect.
\end{enumerate}

The quasi-static boundary element utilized in this work (Appendix \ref{app:QSboundaryelement}) reproduces the first of these mechanisms by way of $\Lambda(q,z)$, whereas its electrodynamic counterpart (Appendix \ref{app:EDboundaryelement}) reproduces all three.  Although plasmonic enhancements are scarcely attainable in metallic probes at infrared frequencies, Secs. \ref{sec:quasistatic} and \ref{sec:electrodynamics} establish the important influence of both the \textit{lightning rod effect} and antenna resonances in near-field spectroscopy.

\begin{figure}[t!]
\centering
  \hspace{-15pt}
   \includegraphics[width=.475\textwidth]{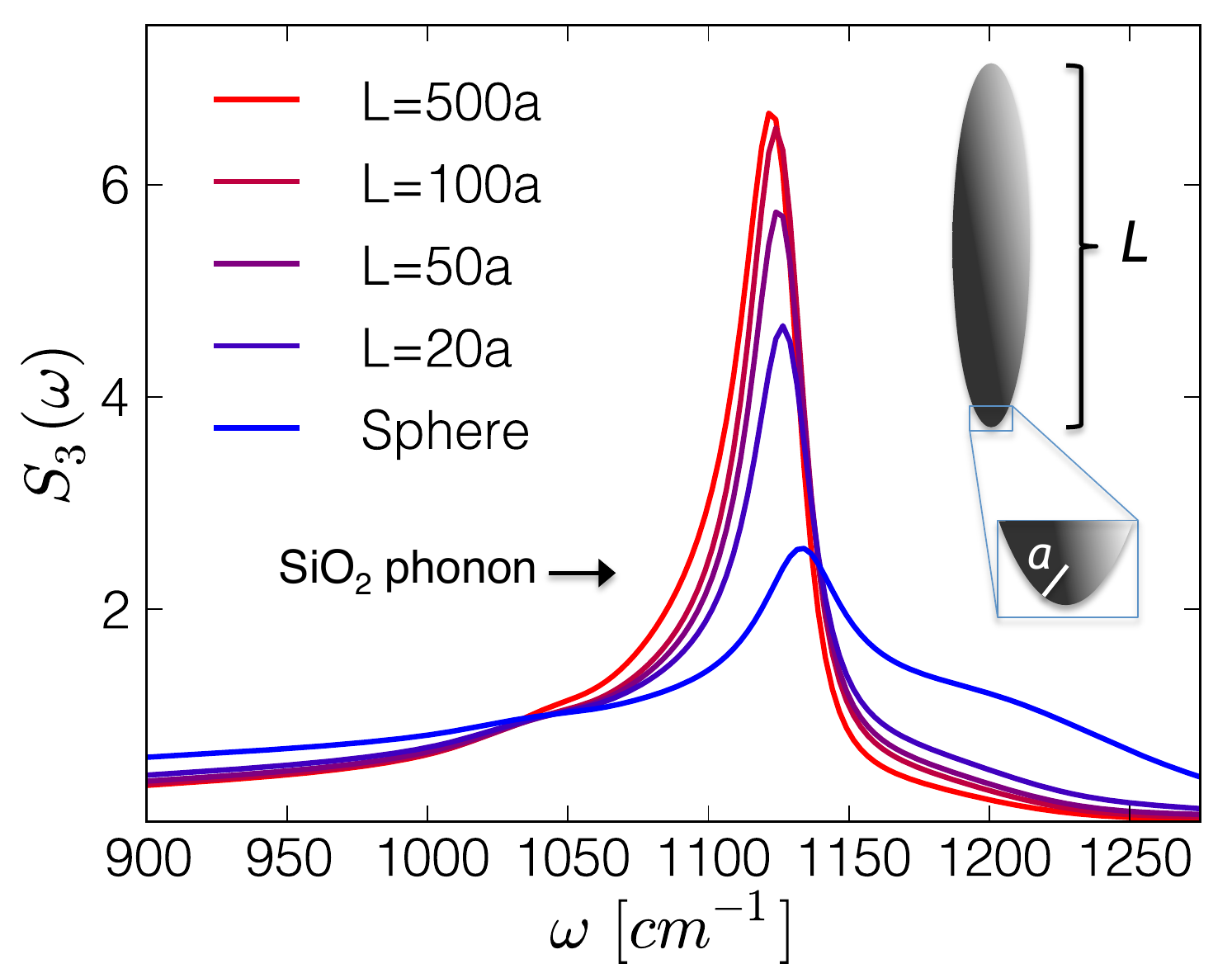}
   \caption{Spectral near-field contrast between the 1130 cm$^{-1}$ surface phonon polariton of SiO$_2$ and silicon (providing normalization) as predicted by the \textit{lightning rod model} for an ellipsoidal probe in the quasi-static approximation.  Contrast increases monotonically beyond experimentally observed levels as the probe length is increased.}
  \label{fig:2}
\end{figure}

\section{The Quasi-static case} \label{sec:quasistatic}
We now apply Eq. \eqref{eq:LRMpolarization} to realistic probe geometries in the quasi-static limit to investigate whether the quasi-static approximation is appropriate for quantitative prediction of near-field contrasts.  By reducing the physical system to electrostatics, this approximation is strictly justified only in treating light-matter interactions at length scales much smaller than the wavelength of light, whereas a typical near-field probe consists of an AFM tip tens of microns in height (Fig. \ref{fig:1}a), comparable to typical wavelengths encountered in infrared near-field spectroscopy.  Consequently, for the assumptions of a quasi-static probe-sample interaction to remain valid, the emergent behavior of a realistic near-field probe must be shown nearly equivalent to those of a deeply sub-wavelength one.

To test this assumption, we consider the near-field interaction between a metallic ellipsoidal probe oriented vertically over a planar sample of 300 nm of thermal silicon dioxide (SiO$_2$) on silicon substrate.  This sample system and model probe geometry were considered in a previous work \cite{Zhang}, demonstrating the capacity of near-field spectroscopy to resolve the $\omega \approx 1130$ cm$^{-1}$ surface optical phonon of thermal oxide films as thin as 2 nm.  We extend the theoretical study presented therein to investigate the influence of the probe length $L$ on the amplitude of experimentally measurable back-scattered near-field signal $S_3(\omega)$ (normalized to silicon) predicted by the \textit{lightning rod model}. The outcome is presented in Fig. \ref{fig:2}.

\begin{figure*}[t!]
\centering
   \includegraphics[width=.9\textwidth]{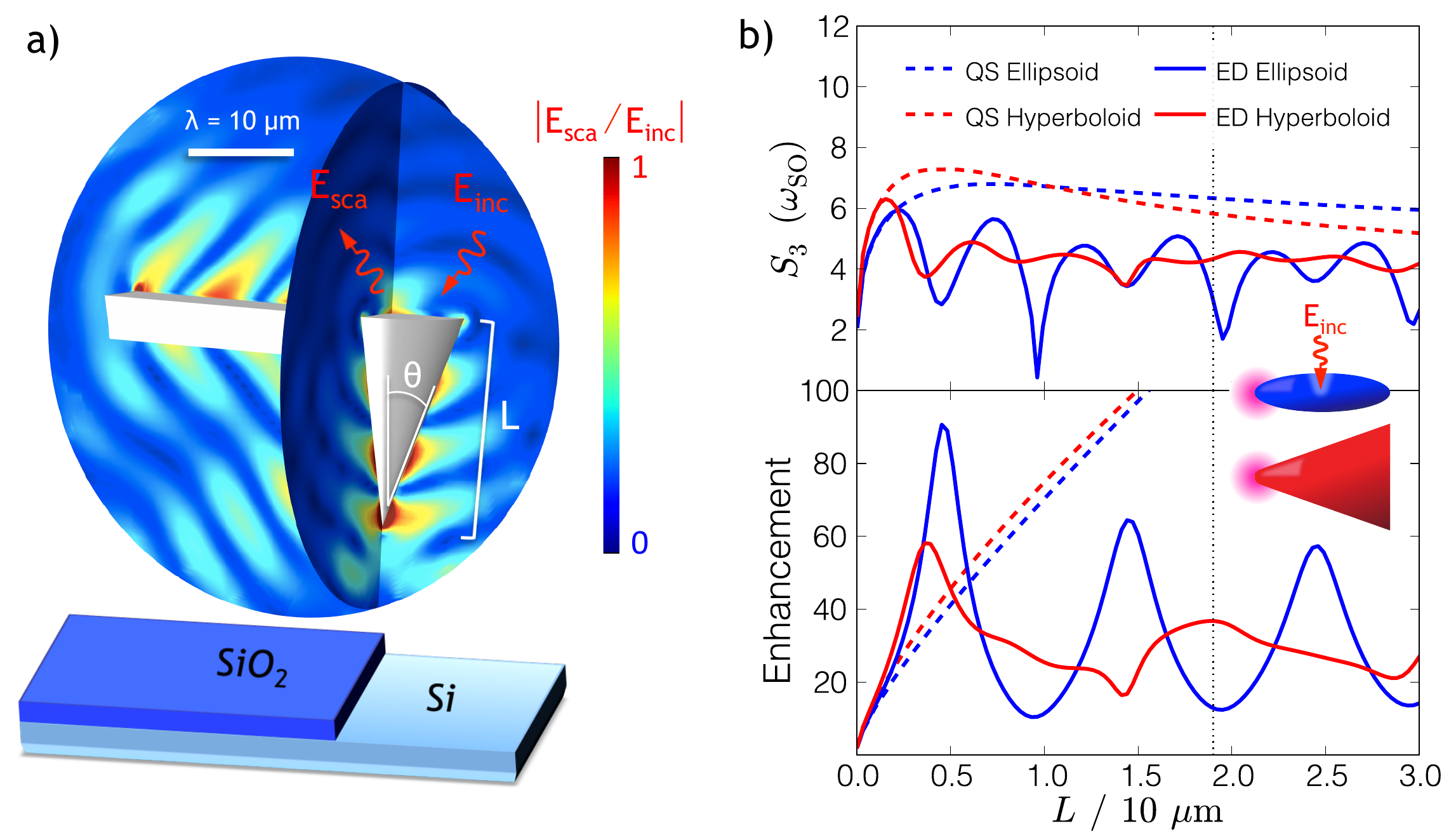}
   \caption{\textbf{a)} Scattered field of a realistic near-field probe geometry under plane wave illumination (incident along the viewing angle) as computed by the finite-element method.  Oscillatory fields near the tip apex are associated with standing wave-like surface charge densities resulting from field retardation.  \textbf{b)} \textbf{Lower panel:} Field enhancement at the tip apex computed quasi-statically (QS) and electrodynamically (ED) for two probe geometries of varying size illuminated perpendicular to their principle axes.  \textbf{Upper panel:} Near-field $S_3$ contrast between SiO$_2$ at the surface optical phonon resonance ($\omega_\mathrm{\,SO}$) and silicon simulated by the \textit{fully retarded lightning rod model}.  The vertical dashed line indicates the length of a typical near-field probe, $L\approx 19\,\mu$m.}
   \label{fig:3}
\end{figure*}

The probe tapping amplitude is 80 nm in these simulated experiments, and the radius of curvature $a$ at the probe apex (equal to the inverse surface concavity) is held constant at 30 nm, typical of experiments with commercially available near-field probes.  The minimum probe-sample distance is taken as $d=0$ nm throughout (\textit{viz.} physical contact, consistent with the established description of tapping mode AFM).  We describe the thin-film optical response with a momentum-dependent Fresnel coefficient \cite{Kildemo} (further discussed in Sec. \ref{sec:thinfilmsio2}) using optical constants of thermal oxide taken from literature\cite{SiO2}.

The probe response function $\Lambda(q,z)$ is computed in the quasi-static approximation once for each probe geometry according to a simple boundary element method.  Mathematical details are provided in Appendix \ref{app:QSboundaryelement}.  Whereas in this work we present calculations only for ideally conducting metallic probes, Appendix \ref{app:QSboundaryelement} presents also the general formulation suitable for application to dielectric probes.  Consequently, the case of a dielectric probe presents no formal difficulty for the model presented here.  However, previously reported models present a suitably simpler description of the ``weak coupling regime" in which externally excited near-fields may be mapped non-perturbatively \cite{StrongWeakCoupling,MappingNanoantennas,PlasmonsPhase}.  We identify this as the regime in which a perturbation expansion of Eq. \eqref{eq:LRMpolarization} is found to converge, and several terms therein might be summed for a sufficient estimate of near-field scattering.

As shown in Fig. \ref{fig:2}, the most dramatic feature of our quasi-static calculations is the strong variation in normalized scattering amplitude with increasing probe length at the probe-sample polariton resonance.  The implication is worrisome: there is no clear rational choice for ``effective probe length" when computing the strength of probe-sample near-field interaction in the quasi-static approximation.  With a free-space wavelength of light $\lambda \sim 10$ $\mu$m, although the largest credibly quasi-static probe length $L\sim \lambda/10$ (or $L \sim 20 a$) provides reasonable qualitative agreement with data acquired by nanoFTIR under identical experimental conditions (Fig. \ref{fig:sio2data}), quantitative agreement is clearly only attainable \textit{a posteriori}, for example by fitting to agreeable values of $L$.   Furthermore, the extreme dependence on probe geometry exhibited here discredits the utility of quantitative ``fits" to experimental data.   The ill-posed description of near-field coupling afforded by this quasi-static treatment lacks clear predictive power.  We are therefore compelled towards a consistent electrodynamic treatment, which as we will show provides an unambiguous description of near-field interactions with wavelength-scale probes -- a case applicable to the vast majority of near-field experiments at infrared and THz frequencies.

\section{\label{sec:electrodynamics} The electrodynamic case: near-field probe as antenna}
Near-field microscopy is occasionally described as an antenna-based technique, in which the antenna-like near-field probe efficiently converts incident light into strongly confined fields at the probe-sample feed-gap\cite{NovotnyReview,KeilmannAntenna,ResonantAntennas,RaschkeReview}.  The antenna's scattering cross section is consequently modulated through strong interactions with the sample surface to provide the nano-resolved optical contrasts of ANSOM \cite{Raschke,Demodulation}.  At a formal level, these considerations leave the mathematical form of the \textit{lightning rod model} unaltered; nevertheless, the probe response function $\Lambda(q,z)$ must encapsulate the probe's role as an antenna, particularly in the probe's response $\Lambda_0(z)$ to illumination fields.

As for any antenna, due to retardation and radiative effects, the field scattered by a near-field probe is manifestly dependent on both its size relative to the free space light wavelength as well as its geometric profile relative to the incident light polarization.  Such electrodynamic effects have been demonstrated experimentally \cite{Polarization,ResonantAntennas}.  To characterize how these length scales influence the observables of near-field spectroscopy, the full electrodynamic responses of two probe geometries were computed numerically as a function of their overall length $L$ relative to the free-space wavelength of incident light.

A \textit{fully retarded} boundary element method taking account of field retardation and radiative forcing (mathematical details provided in Appendix \ref{app:EDboundaryelement}) was used to calculate charge distributions $\Lambda_0(z)$ induced on metallic ellipsoidal and hyperboloidal probe geometries by incident 10 $\mu$m wavelength light ($\omega=1000$ cm$^{-1}$).  We consider here the hyperboloidal geometry to faithfully reflect the cone-like structure of conventional near-field probes which exhibit a taper angle $\theta \approx 20^{\circ}$ relative to their axis in our experiments.  A similar hyperboloid probe geometry was applied previously by Behr and Raschke to explore plasmonic field enhancements \cite{RaschkeHyperboloid}.  However, their fully analytic treatment necessitates a semi-infinite probe geometry treated in the quasi-static approximation, requiring an unconventional field normalization method to obtain finite values for the probe response.  Their formalism also left back-scattering from the ANSOM probe unexplored.  For our examination, we explore the explicit electrodynamics of probes with lengths between $L=60$ nm (rendering a sphere in the ellipsoidal case) and 30 $\mu$m, with the apex curvature radius held constant at $a=30$ nm.

The axisymmetry favored by the \textit{lightning rod model} was maintained throughout these calculations by approximating plane wave illumination by an inwards-propagating cylindrical field bearing a local phase velocity angled towards the tip apex at 60$^\circ$ from the probe axis (see Appendix \ref{app:EDboundaryelement}).  Validity of this axisymmetric approximation was confirmed through comparison of resultant surface charge density profiles with those predicted by full finite-element simulations (\textit{Comsol Multiphysics}), consisting of a realistic metallic probe geometry ($\theta = 20^{\circ}$ and $L=19$ $\mu$m) including AFM cantilever, subject to plane-wave illumination.  Differences in charge density were found to be negligible within microns of the tip apex, suggesting the robustness of key near-field parameters to fine details of the extended probe geometry.  Fig. \ref{fig:3}a displays finite-element predictions for the magnitude of the probe's scattered field $\vec{E}_{sca}$ illustrating the characteristically standing wave-like pattern of charge density along the probe's conical surface, a consequence of field retardation.

\begin{figure*}[t!]
\centering
   \includegraphics[width=.85 \textwidth]{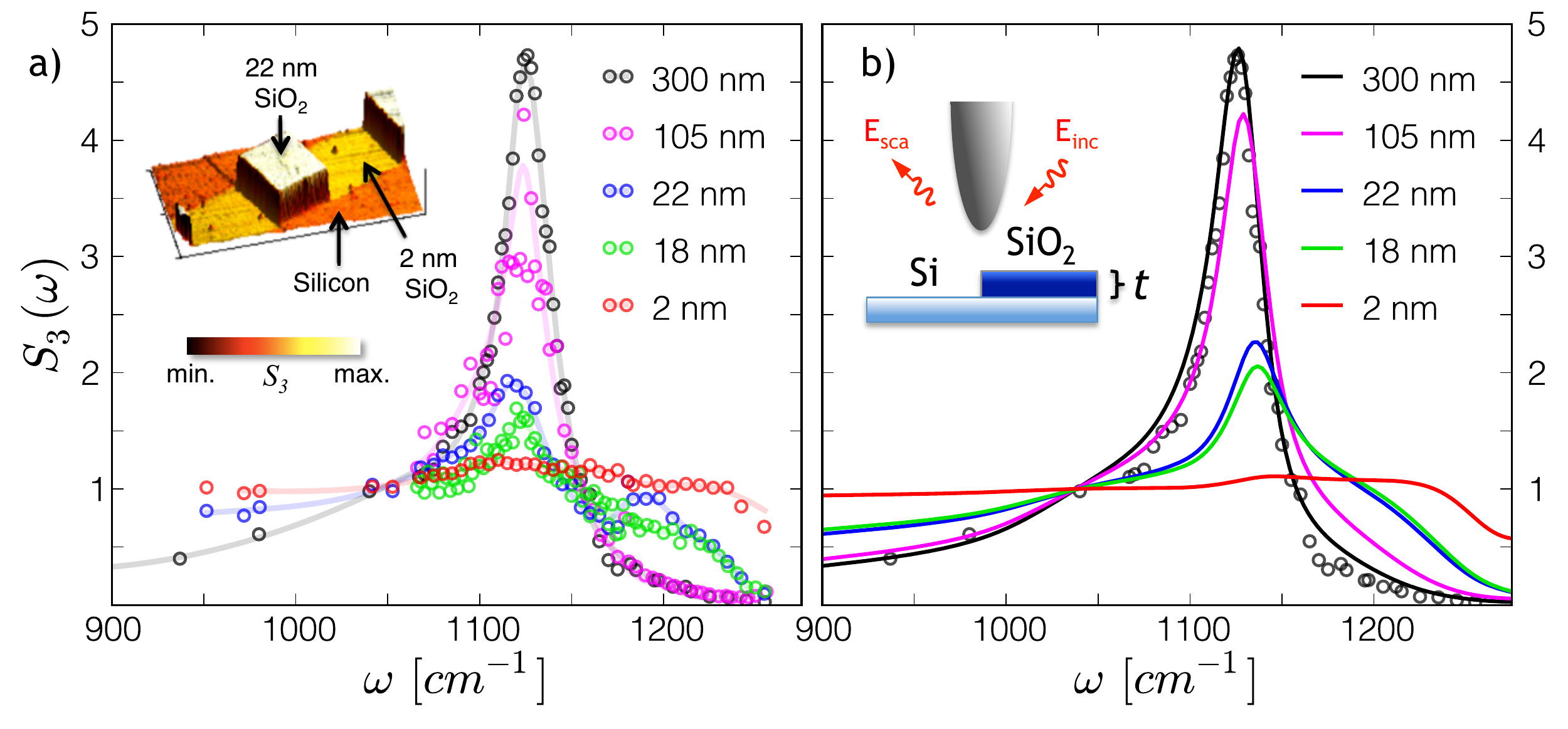}
   \caption{\textbf{a)} Near-field response of SiO$_2$ thin films etched to varying thicknesses on a silicon substrate measured by tunable QCL spectroscopy and normalized to silicon\cite{Zhang} (see text).  The faint curves are provided as guides to the eye.  \textbf{Inset:} Sample near-field signal $S_3$ at $\omega=1130$ cm$^{-1}$ overlaid on simultaneously acquired AFM topography. \textbf{b)} Near-field $S_3$ spectra predicted by the \textit{lightning rod model} using optical constants from literature \cite{SiO2}.  Data points from the 300 nm film are superimposed for point of comparison.  Our model captures the key features of the data; we infer that discrepancies with ultra-thin film data result from substantial variations in optical properties.}
   \label{fig:thinfilmsio2}
\end{figure*}

The resultant field enhancement at the probe apex in the absence of a sample calculated by our fully retarded boundary element method is shown in Fig. \ref{fig:3}b (lower panel) in comparison with the quasi-static case, demonstrating several key phenomena:  First, quasi-static probe geometries exhibit field enhancements that increase monotonically with the geometric ``sharpness" $L/a$ due to the electrostatic lightning rod effect, originating the divergent quasi-static near-field contrast displayed in Fig. \ref{fig:2}.   Second, at lengths $L=m\, \lambda/2$ for odd integers $m \ge 1$, the electrodynamic ellipsoidal probe exhibits resonant enhancement, whereas minima are observed  for $m$ even.  These features signify antenna modes with antisymmetric and symmetric surface charge densities \cite{Biagioni}, respectively, such as those experimentally characterized among similarly elongated near-field probe geometries \cite{ResonantAntennas}.  Due to the axially polarized incident field, Resonant enhancement modes of the hyperboloidal probe are less pronounced and more complicated in character; we discuss them here in no further detail.  Finally, it is clear that quasi-static predictions depart from their electrodynamic counterparts near a probe length $L \sim \lambda/10$, precisely where quasi-static approximations might be expected to falter lacking the antenna enhancement mechanism.  Field retardation halts subsequent increases in enhancement from the quasi-static lightning rod effect, conferring a practical limit to realistically attainable near-field enhancements outside the plasmonic regime.

Similarly, the onset of antenna modes is expected to modulate the intensity of frequency-dependent back-scattered radiation from wavelength-scale near-field probes, opening the possibility to optimally enhance absolute near-field signals through application-driven design of novel probe geometries.  However, the need for a broadband and normalizable probe response is equally crucial for spectroscopy applications \cite{Campanile}.  A typical infrared near-field spectroscopy experiment involves normalizing acquired signals to a reference material that exhibits a nominally flat optical response (\textit{e.g.} gold or undoped silicon) in order to remove the influence of instrumental sensitivities\cite{PhononEnhancement,Zhang,HuthNanoFTIR}, including the probe's frequency-dependent antenna response.  Normalizability of this response is typically assumed, but we predict here for the first time the breakdown of this assumption in the vicinity of strong antenna resonances.

Fig. \ref{fig:3}b (upper panel) displays the result of fully electrodynamic \textit{lightning rod model} predictions for the near-field signal $S_{3\,\mathrm{SO}}$ obtained at the peak probe-sample resonance frequency ($\omega \approx 1130$ cm$^{-1}$) induced by the strong SiO$_2$ surface optical phonon, normalized to the signal from silicon.  Whereas an \textit{increase} in \textit{absolute} back-scattered signal is expected near the onset of a (radiative) antenna mode, this evidently accompanies a remarkable \textit{decrease} in \textit{relative} material contrast.  The effect results from strong cross-talk between the implicit probe response coincident with that of a resonant sample.

The explanation becomes clear when considering that an antenna's resonance can be strongly detuned by its dielectric environment.\cite{DielectricAntennaLoading,PhotoexcitedNanoantenna}  The point dipole model (Eq. \eqref{eq:PDP}) admits interpretation as a dipole interacting with its mirror image projected from the sample surface.  Extending this interpretation to an antenna, the electrodynamic system consists then not of a single antenna, but of a coupled antenna-mirror pair, and it is well established that coupling an antenna with an exact mirror copy induces a resonance red-shift\cite{SchuckBowties,PabloAntenna}.  Whereas the mirror coupling scales with the inverse dimer gap size in the case of physical antenna pairs, this coupling scales with $r_p$ in our case, and could be appropriately considered a case of \textit{dielectric loading}.\cite{MappingNanoantennas,DielectricAntennaLoading}  

Consequently, an $\mathrm{SiO}_2$ film is expected to detune antenna resonances more strongly at $\omega_\mathrm{SO}$ than a $\mathrm{Si}$ substrate, rendering their respective probe back-scattering signals potentially incomparable even when collected at the same frequency, since there is no clear way to normalize out the effect.  Stated another way, interaction with a resonant sample can not only \textit{enhance} the strength of a probe's antenna mode, it can \textit{modify} the antenna behavior outright, driving the probe towards a regime of destructive radiative interference.  Normalized $S_3$ signals calculated for the electrodynamic ellipsoid (Fig. \ref{fig:3}a, solid blue line) therefore resemble a quotient of two resonance functions, oscillatory but shifted versus the light frequency relative to one another.  For this extreme case, we might conclude that fluctuations observed in the frequency-dependent near-field signal radiated from the probe could associate more with variable dielectric loading of the antenna response than with genuine near-field contrast.

Antenna detuning is considerably moderated in the case of the hyperboloidal probe, whose normalized near-field response at frequencies $\lambda < L$ exhibits much weaker dependence on the probe length (or, complementarily, on probing frequency).  The normalization procedure therefore appears sufficient for systematic removal of the probe sensitivity at the 20\% level in the absence of strong antenna resonances.  Furthermore, given the clear asymptotic character of near-field contrast for the broadband hyperboloidal probe, it would appear acceptable to quantitatively model normalized near-field signals from such a probe geometry using electrodynamic charge distributions $\Lambda(q,z)$ computed only for a \textit{single} characteristic frequency.  In the case of weak antenna resonances, this renders implementation of the \textit{fully retarded lightning rod model} no more complex than the quasi-static version.  Therefore, all following calculations presented in this work are electrodynamic and calculated in this fashion unless otherwise indicated.

Nevertheless, this examination tells a cautionary tale concerning the use of strongly resonant probes\cite{ResonantAntennas} for quantitative near-field spectroscopy, wherein convolution of the probe's antenna response may not be easily removed.  However, the resonant enhancement of back-scattered fields by $L\sim \lambda/2$ probes can provide a fortunate trade-off, with encouraging applications to resonantly enhanced THz near-field imaging experiments.

\section{Momentum-dependent light-matter coupling} \label{sec:thinfilmsio2}
To test the \textit{lightning rod model} description of systems exhibiting explicit momentum-dependent light-matter coupling, we consider a thin film of phonon-resonant SiO$_2$ on silicon substrate.  The film thickness $t$ introduces a characteristic length scale to the sample geometry, associated with a characteristic crossover momentum $q\sim t^{-1}$.  Incident evanescent fields exceeding this momentum are reflected much as though bulk SiO$_2$ were present, whereas lower momentum fields can penetrate the film and reflect from the substrate \cite{Zhang}.  With the \textit{lightning rod model} we consider this momentum dependence exactly and directly compare its predictions to near-field spectroscopy measurements performed using the experimental apparatus described in Appendix \ref{sec:setup}.

Mid-infrared near-field images of SiO$_2$ thin films of varying thickness were acquired with a tunable QCL at a probe tapping amplitude of 50 nm, taking signal from the underlying silicon substrate for normalization (Fig. \ref{fig:thinfilmsio2}a).  These data were first presented in an earlier work\cite{Zhang}.  Controlled film thicknesses were produced through selective etching (\textit{NT-MDT Co.}) and confirmed by AFM height measurements acquired simultaneously with the collection of near-field images.  Spectroscopy was obtained from area-averaged near-field contrast levels.

\begin{figure}[t!]
\centering
   \includegraphics[width=.47\textwidth]{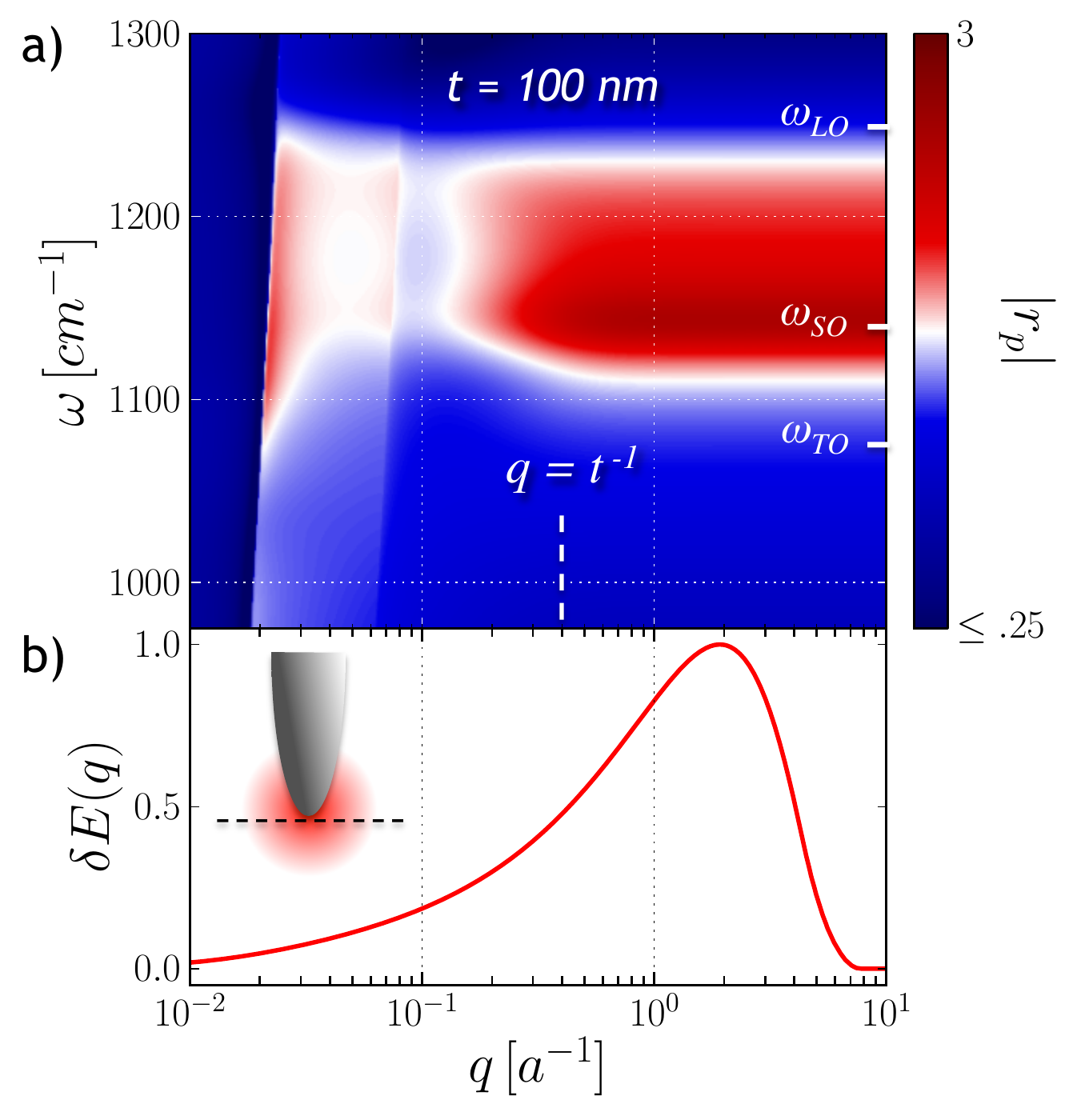}
   \caption{\textbf{a)} Example of strong surface optical phonon dispersion for a 100 nm thick SiO$_2$ film on silicon computed by the Fresnel reflection coefficient $r_p(q,\omega)$ (Eq. \eqref{eq:thinfilmrp}).  \textbf{b)} The momentum-dependent distribution of electric fields at the sample surface (dashed line) calculated by the \textit{lightning rod model} at the tip-sample phonon resonance for a conical tip in full contact.}
   \label{fig:momentumdependence}
\end{figure}

\begin{figure*}[t!]
\centering
   \includegraphics[width=\textwidth]{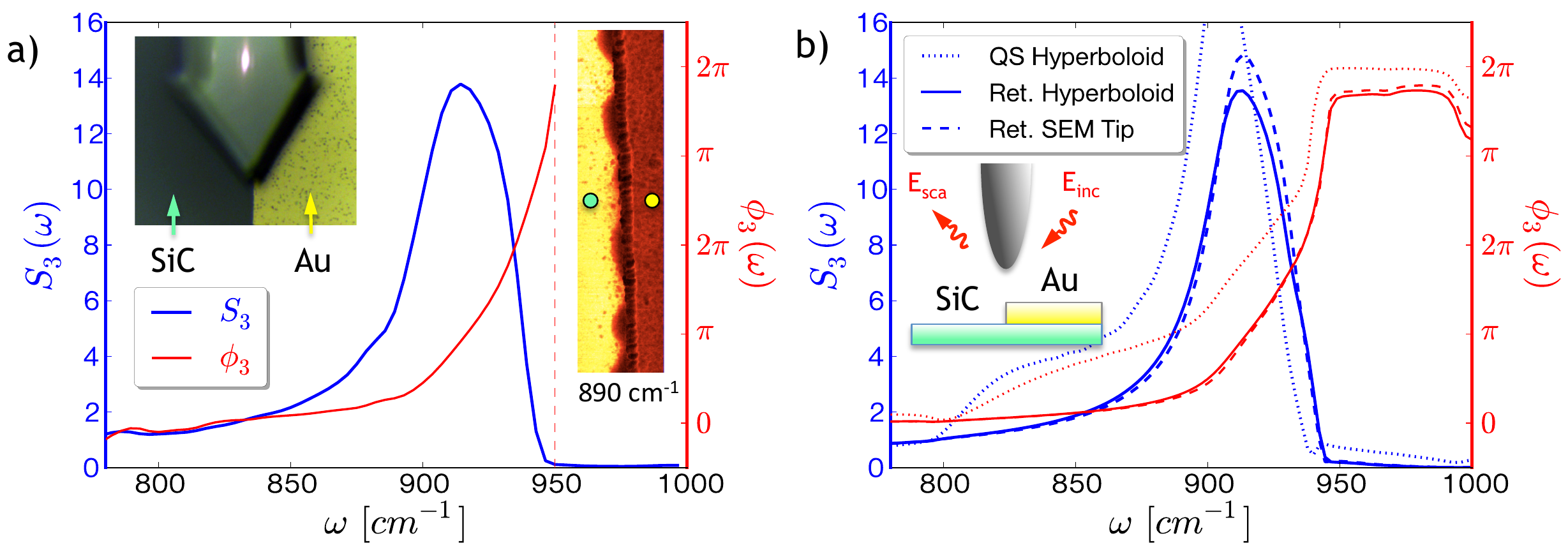}
   \caption{\textbf{a)} Amplitude $S_3$ and phase $\phi_3$ of the back-scattered near-field signal from a 6H SiC crystal, measured in the vicinity of the surface optical phonon and referenced to a surface-deposited gold film, as obtained in a single acquisition by nanoFTIR. \textbf{Left inset:} Visible light image (width 60 $\mu$m) above the near-field probe at the SiC/gold interface. \textbf{Right inset:} PSHet near-field $S_3$ image (width 1 $\mu$m) of the interface with sample and reference nanoFTIR locations indicated.  \textbf{b)} Lightning rod model predictions for near-field signal from SiC using optical constants measured by in-house ellipsometry.  While insensitive to details of probe geometry (see text), fully retarded (\textit{Ret.}) calculations provide superior agreement to the experimental spectra than the quasi-static (\textit{QS}) approximation.}
   \label{fig:sic}
\end{figure*}

Momentum-dependent Fresnel reflection coefficients were used to describe these systems\cite{Kildemo} and provided to the \textit{lightning rod model} in order to predict spectroscopic near-field contrast:
\begin{align}
r_p(q,\omega)&=\frac{\rho_1+\rho_2\,e^{2 i k_{z,1}\, t}}{1+\rho_1 \rho_2\,e^{2 i k_{z,1}\,t}} \label{eq:thinfilmrp} \\
\text{with} \quad \rho_i &\equiv \frac{\epsilon_{i}\,k_{z,i-1}-\epsilon_{i-1}\,k_{z,i}}{\epsilon_{i}\,k_{z,i-1}+\epsilon_{i-1}\,k_{z,i}}, \nonumber \\
\text{and} \quad k_{z,i} &\equiv \sqrt{\epsilon_i\, (\omega/c)^2-q^2}. \nonumber
\end{align}
Here numeric subscripts $0,1,2$ correspond with air, SiO$_2$, and silicon, respectively, $\epsilon_i$ denotes the complex frequency-dependent dielectric function of the relevant material (ellipsometric optical constants for thermal oxide taken from literature \cite{SiO2}), and $t$ denotes the oxide film thickness.

Lightning rod model predictions are presented in Fig. \ref{fig:thinfilmsio2}b for comparison with measured data.  Agreement is superior to that of the simple dipole model and at least as good as earlier quasi-static predictions with an ad-hoc probe geometry \cite{Zhang}.  In contrast to the prediction of a blue-shifting phonon resonance with decreasing film thickness (originating entirely in the Fresnel formula Eq. \ref{eq:thinfilmrp}), experimental data indicate a slight red-shift among ultra-thin films.  This discrepancy should not be counted against our model: although identical optical constants were employed for predictions at all film thicknesses, a growing body of experimental and \textit{ab initio} evidence suggests legitimate phonon confinement effects can modify the intrinsic optical properties of nanostructured samples\cite{Giustino}.  Clear discernment of these effects by near-field spectroscopy opens the possibility for quantitatively evaluating the optical properties of nanostructures that exhibit and utilize \textit{bona fide} three-dimensional confinement\cite{PhononConfinement}.

A clear physical description of the depth sensitivity exhibited in Fig. \ref{fig:thinfilmsio2} proves just as valuable as quantitative agreement.  The onset of a dramatic decrease in near-field signal at the phonon resonance near $t \sim a$ can be understood on the basis of the momentum decomposition of electric fields emitted by the near-field probe.  A straightforward analysis building on Eq. \eqref{eq:LRMcharge} reveals the following decomposition for probe-generated electric fields by their momenta in the plane of the sample (the basis given by Eq. \eqref{eq:qbasis}):
\begin{gather}
\delta E(q_i)\,/\,E_\mathrm{inc}=\left[\mathbf{\Gamma}_{t\rightarrow s}\, \mathbf{\Lambda} \mathbf{G}\, \frac{\vec{\Lambda}_0}{\mathbf{I}-\mathbf{\Lambda} \mathbf{G}} \right]_i\,\delta q\label{eq:tipfields} \\
\text{with}\quad \Big[\mathbf{\Gamma}_{t\rightarrow s}\Big]_{ij} \equiv e^{-q_i d}\,\delta_{ij}, \nonumber
\end{gather}
where $d$ is the tip-sample distance and $\delta E(q)/\delta q$ is understood in the sense of a distribution function.

Fig. \ref{fig:momentumdependence}b displays $\delta E(q)$ calculated on resonance with the SiO$_2$ phonon in comparison with the example dispersion of a 100 nm SiO$_2$ film on silicon shown in Fig. \ref{fig:momentumdependence}a. The surface optical phonon is evident at $\omega_{SO}$, characteristically centered in the \textit{Restrahlen} band between the transverse optical ($\omega_{TO}$) and longitudinal optical ($\omega_{LO}$) phonon frequencies.  Given that our SiO$_2$ forms an amorphous layer, indication of these phonon frequencies is approximate.  Nanoscale thickness introduces considerable momentum dependence in the regime relevant to probe-sample near-field interactions ($q\sim a^{-1}$), effecting a strong phonon response only for momenta $q>t^{-1}$ as mentioned earlier.  The spectroscopic character of the probe-sample near-field response can therefore be inferred from the momentum-space integral of $\delta E(q) \times r_p(q,\omega)$.  Note however the explicit $r_p$- and $d$-dependence of $\delta E(q)$ by way of $\mathbf{G}$ in Eq. \eqref{eq:tipfields} amounts to a near-field response strongly super-linear in the sample's intrinsic surface response.

\section{The strongly resonant limit: silicon carbide} \label{sec:SiC}
We can critically evaluate the generality of the \textit{lightning rod model} formalism through comparison with measurements of crystalline SiC, a strongly resonant material in the mid-infrared owing to an exceedingly strong surface optical phonon at $\omega \approx 950$ cm$^{-1}$.  Here we find the limit at which contingent assumptions for alternative near-field models \cite{PureOpticalContrast,HuthNanoFTIR,Valle,FDPInverse,Esslinger} are expected to break down, since resonant materials can interact non-perturbatively along the entire length of the near-field probe.  This breakdown signals the importance of both probe geometry and field retardation effects.  Lacking these considerations, previous models have dramatically overestimated the near-field contrast generated by SiC \cite{SiCNearField,RaschkeReview}, leaving the estimation of optical properties through quantitative analysis of near-field observables quite ambiguous.

Fig. \ref{fig:sic} displays quantitative agreement between newly presented nanoFTIR spectroscopy of a 6H SiC crystal and \textit{lightning rod model} predictions.  Asymmetry in the observed phonon-induced probe-sample resonance spectrum mimics that of the underlying surface response function $\beta(\omega)$.  To ensure unambiguous comparison between experiment and theory, uniaxial optical constants of our crystal were directly determined by in-house infrared ellipsometry and were found consistent with literature data for similar crystals \cite{SiC}.  A 100 nm gold film was deposited onto the crystal surface to provide a normalization material for nanoFTIR measurements, which were conducted at 60 nm tapping amplitude across the SiC-gold interface.  The right inset of Fig. \ref{fig:sic}a displays strong interfacial contrast in near-field amplitude measured across the interface by pseudoheterodyne (PSHet) imaging \cite{PSHet} with a CO$_2$ laser tuned to 890 cm$^{-1}$, with nanoFTIR acquisition positions indicated.  As confirmed by nanoFTIR, near-field resonance with the SiC surface optical phonon produces a stronger signal than gold across a considerable energy range, 800-940 cm$^{-1}$.  Such strong near-field resonances enable potential technological applications for guiding and switching of confined infrared light within nanostructured polar crystals, as suggested in related work \cite{Phononics}.

Predicted spectra presented in Fig. \ref{fig:sic}b reveal that explicit consideration of field retardation effects according to the findings of Sec. \ref{sec:electrodynamics} (spectra labeled \textit{Ret.}) significantly improves quantitative agreement with experimental spectra in contrast to the quasi-static prediction (labeled \textit{QS}), which drastically overestimates the near-field contrast of SiC up to a factor of 20 over gold.  The \textit{QS} curve additionally reflects an excessive red-shift of the probe-sample resonance peak on account of the overly predominant low-momentum phonon excitations permitted in the quasi-static approximation; these reside at lower energy due to the typical positive group velocity of surface phonon polaritons.  We furthermore explored the influence of particular probe geometries on the predicted near-field spectrum by employing charge distributions $\Lambda(q,z)$ calculated for both the ideal hyperboloidal probe geometry as well as for the actual profile of an used probe tip, obtained from an SEM micrograph (displayed as the blue curve in Fig. \ref{fig:1}a).  The Fig. \ref{fig:sic}b comparison of SiC spectra predicted with these geometries reveals that only essential features of the probe geometry, such as the overall conical shape and taper angle ($\theta \approx 20^{\circ}$) shared by both, are relevant for predicting near-field contrasts at the 10\% level of accuracy.  Further quantitative refinements to near-field spectroscopy will therefore benefit from the standardization of reproducible probe geometries\cite{TipSharpening}.

\section{Nano-resolved extraction of optical constants}
\begin{figure}[t!]
\centering
   \includegraphics[width=.5\textwidth]{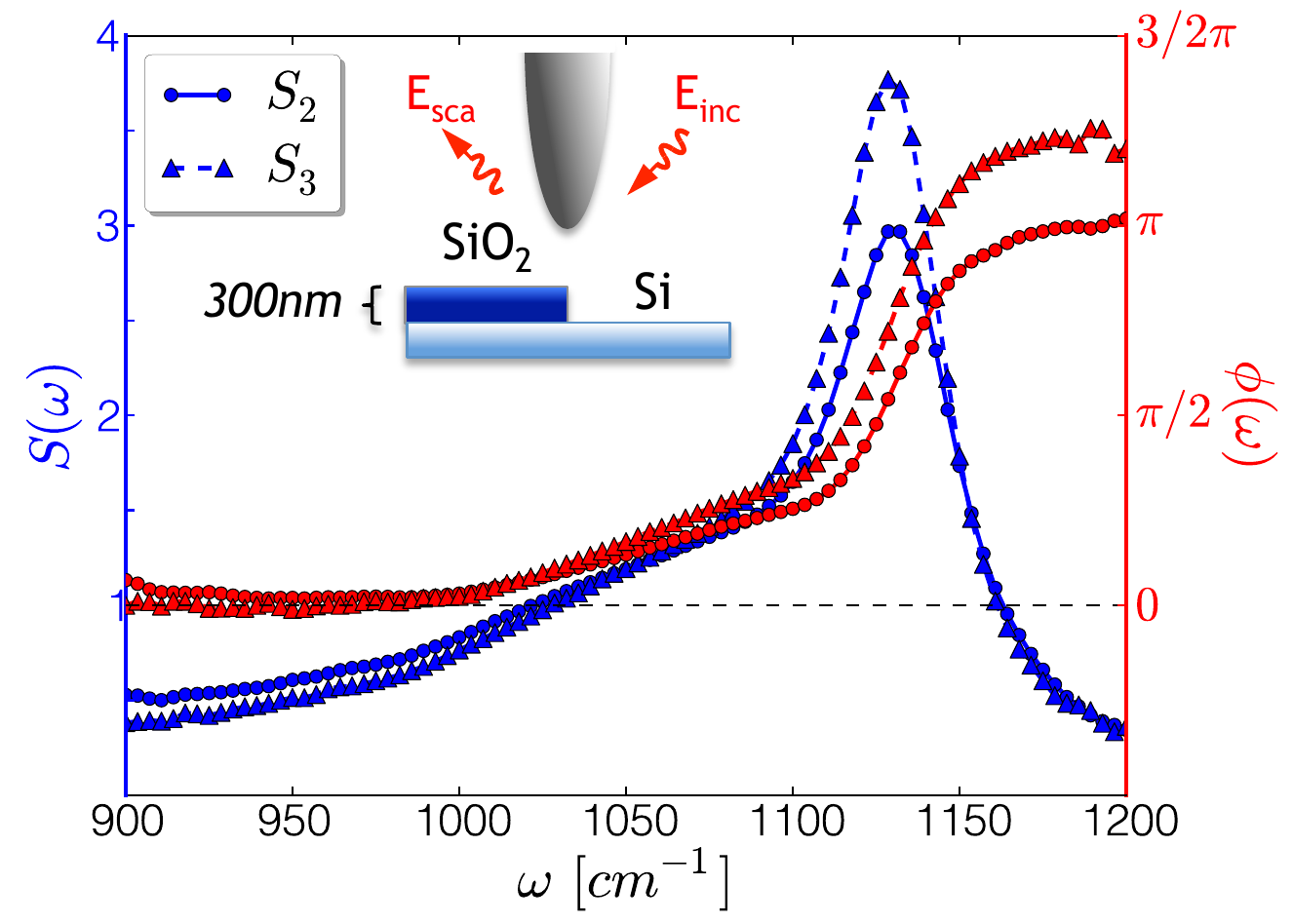}
   \caption{Amplitude $S$ and phase $\phi$ of the back-scattered near-field signal from a 300 nm SiO$_2$ film, measured in the vicinity of the surface optical phonon and referenced to the silicon substrate, as obtained in a single acquisition by nanoFTIR.}
   \label{fig:sio2data}
\end{figure}

Systematic improvements in the light sources and detection methods available for near-field spectroscopy now enable sufficiently high signal-to-noise levels and fast acquisition times for routine, reproducible measurements \cite{AmarieNanoFTIR,BrehmNanoFTIR}.  Fig. \ref{fig:sio2data} displays newly presented nanoFTIR measurements acquired on a 300 nm SiO$_2$ film with silicon used for normalization, displaying both the amplitude $S$ and phase $\phi$ of the probe's back-scattered radiation demodulated at the $2$nd and $3$rd harmonics of the probe frequency, collected at 60 nm tapping amplitude.  Such broadband data are ideally eligible for the quantitative extraction of SiO$_2$ optical constants in the vicinity of the transverse optical phonon ($\omega_{TO}\approx 1075$ cm$^{-1}$).

Using the \textit{lightning rod model}, a method requiring minimal computational effort was developed to solve the inverse problem of near-field spectroscopy, proceeding as follows:  The connection between optical properties of a sample material (\textit{e.g.} the complex dielectric function, $\epsilon=\epsilon_1$ +i$\epsilon_2$) and near-field observables (\textit{e.g.} $S$ and $\phi$, or equivalently the real and imaginary parts of the complex back-scattered signal $s=s_1+i s_2$ at a given harmonic $n\ge2$) is described by a smooth map $\mathrm{NF}: \mathbb{C}\rightarrow \mathbb{C}$, with $\mathbb{C}$ the set of complex numbers.  A ``trajectory" $s(\omega)$ through the space of observable near-field signals therefore corresponds to a trajectory $\epsilon(\omega)$ through the space of possible optical constants.  The uniqueness of this correspondence was confirmed for bulk and layered sample geometries by computing $s=\mathrm{NF}(\epsilon)$ across the parameter range of interest for real materials ($\epsilon_2>0$) and ensuring local invertibility of the map, conditional on the determinant of the Jacobian matrix of $\mathrm{NF}$:
\begin{equation}
\left|J(\epsilon_1,\epsilon_2)\right|>0 \quad \text{with} \quad J(\epsilon_1,\epsilon_2)=\frac{\partial(s_1,s_2)}{\partial(\epsilon_1,\epsilon_2)}.
\end{equation}

\begin{figure*}[t!]
\centering
   \includegraphics[width=.8 \textwidth]{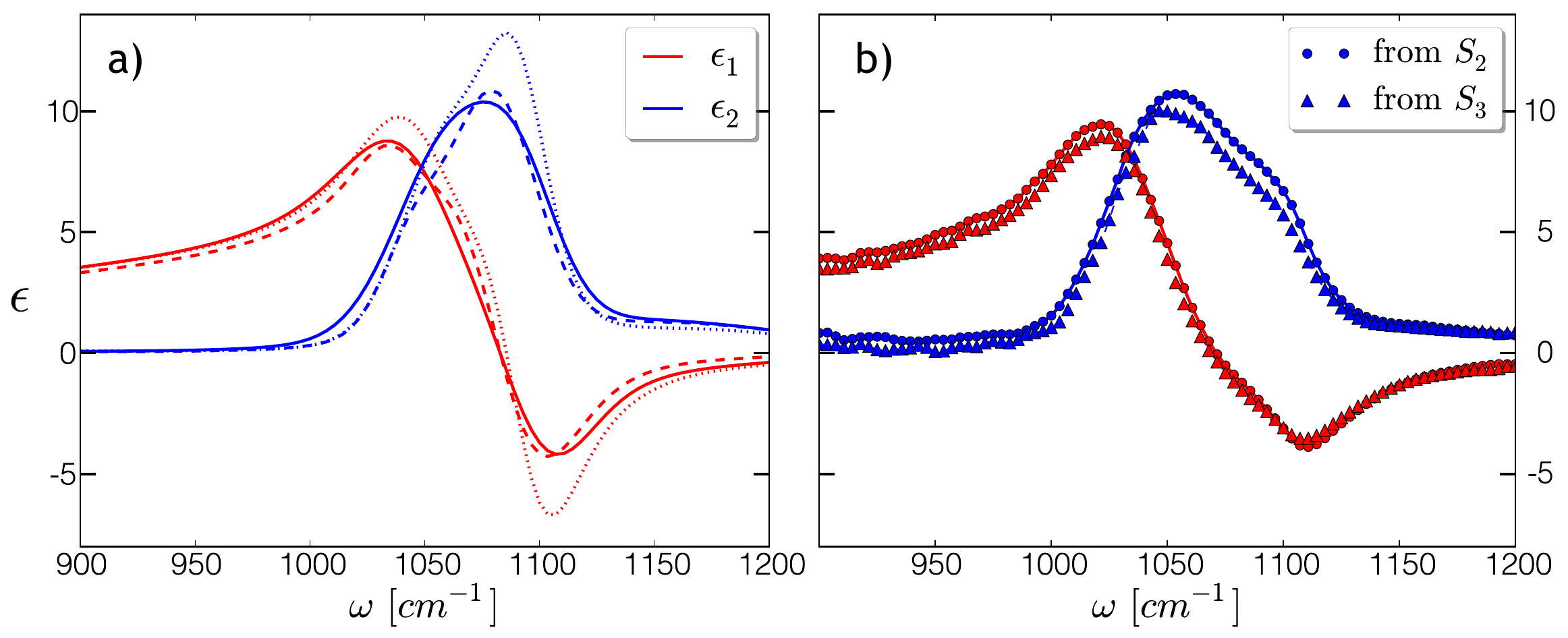}
   \caption{\textbf{a)} Typical variation in optical constants among thermal oxide thin films taken from literature ellipsometry.  Pairs of red and blue curves with identical line style are associated with distinct thin film samples.\cite{SiO2}  \textbf{b)} Optical constants of a 300 nm SiO$_2$ film extracted from near-field spectra $S_2(\omega)$ and $S_3(\omega)$ following the method of Eq. \eqref{eq:dampedoscillator}.}
   \label{fig:opticalconstantextraction}
\end{figure*}

Because parameters internal to the operation of $\mathrm{NF}$ are frequently variable (\textit{e.g.} sample thickness, tip radius of the probe, tapping amplitude), instead of establishing the inverse map $\mathrm{NF}^{-1}$ as a ``look-up table" by brute computation, we instead introduce a method for nucleated growth of the trajectory $\epsilon(\omega)$ which optimizes consistency with the forward mapping $s=\mathrm{NF}(\epsilon)$ beginning at some initial frequency $\omega_0$.  We re-imagine the problem as a particle navigating $\epsilon$-space under the influence of external forces penalizing displacements $\delta s$ from measured signal values $s(\omega)$.  The trajectory $\epsilon(\omega)$ for such a particle solves, for example, the equation of motion for a damped harmonic oscillator equilibrating to $s=\mathrm{NF}(\epsilon)$:
\begin{gather}
\frac{d^2}{d\omega^2} \delta s+2 \zeta\, \Omega \, \frac{d}{d\omega} \delta s+\Omega^2\,\delta s=0 \label{eq:dampedoscillator} \\
\text{with}\quad \delta s(\omega) \equiv s(\omega)-\mathrm{NF}\big(\epsilon(\omega)\big). \nonumber
\end{gather}
Here $\zeta$ denotes a damping constant tuned to induce critical damping ($\zeta$=1), and $\Omega$ is a force constant ensuring decay to equilibrium over an interval $\delta \omega=2 \pi/\Omega$ comparable to the frequency resolution of measurement.  This equation of motion enables adiabatic tracking of experimentally observed signal values while both penalizing deviations $\delta s$ and dissipating their energy.  Eq. \eqref{eq:dampedoscillator} may alternatively be parametrized by an auxiliary independent variable $x$ for which $\omega(x)$ increments only when $\left|\delta s(\omega)\right|<\delta s_\mathrm{thresh}$, a threshold value ensuring system equilibration arbitrarily close to the measured signal value at each $\omega$.  This also ensures solutions to Eq. \eqref{eq:dampedoscillator} are relatively insensitive to the ``guessed" initial condition $\epsilon(\omega_0)$, amounting to a robust relaxation method.

Our inversion of measured $s(\omega)$ consists of numerically solving Eq. \eqref{eq:dampedoscillator} for $\epsilon$ by finite difference techniques \cite{Kiusalaas}.  This requires at least five evaluations of $\mathrm{NF}$ per $\omega$- or $x$-step in order to estimate local first and second derivatives of $\mathrm{NF}$ with respect to real and imaginary parts of $\epsilon$.  Although consequently the procedure is more computationally costly than forward evaluation by the \textit{lightning rod model}, it is at least as efficient in principle as global nonlinear least-squares methods (\textit{e.g.} Levenberg-Marquardt \cite{Marquardt}) and often considerably faster, furthermore requiring no \textit{a priori} knowledge for the form of the fitting function.  This is considerably advantageous in cases where spectra are not available in a sufficiently wide frequency range to permit well-determined fitting to $\epsilon(\omega)$ by Kramers-Kronig-consistent oscillators\cite{Kuzmenko}.

We applied our inversion technique to the spectroscopic data displayed in Fig. \ref{fig:sio2data} by parametrizing $\mathrm{NF}$ with the reflection coefficient of an ``unknown" 300 nm layer (film thickness determined by AFM) on silicon substrate.  For mapping the film's optical constants $\epsilon_\mathrm{film}(\omega)=\epsilon_1(\omega)+i\,\epsilon_2(\omega)$ to a measurable, normalized near-field spectrum $s_n(\omega)$, the form for $\mathrm{NF}$ used here is that given by the \textit{lightning rod model}, namely:
\begin{gather}
\mathrm{NF}\big(\epsilon_1(\omega),\epsilon_2(\omega)\big)=s^\mathrm{\,film}_n(\omega) / s^\mathrm{Si}_n
\end{gather}
with
\begin{align}
s^\mathrm{\,film}_n(\omega) &= \int_{-\pi/\Omega}^{\pi/\Omega} dt\,\sin \left(n\Omega t \right) E_\mathrm{\,rad}^\mathrm{\,film}(d,\omega), \\
E_\mathrm{\,rad}^\mathrm{\,film}(d,\omega)&=\vec{e}_\mathrm{rad} \cdot \mathbf{G}_\mathrm{\,film}(d,\omega) \, \frac{\vec{\Lambda}_0}{\mathbf{I}-\mathbf{\Lambda} \mathbf{G}_\mathrm{\,film}(d,\omega)},
\end{align}
and $d=A \left(1+\sin\left(\Omega t\right) \right)$.  Describing the near-field response of the film, $\mathbf{G}_\mathrm{\,film}(d,\omega)$ is given by Eq. \eqref{eq:G} in terms of the film reflection coefficient $r_p^\mathrm{\,film}(q,\omega)$, which is in turn a function of $\epsilon_\mathrm{film}(\omega)$ via Eq. \eqref{eq:thinfilmrp}.  The silicon normalization signal $s^\mathrm{Si}_n$ is computed analogously, but using the reflection coefficient for a bulk surface with frequency-independent dielectric constant $\epsilon_\mathrm{Si} \approx  11.7$.  All other parameters are defined as detailed in Sec. \ref{sec:model}.

 In Fig. \ref{fig:opticalconstantextraction} we present the favorable comparison of our extracted $\epsilon_\mathrm{film}(\omega)$ with typical literature optical constants for three thermal oxide films measured by conventional infrared ellipsometry\cite{SiO2}.  Fig. \ref{fig:opticalconstantextraction}a makes clear the typical variation in optical constants expected among oxide films grown even under nominally fixed conditions.  Furthermore, our extraction technique produced virtually identical output when conducted on both $2$nd and $3$rd harmonic near-field spectra ($s_2(\omega)$ and $s_3(\omega)$), attesting to the internal consistency of the \textit{lightning rod model}.  

Although near-field inversion has been very recently demonstrated on measurements of prepared polymers, the existing technique relies on a polynomial expansion in $\beta$ strictly limited to weakly resonant samples, \textit{vis.} the perturbative limit of Eq. \eqref{eq:LRMpolarization}, and employs a model with tunable \textit{ad hoc} parameters \cite{Cvitkovic,FDPInverse}.  Our procedure removes both shortcomings.  These advantages make Eq. \eqref{eq:dampedoscillator} a suitable technique for the unconditional on-line analysis of near-field spectroscopy data in a diagnostic setting.  Combining for the first time the powerful nanoFTIR instrumentation with a quantitative inversion methodology unlimited by sample characteristics, this procedure makes possible potent new applications of nano-spectroscopy to the quantitative optical study of phase-separated materials\cite{RaschkeReview,Qazilbash} and nano-engineered devices\cite{FeiGraphenePlasmons,ChenGraphenePlasmons}, as well for the nano-resolved chemical identification of structures in biological or surface science applications\cite{AmarieNanoFTIR,HuthNanoFTIR,FDPInverse,AmarieBones}.

\section{Conclusions and Outlook}
The \textit{lightning rod model} provides a general quantitative formalism for predicting and interpreting the experimental observables of near-field spectroscopy.  Simplified descriptions of the probe-sample near-field interaction such as the point dipole model can be obtained as special cases resulting from convenient though unnecessary physical assumptions.  In particular, the choice of effective probe length $L$\cite{Cvitkovic,FDPInverse} was shown to be \textit{ad hoc} in the quasi-static approximation, and consequently susceptible to dubious \textit{a posteriori} fitting to experimental data.

We find a fully electrodynamic treatment renders the effective length construct unnecessary, since field retardation effects modify the distribution of probe charge interacting with the sample.  While this provides a resolution to problems of convergence inherent to the quasi-static treatment, sample-induced dielectric loading of strong antenna resonances (\textit{e.g.} for the long ellipsoidal probe) was found to deceivingly modulate relative material contrasts predicted in the vicinity of sample resonances, such as the surface optical phonon of SiO$_2$, an important caveat and consideration for the rational design of optimized spectroscopic probes \cite{ResonantAntennas}.  Nevertheless, fine details of the probe geometry for realistic conical probe geometries are predicted to impact observable near-field material contrasts at or below a 10\% level of variation.

Using the \textit{fully retarded lightning rod model} with a realistic probe geometry, we obtain quantitatively predictive agreement compared both with tunable QCL near-field spectroscopy of SiO$_2$ films with varying thickness and with newly presented nanoFTIR spectroscopy measurements of the strongly resonant polar material SiC.  This exhibits our model's proper momentum-space description of the probe-sample optical interaction, as well as its suitability for the truly quantitative description of strongly resonant near-field interactions, in contrast with the capabilities or implementations of the alternative models heretofore demonstrated.\cite{SiCNearField,Esslinger,Valle}

Finally, we present a deterministic method to \textit{invert} the \textit{lightning rod model} without recourse to \textit{ad hoc} parameters or over-simplifications.  This rather general technique flexibly solves the inverse problem of near-field spectroscopy at a computational cost significantly lower than exhaustive lookup-tables or oscillator fitting methods, offering exciting opportunities for the on-line interpretation of nano-resolved near-field spectra acquired in a diagnostic setting.  We envision the inverse \textit{lightning rod model} employed quantitatively for deeply sub-wavelength optical studies of naturally or artificially heterogeneous and phase-separated materials, promising further novel applications to systems like energy storage nanostructures,\cite{EnergyNanostructures} transition metal oxide heterostructures,\cite{OxideHeterostructures} and single- or multi-layered graphene plasmonic devices \cite{PhononEnhancement,FeiGraphenePlasmons}.

There remain outstanding challenges for the present model, including its extension to cases where deviations from axisymmetry are crucial, as for \textit{s}-polarization of incident light, or for probe geometries with strong rotational asymmetry.  We envision an expansion of our boundary element methods and of the lightning rod model into basis components with differing rotational ``quantum numbers"\cite{AizpuruaDielectrics} to capture the features of irrotational geometries in a computationally inexpensive fashion.  Furthermore, the explicit application of our model to dielectric probes, particularly in the plasmonic regime, is an undertaking of great potential interest for which the extension of our electrodynamic boundary element (Appendix \ref{app:EDboundaryelement}) to materials of non-negligible skin depth might play a crucial role.  However, even at its present stage the quantitative scattering formalism presented here also lays a solid foundation for the rational analysis and optimization of tip-enhanced optical phenomena in an ever-growing number of exciting experimental applications, including single-molecule Raman spectroscopy\cite{RaschkeVibration,SingleMoleculeRaman} and other novel partnerings of optics with scanning tunneling microscopy.\cite{QiuFluorescence,CrommieInfrared}

\begin{acknowledgments}
This work was supported through NASA grants NNX08AI15G and NNX11AF24G.  Alexander S. McLeod acknowledges support from a U. S. Department of Energy Office of Science Graduate Fellowship.
\end{acknowledgments}

\appendix

\section{Experimental methods} \label{sec:setup}
In the following sections, we apply the \textit{lightning rod model} in comparison with near-field spectra measured for SiO$_2$ thin films and SiC, acquired with the following experimental apparatus.  Infrared nano-imaging and nano-spectroscopy measurements were performed with a NeaSNOM scanning near-field optical microscope (\textit{Neaspec GmBH}) by scanning a platinum silicide AFM probe (PtSi-NCH, \textit{NanoAndMore USA}; cantilever resonance frequency 300 kHz, nominal radius of curvature 20-30 nanometers) in tapping mode over the sample while illuminating with a focused infrared laser beam, resulting in back-scattered radiation modulated at the probe tapping frequency $\Omega$ and harmonics thereof.  In our pseudo-heterodyne detection setup, this back-scattered radiation interferes at a mercury-cadmium-telluride detector (\textit{Kolmar Technologies Inc.}) with a reference beam whose phase is continuously modulated by reflection from a mirror piezoelectrically oscillated at a frequency $\delta \Omega$ ($\approx 300$ Hz). Demodulation of the detector signal at frequency side-bands $n \Omega \pm m\, \delta \Omega$ for integral $m$ supplies the background-free amplitude $S_n$ and phase $\phi_n$ of the infrared signal at harmonics $n$ of the probe's tapping frequency.\cite{Demodulation,GomezPSHet,PSHet}

The super-linear dependence of near-field interactions versus the tip-sample separation distance implies that, in the case of harmonic tapping motion, signal harmonics at $n \ge 2$ are directly attributable to near-field polarization of the tip\cite{Raschke}. Contrasts in near-field signal intensity and phase at these harmonics therefore correspond to variations in local optical properties of the sample\cite{PureOpticalContrast}. Tunable fixed-frequency CW quantum cascade lasers (QCLs, \textit{Daylight Solutions Inc.}) and a tunable CO$_2$ laser (\textit{Access Laser Co.}) were used for imaging and spectroscopy of SiO$_2$ films and SiC, respectively.

NanoFTIR spectroscopy\cite{AmarieNanoFTIR}\cite{BrehmNanoFTIR} was enabled by illumination from a broadband mid-infrared laser producing tunable radiance across the frequency range 700-2400 cm$^{-1}$. This coherent mid-infrared illumination is generated through the nonlinear difference-frequency combination of beams from two near-infrared erbium-doped fiber lasers -- one at 5400 cm$^{-1}$ and the other a tunable supercontinuum near-infrared laser (\textit{TOPTICA Photonics Inc.}) -- resulting in ~100 fs pulses at a repetition rate of 40 MHz. An asymmetric Michelson interferometer with 1.5 millimeter travel range translating mirror enables collection of demodulated near-field amplitude $S_n(\omega)$ and phase $\phi_n(\omega)$ spectra with 3 cm$^{-1}$ resolution.

\section{Resolution of the field from a charged ring into evanescent waves} \label{app:fielddecomposition}
The $xy$-plane Fourier decomposition of the Coulomb field of a point charge $Q$ located at the origin is well known\cite{FieldDecomposition}:
\begin{eqnarray}
\vec{E}(\vec{r})&=&-\frac{Q}{2\pi} \iint_{-\infty}^\infty dk_x\,dk_y\, \left(i\frac{k_x\hat{x}+k_y\hat{y}}{q}+\hat{z}\right) \nonumber \\
	&\;&\quad\quad\quad\quad\quad\quad \times e^{i \left(k_x x+k_y y\right)+q z} \nonumber \\
	&=&-Q \int_0^\infty dq\,q\,\left(J_0(q \rho)\hat{z}+J_1(q\rho)\hat{\rho}\right) e^{q z}
\end{eqnarray}
for $z<0$ and with $q\equiv\sqrt{k_x^2+k_y^2}$.  This decomposition can be applied similarly to a ring of charge with radius $\mathcal{R}$, centered in a plane through the origin with $z$-axis normal:
\begin{eqnarray}
\vec{E}_\mathcal{R}(\vec{r})&=&\frac{Q}{4\pi^2} \int_0^{2\pi} d\phi' \int_{0}^\infty dq \int_0^{2\pi} d\phi \;  \vec{\mathcal{E}}_\mathcal{R}(q,\vec{r},\phi') \nonumber \\
\vec{\mathcal{E}}_\mathcal{R} &\equiv& -\left(\hat{z}+ i\frac{k_x\hat{x}+k_y\hat{y}}{q}\right)\,e^{i q \left(\rho \cos{\phi}-\mathcal{R}\cos{(\phi-\phi')}\right)+q z}. \nonumber
\end{eqnarray}
Here $\phi'$ is an angular integration variable about the circumference of the ring.  We obtain
\begin{equation} \label{eq:chargedring}
\vec{E}_\mathcal{R}(\vec{r})=-Q \int_0^\infty dq\,q\,J_0(q \mathcal{R})\left(J_0(q \rho)\hat{z}+J_1(q\rho)\hat{\rho}\right)e^{q z}.
\end{equation}

The total field is thus a sum of axisymmetric $p$-polarized evanescent waves weighted by the geometry-induced prefactor $q\,J_0(q\mathcal{R})$.  Eq. \eqref{eq:chargedring} constitutes the central result of this section.

\section{Quasi-static boundary element method for an axisymmetric dielectric and conductor} \label{app:QSboundaryelement}
A tractable electrostatic boundary element method applicable to systems of homogeneous dielectrics can be developed as follows \cite{AizpuruaDielectrics}.  Gauss's law constrains the density of bound charge $\rho_b$ at the boundaries between dielectric media as:
\begin{gather}
\nabla \cdot \vec{E} = \nabla \cdot \frac{\vec{D}}{\epsilon}  = 4\pi \rho_b \nonumber \\
	\therefore \quad 4\pi \rho_b=\delta_{S} \left(\frac{1}{\epsilon_2}-\frac{1}{\epsilon_1}\right) \hat{n}_{12} \cdot \vec{D} \label{eq:GaussDielectric}
\end{gather}
Eq. \ref{eq:GaussDielectric} follows in the case that free charge is absent at dielectric boundaries such that $\nabla \cdot \vec{D}=0$, and $\delta_S$ is a surface Dirac delta function associated with the boundary between media of dielectric constant $\epsilon_1$ and $\epsilon_2$, with $\hat{n}_{12}$ the unit vector perpendicular to this boundary oriented from medium 1 to medium 2.  Continuity of the surface normal displacement field across the dielectric interface permits its evaluation at positions $\vec{r}$ along the boundary as a limit taken infinitesimally inside medium 2:
\begin{equation}
\hat{n}_{12} \cdot \vec{D}\left(\vec{r}\right)=\epsilon_2 \lim_{\;t\rightarrow 0^+} -\hat{n}_{12} \cdot \nabla V\left(\vec{r}+t\, \hat{n}_{12}\right). \label{eq:NormalDisplacement}
\end{equation}
The scalar potential $V(\vec{r})$ finds contributions from both incident (external) fields $\vec{E}_\mathrm{inc}$, originating as from distant free charges, as well as from bound charges at the dielectric boundary.  Taking the bound charge $\rho_b$ as the product of a surface density $\sigma_Q$ (distinguished from electrical conductivity $\sigma$) with the surface Dirac delta function, the latter contribution comprises a surface integral on the boundary $S$:
\begin{equation}
V_b(\vec{r})=\int_S dS'\, \frac{\sigma_Q(\vec{r}\prime)}{\left|\vec{r}-\vec{r}\prime\right|}
\end{equation}
Evaluating the discontinuous surface normal electric field $-\hat{n}_{12}\cdot \nabla V_b$ from this contribution involves:
\begin{gather}
\lim_{\;t\rightarrow 0^+} -\hat{n}_{12} \cdot \nabla \left(1/ \left|\vec{r}+t\, \hat{n}_{12}-\vec{r}\prime \right|\right) =2\pi \delta(\vec{r}-\vec{r}\prime)-F(\vec{r},\vec{r}\prime) \nonumber \\
\text{with}\quad F(\vec{r},\vec{r}\prime) \equiv -\frac{\hat{n}_{12} \cdot \left(\vec{r}-\vec{r}\prime\right)}{\left|\vec{r}-\vec{r}\prime\right|^3}. \label{eq:KernelIdentity}
\end{gather}
Gauss's law (Eq. \ref{eq:GaussDielectric}) therefore yields an integral equation in the surface bound charge density $\sigma_Q(\vec{r})$:
\begin{gather}
4\pi \frac{\epsilon_1 \epsilon_2}{\epsilon_1-\epsilon_2} \sigma_Q(\vec{r})=\epsilon_2 \bigg[ \hat{n}_{12}\cdot \vec{E}_\mathrm{inc}(\vec{r})+2\pi \sigma_Q(\vec{r}) \quad\quad \nonumber \\
\hspace{1.4in} -\int_S dS' \,F(\vec{r},\vec{r}\prime)\,\sigma_Q(\vec{r}\prime)\bigg], \label{eq:DielectricBEM1}
\end{gather}
which upon consolidation yields:
\begin{equation}
2\pi \frac{\epsilon_1+\epsilon_2}{\epsilon_1-\epsilon_2} \sigma_Q(\vec{r})=\hat{n}_{12}\cdot \vec{E}_\mathrm{inc}-\int_S dS'\,F(\vec{r},\vec{r}\prime)\,\sigma_Q(\vec{r}\prime). \label{eq:DielectricBEM2}
\end{equation}
Without loss of generality, this equation can be utilized to pre-compute the quasi-electrostatic response of an axisymmetric body of dielectric constant $\epsilon_2$ to incident fields, taking $\epsilon_1=1$ as air, parametrizing the integral kernel $F$ by axial and surface radial coordinates $z$ and $\mathcal{R}_z$, respectively, and expressing $\hat{n}_{12}$ through axial derivatives of the latter.

However, to unambiguously present our method of solution to equations like Eq. \ref{eq:DielectricBEM2} and to promote its application for the description of metallic near-field probes, we confine our attention specifically to the ideally conducting limit, wherein $\epsilon_2$ is divergent.  For Eq. \ref{eq:GaussDielectric} to hold with finite normal displacement in Eq. \ref{eq:NormalDisplacement} therefore requires a vanishing normal gradient of the total potential $V$ just inside the probe surface.  Lacking free or bound charges within its volume, the probe interior and surface therefore reside at constant total potential, signifying zero internal field and perfect screening by the surface:
\begin{equation} \label{eq:QSforcebalance}
V_\mathrm{inc}(\vec{r})+V_{b}(\vec{r}) = V_0\quad  \text{on}\; S.
\end{equation}
This criterion follows equivalently from Eq. \eqref{eq:DielectricBEM2} in the limit $\epsilon_2 \gg \epsilon_1$ through reverse application of Eq. \eqref{eq:KernelIdentity}.

The incident potential of an axisymmetric evanescent field is given in cylindrical coordinates $\rho$, $\phi$, and $z$ by $V_\mathrm{inc}(\vec{r}) = J_0(q \rho)/q\,e^{-q z}$.  The potential $V_{b}$ is generated by the surface charge density $\sigma_Q(\vec{r})$, which may be divided into a continuum of rings, each with charge $dQ=\lambda_Q(z)\, dz$ and radius $\mathcal{R}_{z}$:
\begin{align}
V_{b}(\vec{r})&=\int_S dS'\, \frac{\sigma_Q(\vec{r}\prime)}{|\vec{r}\prime-\vec{r}|} \nonumber \\
	&=\int_{0}^{\infty} dz' \, \Phi(\vec{r},z')\, \lambda_Q(z') \label{eq:Vobj} \\
	\Phi&\equiv \int_{0}^{2 \pi} \frac{d\phi'}{2\pi} \frac{1}{\sqrt{(z'-z)^2+\rho^2+\mathcal{R}_{z'}^2-2\rho \mathcal{R}_{z'} \cos{\phi'}}} \nonumber \\
	&=\frac{2 K(-\frac{4 \,\rho\, \mathcal{R}_{z'}}{(\rho-\mathcal{R}_{z'})^2+(z-z')^2})}
	    	         {\pi \sqrt{(\rho-\mathcal{R}_{z'})^2+(z-z')^2}}.
\end{align}
Here $\Phi$ constitutes the Coulomb kernel for a ring of charge, and $K(\ldots)$ denotes the elliptic integral of the first kind.  Evaluating $V_{obj}$ at the boundary of the object ($\rho =\mathcal{R}_z$) and discretizing $z$ in Eqs. \eqref{eq:QSforcebalance} and \eqref{eq:Vobj} as by Gauss-Legendre quadrature, we obtain the linear system
\begin{gather}
\mathbf{M}\, \vec{\lambda}_Q=V_0-\vec{V}_\mathrm{inc} \label{eq:QSlinearsystem} \\
\text{with}\quad \mathbf{M}_{ij} \equiv \Phi(z_i,z_j)\,\delta z_j. \nonumber
\end{gather}
Vectors denote evaluation at positions $\{z_i,\mathcal{R}(z_i)\}$.  The condition of overall charge neutrality fixes the value of $V_0$:
\begin{gather}
\sum_i \vec{\lambda}_Q\,\delta z_i=0=\sum_i \delta z_i \,\mathbf{M}^{-1}\left( V_0-\vec{V}_\mathrm{inc} \right) \nonumber \\
\therefore \quad V_0=\frac{\sum_i \delta z_i \left[\mathbf{M}^{-1}\, \vec{V}_\mathrm{inc}\right]_i }
 	{\sum_i \delta z_i \left[\mathbf{M}^{-1}\,\vec{I}\right]_i }. \label{eq:QSchargeneutrality}
\end{gather}
Here $\vec{I}$ denotes a vector with all entries unity.  While Eq. \eqref{eq:QSlinearsystem} would appear to be directly solvable, such Fredholm integral equations of the first kind are notoriously ill-conditioned.  Consequently, we adopt regularization methods \cite{Twomey}\cite{Phillips} to invert the integral operator (matrix) $\mathbf{M}$, yielding smooth functions $\lambda(z)$ in accord with standard quasi-static solutions for well-studied geometries like the conducting sphere and ellipsoid.  (It is worth noting that, since Eq. \eqref{eq:DielectricBEM2} presents a well-conditioned Fredholm integral equation of the second kind, no such regularization of the solution is required in the case of a dielectric solid.) Once the inverse operator $\mathbf{M}^{-1}$ has been computed for a given geometry, calculation of $\lambda(z)$ for arbitrary $V_\mathrm{inc}(\vec{r})$ is fast and trivial.

For an axisymmetric system, Eqs. \eqref{eq:QSlinearsystem} and \eqref{eq:QSchargeneutrality} together with this solution method are sufficient to calculate the linear charge density induced on a conducting body due to an incident quasi-static field, and constitute the central result of this section.  In practice, the converged calculation of $\mathbf{M}^{-1}$ for a particular axisymmetric geometry takes no longer than a few tens of seconds on a single 2.7 GHz processor.   Calculation of $\lambda(z)$ for a range of $q$ values sufficient for converged \textit{lightning rod model} calculations requires only several seconds using the same processor.

\section{Electrodynamic boundary element method for an axisymmetric conductor} \label{app:EDboundaryelement}
As in the quasi-static case, the charge distribution induced on a nearly perfectly conducting object by an incident electrodynamic field oscillating at frequency $\omega$ resides exclusively at the object's surface.  To compute this distribution, we begin with detailed force balance at the boundary $S$ along directions tangential to the surface.  Assuming axisymmetry, we need only consider without loss of generality the surface tangential directions $\hat{\xi}$ orthogonal to $\hat{\phi}$ that possess positive $\hat{z}$-component:
\begin{equation}
\hat{\xi} \cdot \left(\vec{E}_\mathrm{inc}+\vec{E}_{obj}\right)=\vec{0} \quad \text{on}\quad S.
\end{equation}
Since $\vec{E}=-\nabla V+i\omega \vec{A}$ for scalar and vector potentials $V$ and $\vec{A}$, we have
\begin{eqnarray}
E_{inc\,\xi}(\vec{r})&=&\int_S \left(\partial_\xi \,dV_{obj}(\vec{r})- i\omega \,\hat{\xi} \cdot d\vec{A}_{obj}(\vec{r})\right) \quad \text{on}\quad S \nonumber \\
	&=&\int_0^L dz' \Big[\partial_\xi \Phi(z,z')\,\lambda_Q(z') \nonumber \\
	&\quad&\quad\quad\quad\quad-i \omega\,\mathcal{A}_\xi(z,z') \,I(z') \Big],
\end{eqnarray}
where we have parametrized points on $S$ by the object's axial coordinate $0<z<L$; meanwhile $\lambda_Q(z)\equiv dQ/dz$ denotes the linear charge density and $I(z)$ denotes the total current passing along the object surface through a $\hat{z}$-normal plane at $z$.  $\Phi$ and $\mathcal{A}_\xi$ denote integration kernels for the scalar and vector potentials, respectively.

The continuity equation for charge implies $\partial_z I(z)=i \omega\,\lambda_Q(z)$, and since current is forbidden to flow from the hypothetically isolated object, integration by parts  yields:
\begin{gather}
E_{inc\,\xi}(z)=\int_0^L dz'\lambda_Q(z')  \Big[\partial_\xi \Phi(z,z') \nonumber \\
\hspace{120pt} -\omega^2 \int_0^{z'} d\mathfrak{z}'\,\mathcal{A}_\xi(z,\mathfrak{z}') \Big] \label{eq:forcebalancehalfway}
\end{gather}
In terms of the azimuthal angle $\phi$ and surface radial coordinate at $z$ denoted $\mathcal{R}_z$, the scalar potential kernel is given by
\begin{gather}
\Phi(z,z')=\int_0^{2\pi} \frac{d\phi'}{2\pi}\, \frac{e^{i \omega/c\,\Delta(z,z',\phi)}}{\Delta(z,z',\phi)} \label{eq:EDscalarpotentialkernel} \\
\text{with} \quad \Delta\equiv \sqrt{(z-z')^2+\mathcal{R}_z^2+\mathcal{R}_{z'}^2-2\, \mathcal{R}_z \mathcal{R}_{z'} \cos{\phi}}, \nonumber
\end{gather}
which may be computed straightforwardly for a given object geometry by adaptive quadrature.  The exponential phase ensures the integrand is evaluated at retarded time.  The vector potential kernel may be established from
\begin{gather}
\vec{A}(\vec{r}) =\frac{1}{c^2} \int dS' \frac{\vec{K}(\vec{r}\prime)}{|\vec{r}-\vec{r}\prime|} \,e^{i \omega/c\,|\vec{r}-\vec{r}\prime|} \\
\text{and}\quad \mathcal{A}_\xi(z,z')\equiv \hat{\xi}_z \cdot \frac{d\vec{A}}{dz'}(z),
\end{gather}
with $\vec{K}(z)\equiv I(z)/{2\pi \mathcal{R}_z}\, \hat{\xi}$ denoting the local surface current.  Noting that $dS'=2\pi \mathcal{R}_z' \sqrt{1+\partial_{z'}\mathcal{R}_z'} \,dz'$ and that the direction of $\hat{\xi}$ is manifestly $z$- and $\phi$-dependent (expressed here as $\hat{\xi}_{z \phi}$), we obtain
\begin{equation}
\frac{dA_\xi}{dz'}(z)=\sqrt{1+\partial_{z'}\mathcal{R}_z'}\, \frac{I(z')}{c^2} \int_0^{2\pi} \frac{d\phi'}{2\pi} \,\frac{\hat{\xi}_{z \phi} \cdot \hat{\xi}_{z' \phi'}}{|\vec{r}-\vec{r'}|} \, e^{i \,\ldots}, \nonumber
\end{equation}
with the exponential factor unchanged.  The $\phi$-dependence of $A_\xi$ is rendered moot on account of axisymmetry, and so is suppressed.

Finally, since the surface tangential unit vector at height $z$ and azimuthal angle $\phi$ is expressed in terms of the radial coordinate $\mathcal{R}_z$ and the radial unit vector $\hat{\rho}_\phi$ as
\begin{equation}
\hat{\xi}_{z \phi}=\frac{\partial_z \mathcal{R}_z\, \hat{\rho}_\phi+\hat{z}}{\sqrt{1+\partial_z \mathcal{R}_z^2}}, \quad \mathrm{with} \quad \hat{\rho}_\phi \cdot \hat{\rho}_{\phi'}=\cos(\phi-\phi'), \nonumber
\end{equation}
we obtain the vector potential kernel as:
\begin{align}
\mathcal{A}_\xi(z,z')=\frac{1}{c^2}\Bigg[&\int_0^{2\pi} \frac{d\phi'}{2\pi} \frac{e^{i \omega/c\,\Delta(z,z',\phi')}}{\Delta(z,z',\phi')} + \nonumber \\
	&\partial_z \mathcal{R}_z\, \partial_{z'} \mathcal{R}_{z'} \int_0^{2\pi} \frac{d\phi'}{2\pi} \frac{e^{i \omega/c\,\Delta(z,z',\phi')}}{\Delta(z,z',\phi')}\cos\phi' \Bigg] \nonumber \\
	&\quad\times \frac{1}{\sqrt{1+\partial_z \mathcal{R}_z^2}}. \label{eq:EDvectorpotentialkernel}
\end{align}
Here $\Delta$ is defined as in Eq. \eqref{eq:EDscalarpotentialkernel}, and note that the first term within brackets in fact equates with the scalar potential kernel.  Only the second term must be computed anew, and similarly by adaptive quadrature.

We now define a convenient quasi-potential function for the incident field:
\begin{align}
V_\mathrm{inc}(z) &\equiv -\int dz \sqrt{1+\partial_z \mathcal{R}_z^2}\;\, \hat{\xi}_z \cdot \vec{E}_\mathrm{inc}(z) \nonumber \\
	&=-\int dz\,\left(\partial_z \mathcal{R}_z\,E_{inc\,\rho}+E_{inc\,z}\right). \label{eq:quasipotential}
\end{align}
Proceeding with Eq. \eqref{eq:forcebalancehalfway}, we relabel $z\rightarrow \mathfrak{z}$ before applying the operation $\int_0^{\xi_z} d\xi = \int_0^z d\mathfrak{z} \sqrt{1+\partial_\mathfrak{z} \mathcal{R}_\mathfrak{z}^2}$ to both sides, resulting in:
\begin{align}
\int_0^L dz' \lambda_Q(z') & \left[\Phi(z,z')-\frac{\omega^2}{c^2}\, \int_0^z d\mathfrak{z} \int_0^{z'} d\mathfrak{z}'\,\,\mathcal{\bar{A}}_\xi(\mathfrak{z},\mathfrak{z}') \right] \nonumber \\
	&\quad\quad =V_0-V_\mathrm{inc}(z). \label{eq:forcebalanceED}
\end{align}
Here we have applied Eq. \eqref{eq:quasipotential} and taken $V_0$ as a constant of integration.  Furthermore we have defined a new vector potential kernel $\mathcal{\bar{A}}_\xi(z,z') \equiv \sqrt{1+\partial_z \mathcal{R}_z^2}\, \mathcal{A}_\xi (z,z')$, which is now symmetric in its two arguments.  Note that the first term in brackets in Eq. \eqref{eq:forcebalanceED} accounts for the retarded Coulomb force among surface charges, whereas the second term describes radiative forces with strength of order $\mathcal{O}^2( L/\lambda)$ produced by conduction currents, where $\lambda$ the free-space wavelength of light.

As in Appendix \ref{app:QSboundaryelement}, discretizing $z$ results in a linear system
\begin{equation} \label{eq:EDlinearsystem}
\left[\mathbf{\Phi}-\frac{\omega^2}{c^2} \overline{\mathbf{W}}^T \mathbf{\mathcal{\bar{A}}}\, \overline{\mathbf{W}} \right] \mathbf{W} \vec{\lambda}_Q = V_0-\vec{V}_\mathrm{inc},
\end{equation}
where $\mathbf{\Phi}_{ij}\equiv\Phi(z_i,z_j)$, $\mathbf{W}\equiv \mathrm{diag}\{\delta z_i\}$, $\bar{\mathcal{A}}_{ij}\equiv\bar{\mathcal{A}}_\xi(z_i,z_j)$, $\mathbf{\overline{W}}_{ij}\equiv \delta z_i \,\theta(j-i)$, and $\theta(\ldots)$ denotes the Heaviside unit step function.  The superscript $T$ denotes matrix transpose.  Vectors again denote evaluation at axial and radial coordinates $\{z_i,\mathcal{R}(z_i)\}$ along the object surface.  Self-consistency requires a value of $V_0$ ensuring charge neutrality.  Taking $\mathbf{M}$ to be the full integral operator preceding $\vec{\lambda}_Q$ in the linear system above, $V_0$ is again given by Eq. \eqref{eq:QSchargeneutrality}, and $\vec{\lambda}_Q$ is obtained via inversion of $\mathbf{M}$.  Note that the particular selections of lower integration bounds on $V_\mathrm{inc}$ in Eq. \eqref{eq:quasipotential} and $\mathcal{\bar{A}}$ in Eq. \eqref{eq:forcebalanceED} are naturally rendered arbitrary when this condition is satisfied.  As in the quasi-static case, once $\mathbf{M}^{-1}$ has been computed for a given geometry (less than one minute of computation on a 2.7 GHz processor), the calculation of $\lambda_Q(z)$ for arbitrary $\vec{E}_\mathrm{inc}(\vec{r})$ is both fast and trivial (several milliseconds).  To emulate plane wave illumination from an inclination angle $\theta$ with respect to the $z$-axis, in this work we substitute the axisymmetric analog
\begin{gather}
\vec{E}_\mathrm{inc}(\vec{r})=\Big(J_0(q \rho)\, \hat{z}+i\frac{\sqrt{k^2-q^2}}{q} J_1(q \rho)\,\hat{\rho}\Big) \nonumber \\
\quad\quad \quad \quad \times \,e^{-i \sqrt{k^2-q^2} z} \\
	\text{with}\quad q \equiv k\,\sin{\theta}\quad\text{and}\quad k\equiv \omega/c. \nonumber
\end{gather}
This field profile equates with a rotational sum of $\theta$-directed plane waves inbound from all azimuthal angles $\phi$.

For an axisymmetric system, Eqs. \eqref{eq:EDscalarpotentialkernel}, \eqref{eq:EDvectorpotentialkernel}, and \eqref{eq:EDlinearsystem} are sufficient to calculate the linear charge density induced on a conducting body due to an incident electrodynamic field, and constitute the central result of this section.  In practice, the converged electrodynamic calculation of $\mathbf{M}^{-1}$ for a particular axisymmetric geometry takes only twice as long as for the quasi-static case.

\section{Radiation from an axisymmetric conductor} \label{app:radiation}
The far-field radiation profile from an arbitrary current distribution can be obtained by integrating the far-field contribution $\overleftrightarrow{G}_{\mathrm{FF}}$ to the Green's dyadic function $\overleftrightarrow{G}$ \cite{NovotnyOptics} from infinitesimal current elements at positions $\vec{r}\prime$, here for demonstration considered oriented along the $\hat{z}$-direction, as
\begin{gather}
d\vec{E}_\mathrm{rad}(\vec{r})=\frac{i\omega}{4\pi}\vec{G}_{\mathrm{FF},z}(\vec{r},\vec{r}\prime)\,j_z(\vec{r}\prime)\, dV' \label{eq:Gfarfield} \\
\text{with}\quad \vec{G}_{\mathrm{FF},z}(\vec{r},\vec{r}\prime) \equiv -\frac{1}{c^2} \,\frac{e^{i \omega/c\,|\vec{r}-\vec{r}\prime|}}{|\vec{r}-\vec{r}\prime|}\,\sin{\theta}\,\hat{\theta}, \nonumber
\end{gather}
exhibiting the familiar field profile of a radiating dipole.  Here $\theta$ denotes the inclination angle of the observation point $\vec{r}$ from the $z$-axis in a spherical coordinate system.  

\begin{figure}[t!]
\centering
   \includegraphics[width=0.5\textwidth]{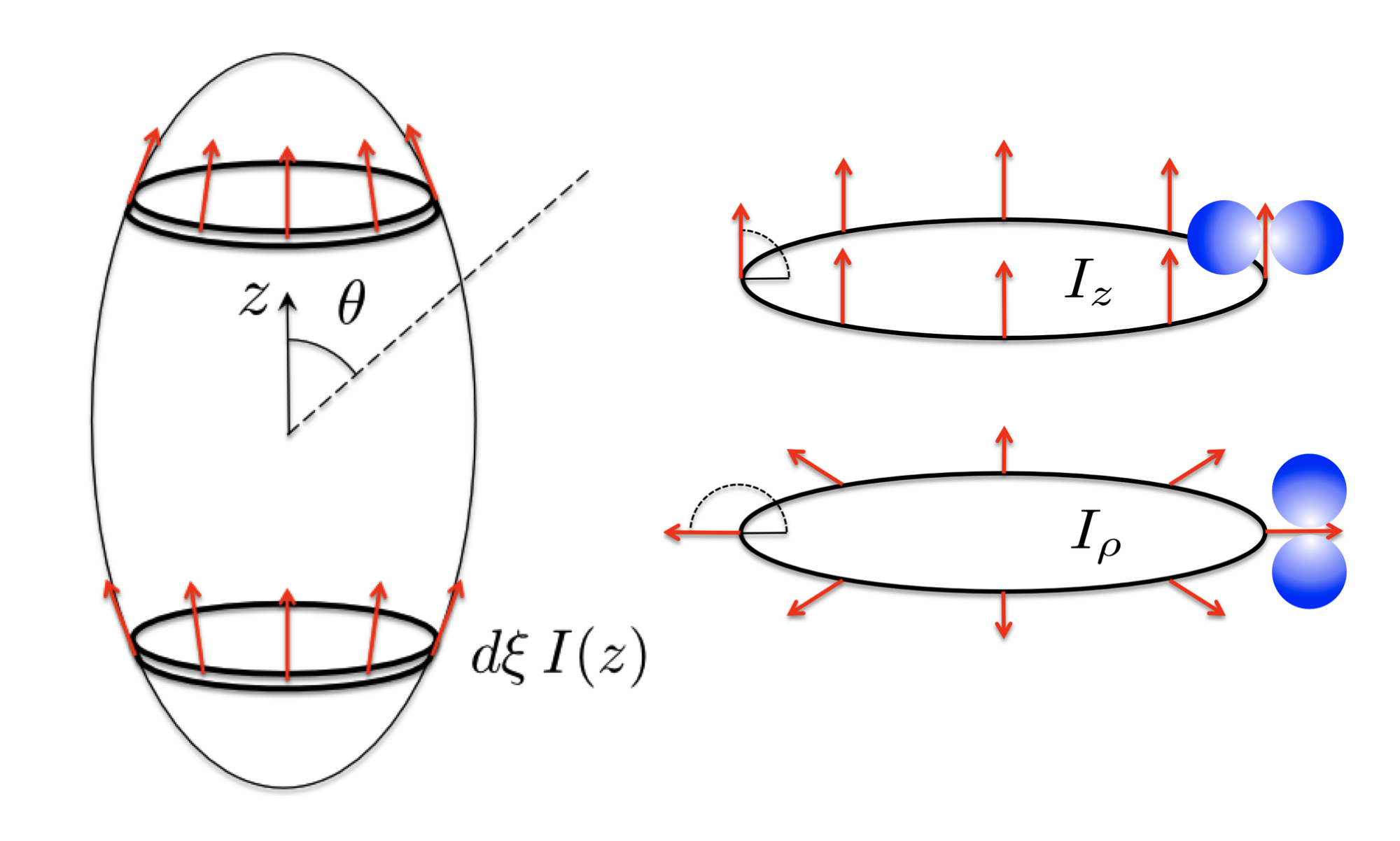}
   \caption{(Left) The field radiated from an axisymmetric body to an observation point at inclination angle $\theta$ is constructed from the contributions of currents (shown in red) through infinitesimal surface annuli.  (Right) The two independent polarizations composing any annular axisymmetric current distribution, with associated angular radiation profiles (Eq. \ref{eq:Gfarfield}) shown schematically in blue.}
   \label{fig:currentcomponents}
\end{figure}

The dimension of a nearly perfect conductor is by definition much greater than the magnetic skin depth of the constituent material.  Consequently, volume integration reduces to an integral over surface current contributions $dS\, \vec{K}(\vec{r}$).  In an axisymmetric object, these contributions are associated with surface annuli located at axial coordinates $z$ and radii $\mathcal{R}_z$, for which $dS=2\pi \mathcal{R}_z\,d\xi$ with $d\xi \equiv \sqrt{1+\partial_z \mathcal{R}_z^2}\, dz$.  We first evaluate the radiated field from such an annulus, considering contributions from the two independently allowed polarizations of axisymmetric current separately, as depicted in Fig. \ref{fig:currentcomponents}.

We define the total current as $I_\alpha(z)=2\pi \mathcal{R}_z\,K_\alpha(z)$ for polarizations $\alpha=z,\rho$.  For the $z$-polarized contribution, integrating Eq. \ref{eq:Gfarfield} through an annulus about azimuthal angle $\phi$ obtains
\begin{align}
d\vec{E}_{\mathrm{rad},z}&= -\frac{i\omega}{4\pi c^2}  \frac{e^{i\omega/c\, \delta r(z)}}{\delta r(z)}\,  \sin{\theta} \,K_z(z)\,d\xi \nonumber \\
	&\quad \times \int_0^{2\pi} d\phi\, \mathcal{R}_z \exp\left(i \omega/c \,\mathcal{R}_z \cos{\phi}\, \sin{\theta}\right) \hat{\theta} \nonumber \\
	&= -\frac{i\omega}{4\pi c^2} \, \frac{e^{i\omega/c\, \delta r(z)}}{\delta r(z)}  \sin{\theta}\,I_z(z)\, d\xi \nonumber \\
	&\quad\quad\quad \times J_0(\omega/c\,\mathcal{R}_z  \sin{\theta})\, \hat{\theta}. \label{eq:radiationzcontribution}
\end{align}
Here $\delta r(z)$ denotes the distance from the center of the $z$-located annulus to the observation point, and we have applied the approximation $|\vec{r}-\vec{r}\prime|^{-1} \approx \delta r(z)^{-1}$ valid for $\delta r(z) \gg \mathcal{R}_z$.  An elementary analysis accounting for rotation of the radiant polarization vector in the integrand of the $\rho$-polarized contribution similarly results in:
\begin{align}
d\vec{E}_{\mathrm{rad},\rho}&=\frac{\omega}{4\pi c^2} \,\frac{e^{i\omega/c\, \delta r(z)}}{\delta r(z)}\, \cos{\theta}\,I_\rho(z)\,d\xi \nonumber \\
	&\quad\quad \times J_1(\omega/c\,\mathcal{R}_z  \sin{\theta})\, \hat{\theta}. \label{eq:radiationrcontribution}
\end{align}

Current $I(z)$ flows on the surface of an axisymmetric conductor along a surface tangent vector
\begin{equation}
\hat{\xi}=\frac{\partial_z \mathcal{R}_z\, \hat{\rho}+\hat{z}}{\sqrt{1+\partial_z \mathcal{R}_z^2}}.\nonumber
\end{equation}
Consequently, the total radiation from the axisymmetric body of length $L$ is given by a commensurate sum of $\hat{z}$- and $\hat{\rho}$-polarized contributions:
\begin{align}
\vec{E}_\mathrm{rad}(\theta)&=-\frac{i\omega}{4\pi c^2}\,\frac{e^{i\omega/c \,\Delta r}}{\Delta r} \int_0^L dz\,\mathcal{E}(z,\theta)\,I(z)\,\hat{\theta} \label{eq:radiation} \\
\text{with}\quad \mathcal{E}&=e^{-i\omega/c \, z \cos\theta}\Big[\sin{\theta}\,J_0(\omega/c\,\mathcal{R}_z \sin\theta) \nonumber \\
	&\quad\quad +i\, \partial_z \mathcal{R}_z \cos{\theta}\,J_1(\omega/c\,\mathcal{R}_z \sin\theta)\Big].\label{eq:radiationkernel}
\end{align}
Note the factor $1/\sqrt{1+\partial_z\mathcal{R}_z^2}$ has been absorbed by the integration measure $dz$.  Here $\Delta r \equiv \delta r(z)+z \cos{\theta}$ is the distance from one apex of the object (at $z=0$) to the observation point, and we have applied the approximation $\delta r(z)^{-1} \approx \Delta r^{-1}$ appropriate for distances $\Delta r \gg L$.  After applying the continuity equation for charge $\partial_z I(z)=i \omega\,\lambda_Q(z)$ (with $\lambda_Q$ the linear charge density) together with the fact that the current vanishes at the extrema of a hypothetically isolated body (at $z=0,\,L$), integration by parts yields:
\begin{equation}
\vec{E}_\mathrm{rad}(\theta)=-\frac{\omega^2}{4\pi c^2}\,\frac{e^{i\omega/c\, \Delta r}}{\Delta r} \int_0^L dz\,\lambda_Q(z) \int_0^z dz'\mathcal{E}(z',\theta)\,\hat{\theta}. \label{eq:radiationequation}
\end{equation}

Provided an electrodynamically consistent charge distribution $\lambda_Q(z)$ calculated at frequency $\omega$, Eq. \eqref{eq:radiationequation} can be evaluated straightforwardly for a given object geometry by quadrature.  In the notation of Appendix \ref{app:EDboundaryelement}, the complex amplitude of the $\hat{\theta}$-polarized radiation field becomes:
\begin{equation}
E_\mathrm{rad}(\theta)=-\frac{\omega^2}{4\pi c^2}\,\frac{e^{i\omega/c\, \Delta r}}{\Delta r}\,\vec{\lambda}_Q^T\,\mathbf{W}\,\mathbf{\overline{W}}\,\vec{\mathcal{E}}_\theta . \label{eq:radiatedfield}
\end{equation}
Here the superscript $T$ denotes vector transpose, and $\vec{\mathcal{E}}_\theta$ indicates evaluation of $\mathcal{E}(z,\theta)$ at the chosen observation angle $\theta$.  Together with Eq. \eqref{eq:radiationkernel}, this expression is sufficient to compute the electric field radiated from a conducting axisymmetric system, and constitutes the central result of this section.

Projected onto a detector sensitive to $\hat{\theta}$-polarized light, the radiation contributions $\left[\vec{e}_\mathrm{rad}\right]_i$ utilized in Sec. \ref{sec:model} are computed by expressing each single-momentum probe response function (linear charge density) $\Lambda(q_i,z)$ in discretized real space representation $\big[\vec{\lambda}_Q\big]_j\equiv \Lambda(q_i,z_j)$ and applying Eq. \eqref{eq:radiatedfield}, taking $\theta \approx 60^\circ$ relative to the $z$-axis of the near-field probe as the typical collection angle of experimental detection optics.

\nocite{*}
\bibliography{LightningRodModel_SubmissionPRB}

\begin{thebibliography}{79}%
\makeatletter
\providecommand \@ifxundefined [1]{%
 \@ifx{#1\undefined}
}%
\providecommand \@ifnum [1]{%
 \ifnum #1\expandafter \@firstoftwo
 \else \expandafter \@secondoftwo
 \fi
}%
\providecommand \@ifx [1]{%
 \ifx #1\expandafter \@firstoftwo
 \else \expandafter \@secondoftwo
 \fi
}%
\providecommand \natexlab [1]{#1}%
\providecommand \enquote  [1]{``#1''}%
\providecommand \bibnamefont  [1]{#1}%
\providecommand \bibfnamefont [1]{#1}%
\providecommand \citenamefont [1]{#1}%
\providecommand \href@noop [0]{\@secondoftwo}%
\providecommand \href [0]{\begingroup \@sanitize@url \@href}%
\providecommand \@href[1]{\@@startlink{#1}\@@href}%
\providecommand \@@href[1]{\endgroup#1\@@endlink}%
\providecommand \@sanitize@url [0]{\catcode `\\12\catcode `\$12\catcode
  `\&12\catcode `\#12\catcode `\^12\catcode `\_12\catcode `\%12\relax}%
\providecommand \@@startlink[1]{}%
\providecommand \@@endlink[0]{}%
\providecommand \url  [0]{\begingroup\@sanitize@url \@url }%
\providecommand \@url [1]{\endgroup\@href {#1}{\urlprefix }}%
\providecommand \urlprefix  [0]{URL }%
\providecommand \Eprint [0]{\href }%
\providecommand \doibase [0]{http://dx.doi.org/}%
\providecommand \selectlanguage [0]{\@gobble}%
\providecommand \bibinfo  [0]{\@secondoftwo}%
\providecommand \bibfield  [0]{\@secondoftwo}%
\providecommand \translation [1]{[#1]}%
\providecommand \BibitemOpen [0]{}%
\providecommand \bibitemStop [0]{}%
\providecommand \bibitemNoStop [0]{.\EOS\space}%
\providecommand \EOS [0]{\spacefactor3000\relax}%
\providecommand \BibitemShut  [1]{\csname bibitem#1\endcsname}%
\let\auto@bib@innerbib\@empty
\bibitem [{\citenamefont {Novotny}(2007)}]{NovotnyHistory}%
  \BibitemOpen
  \bibfield  {author} {\bibinfo {author} {\bibfnamefont {L.}~\bibnamefont
  {Novotny}},\ }\href@noop {} {\bibfield  {journal} {\bibinfo  {journal}
  {Progress in Optics}\ }\textbf {\bibinfo {volume} {50}},\ \bibinfo {pages}
  {137} (\bibinfo {year} {2007})}\BibitemShut {NoStop}%
\bibitem [{\citenamefont {Fischer}\ and\ \citenamefont {Pohl}(1989)}]{Pohl}%
  \BibitemOpen
  \bibfield  {author} {\bibinfo {author} {\bibfnamefont {U.~C.}\ \bibnamefont
  {Fischer}}\ and\ \bibinfo {author} {\bibfnamefont {D.~W.}\ \bibnamefont
  {Pohl}},\ }\href {\doibase 10.1103/PhysRevLett.62.458} {\bibfield  {journal}
  {\bibinfo  {journal} {Phys. Rev. Lett.}\ }\textbf {\bibinfo {volume} {62}},\
  \bibinfo {pages} {458} (\bibinfo {year} {1989})}\BibitemShut {NoStop}%
\bibitem [{\citenamefont {Inouye}\ and\ \citenamefont {Kawata}(1994)}]{Inouye}%
  \BibitemOpen
  \bibfield  {author} {\bibinfo {author} {\bibfnamefont {Y.}~\bibnamefont
  {Inouye}}\ and\ \bibinfo {author} {\bibfnamefont {S.}~\bibnamefont
  {Kawata}},\ }\href {\doibase 10.1364/OL.19.000159} {\bibfield  {journal}
  {\bibinfo  {journal} {Opt. Lett.}\ }\textbf {\bibinfo {volume} {19}},\
  \bibinfo {pages} {159} (\bibinfo {year} {1994})}\BibitemShut {NoStop}%
\bibitem [{\citenamefont {Knoll}\ and\ \citenamefont
  {Keilmann}(2000)}]{Demodulation}%
  \BibitemOpen
  \bibfield  {author} {\bibinfo {author} {\bibfnamefont {B.}~\bibnamefont
  {Knoll}}\ and\ \bibinfo {author} {\bibfnamefont {F.}~\bibnamefont
  {Keilmann}},\ }\href {\doibase 10.1016/S0030-4018(00)00826-9} {\bibfield
  {journal} {\bibinfo  {journal} {Optics Communications}\ }\textbf {\bibinfo
  {volume} {182}},\ \bibinfo {pages} {321 } (\bibinfo {year}
  {2000})}\BibitemShut {NoStop}%
\bibitem [{\citenamefont {Keilmann}\ and\ \citenamefont
  {Hillenbrand}(2004)}]{ElasticScattering}%
  \BibitemOpen
  \bibfield  {author} {\bibinfo {author} {\bibfnamefont {F.}~\bibnamefont
  {Keilmann}}\ and\ \bibinfo {author} {\bibfnamefont {R.}~\bibnamefont
  {Hillenbrand}},\ }\href {\doibase 10.1098/rsta.2003.1347} {\bibfield
  {journal} {\bibinfo  {journal} {Philosophical Transactions of the Royal
  Society of London. Series A: Mathematical, Physical and Engineering
  Sciences}\ }\textbf {\bibinfo {volume} {362}},\ \bibinfo {pages} {787}
  (\bibinfo {year} {2004})}\BibitemShut {NoStop}%
\bibitem [{\citenamefont {Huber}\ \emph {et~al.}(2008)\citenamefont {Huber},
  \citenamefont {Keilmann}, \citenamefont {Wittborn}, \citenamefont
  {Aizpurua},\ and\ \citenamefont {Hillenbrand}}]{THzNearField}%
  \BibitemOpen
  \bibfield  {author} {\bibinfo {author} {\bibfnamefont {A.~J.}\ \bibnamefont
  {Huber}}, \bibinfo {author} {\bibfnamefont {F.}~\bibnamefont {Keilmann}},
  \bibinfo {author} {\bibfnamefont {J.}~\bibnamefont {Wittborn}}, \bibinfo
  {author} {\bibfnamefont {J.}~\bibnamefont {Aizpurua}}, \ and\ \bibinfo
  {author} {\bibfnamefont {R.}~\bibnamefont {Hillenbrand}},\ }\href {\doibase
  10.1021/nl802086x} {\bibfield  {journal} {\bibinfo  {journal} {Nano Letters}\
  }\textbf {\bibinfo {volume} {8}},\ \bibinfo {pages} {3766} (\bibinfo {year}
  {2008})},\ \bibinfo {note} {pMID: 18837565},\ \Eprint
  {http://arxiv.org/abs/http://pubs.acs.org/doi/pdf/10.1021/nl802086x}
  {http://pubs.acs.org/doi/pdf/10.1021/nl802086x} \BibitemShut {NoStop}%
\bibitem [{\citenamefont {Amarie}\ and\ \citenamefont
  {Keilmann}(2011)}]{AmarieNanoFTIR}%
  \BibitemOpen
  \bibfield  {author} {\bibinfo {author} {\bibfnamefont {S.}~\bibnamefont
  {Amarie}}\ and\ \bibinfo {author} {\bibfnamefont {F.}~\bibnamefont
  {Keilmann}},\ }\href {\doibase 10.1103/PhysRevB.83.045404} {\bibfield
  {journal} {\bibinfo  {journal} {Phys. Rev. B}\ }\textbf {\bibinfo {volume}
  {83}},\ \bibinfo {pages} {045404} (\bibinfo {year} {2011})}\BibitemShut
  {NoStop}%
\bibitem [{\citenamefont {Brehm}\ \emph {et~al.}(2006)\citenamefont {Brehm},
  \citenamefont {Schliesser},\ and\ \citenamefont {Keilmann}}]{BrehmNanoFTIR}%
  \BibitemOpen
  \bibfield  {author} {\bibinfo {author} {\bibfnamefont {M.}~\bibnamefont
  {Brehm}}, \bibinfo {author} {\bibfnamefont {A.}~\bibnamefont {Schliesser}}, \
  and\ \bibinfo {author} {\bibfnamefont {F.}~\bibnamefont {Keilmann}},\ }\href
  {\doibase 10.1364/OE.14.011222} {\bibfield  {journal} {\bibinfo  {journal}
  {Opt. Express}\ }\textbf {\bibinfo {volume} {14}},\ \bibinfo {pages} {11222}
  (\bibinfo {year} {2006})}\BibitemShut {NoStop}%
\bibitem [{\citenamefont {Huth}\ \emph {et~al.}(2012)\citenamefont {Huth},
  \citenamefont {Govyadinov}, \citenamefont {Amarie}, \citenamefont {Nuansing},
  \citenamefont {Keilmann},\ and\ \citenamefont {Hillenbrand}}]{HuthNanoFTIR}%
  \BibitemOpen
  \bibfield  {author} {\bibinfo {author} {\bibfnamefont {F.}~\bibnamefont
  {Huth}}, \bibinfo {author} {\bibfnamefont {A.}~\bibnamefont {Govyadinov}},
  \bibinfo {author} {\bibfnamefont {S.}~\bibnamefont {Amarie}}, \bibinfo
  {author} {\bibfnamefont {W.}~\bibnamefont {Nuansing}}, \bibinfo {author}
  {\bibfnamefont {F.}~\bibnamefont {Keilmann}}, \ and\ \bibinfo {author}
  {\bibfnamefont {R.}~\bibnamefont {Hillenbrand}},\ }\href {\doibase
  10.1021/nl301159v} {\bibfield  {journal} {\bibinfo  {journal} {Nano Letters}\
  }\textbf {\bibinfo {volume} {12}},\ \bibinfo {pages} {3973} (\bibinfo {year}
  {2012})},\ \Eprint
  {http://arxiv.org/abs/http://pubs.acs.org/doi/pdf/10.1021/nl301159v}
  {http://pubs.acs.org/doi/pdf/10.1021/nl301159v} \BibitemShut {NoStop}%
\bibitem [{\citenamefont {Gomez}\ \emph {et~al.}(2006)\citenamefont {Gomez},
  \citenamefont {Bachelot}, \citenamefont {Bouhelier}, \citenamefont
  {Wiederrecht}, \citenamefont {hui Chang}, \citenamefont {Gray}, \citenamefont
  {Hua}, \citenamefont {Jeon}, \citenamefont {Rogers}, \citenamefont {Castro},
  \citenamefont {Blaize}, \citenamefont {Stefanon}, \citenamefont {Lerondel},\
  and\ \citenamefont {Royer}}]{GomezPSHet}%
  \BibitemOpen
  \bibfield  {author} {\bibinfo {author} {\bibfnamefont {L.}~\bibnamefont
  {Gomez}}, \bibinfo {author} {\bibfnamefont {R.}~\bibnamefont {Bachelot}},
  \bibinfo {author} {\bibfnamefont {A.}~\bibnamefont {Bouhelier}}, \bibinfo
  {author} {\bibfnamefont {G.~P.}\ \bibnamefont {Wiederrecht}}, \bibinfo
  {author} {\bibfnamefont {S.}~\bibnamefont {hui Chang}}, \bibinfo {author}
  {\bibfnamefont {S.~K.}\ \bibnamefont {Gray}}, \bibinfo {author}
  {\bibfnamefont {F.}~\bibnamefont {Hua}}, \bibinfo {author} {\bibfnamefont
  {S.}~\bibnamefont {Jeon}}, \bibinfo {author} {\bibfnamefont {J.~A.}\
  \bibnamefont {Rogers}}, \bibinfo {author} {\bibfnamefont {M.~E.}\
  \bibnamefont {Castro}}, \bibinfo {author} {\bibfnamefont {S.}~\bibnamefont
  {Blaize}}, \bibinfo {author} {\bibfnamefont {I.}~\bibnamefont {Stefanon}},
  \bibinfo {author} {\bibfnamefont {G.}~\bibnamefont {Lerondel}}, \ and\
  \bibinfo {author} {\bibfnamefont {P.}~\bibnamefont {Royer}},\ }\href
  {\doibase 10.1364/JOSAB.23.000823} {\bibfield  {journal} {\bibinfo  {journal}
  {J. Opt. Soc. Am. B}\ }\textbf {\bibinfo {volume} {23}},\ \bibinfo {pages}
  {823} (\bibinfo {year} {2006})}\BibitemShut {NoStop}%
\bibitem [{\citenamefont {Ocelic}\ \emph {et~al.}(2006)\citenamefont {Ocelic},
  \citenamefont {Huber},\ and\ \citenamefont {Hillenbrand}}]{PSHet}%
  \BibitemOpen
  \bibfield  {author} {\bibinfo {author} {\bibfnamefont {N.}~\bibnamefont
  {Ocelic}}, \bibinfo {author} {\bibfnamefont {A.}~\bibnamefont {Huber}}, \
  and\ \bibinfo {author} {\bibfnamefont {R.}~\bibnamefont {Hillenbrand}},\
  }\href {\doibase 10.1063/1.2348781} {\bibfield  {journal} {\bibinfo
  {journal} {Applied Physics Letters}\ }\textbf {\bibinfo {volume} {89}},\
  \bibinfo {eid} {101124} (\bibinfo {year} {2006})}\BibitemShut {NoStop}%
\bibitem [{\citenamefont {Hillenbrand}\ and\ \citenamefont
  {Keilmann}(2000)}]{ComplexOpticalConstants}%
  \BibitemOpen
  \bibfield  {author} {\bibinfo {author} {\bibfnamefont {R.}~\bibnamefont
  {Hillenbrand}}\ and\ \bibinfo {author} {\bibfnamefont {F.}~\bibnamefont
  {Keilmann}},\ }\href {\doibase 10.1103/PhysRevLett.85.3029} {\bibfield
  {journal} {\bibinfo  {journal} {Phys. Rev. Lett.}\ }\textbf {\bibinfo
  {volume} {85}},\ \bibinfo {pages} {3029} (\bibinfo {year}
  {2000})}\BibitemShut {NoStop}%
\bibitem [{\citenamefont {Schnell}\ \emph {et~al.}(2010)\citenamefont
  {Schnell}, \citenamefont {Garcia-Etxarri}, \citenamefont {Huber},
  \citenamefont {Crozier}, \citenamefont {Borisov}, \citenamefont {Aizpurua},\
  and\ \citenamefont {Hillenbrand}}]{PlasmonsPhase}%
  \BibitemOpen
  \bibfield  {author} {\bibinfo {author} {\bibfnamefont {M.}~\bibnamefont
  {Schnell}}, \bibinfo {author} {\bibfnamefont {A.}~\bibnamefont
  {Garcia-Etxarri}}, \bibinfo {author} {\bibfnamefont {A.~J.}\ \bibnamefont
  {Huber}}, \bibinfo {author} {\bibfnamefont {K.~B.}\ \bibnamefont {Crozier}},
  \bibinfo {author} {\bibfnamefont {A.}~\bibnamefont {Borisov}}, \bibinfo
  {author} {\bibfnamefont {J.}~\bibnamefont {Aizpurua}}, \ and\ \bibinfo
  {author} {\bibfnamefont {R.}~\bibnamefont {Hillenbrand}},\ }\href {\doibase
  10.1021/jp909252z} {\bibfield  {journal} {\bibinfo  {journal} {The Journal of
  Physical Chemistry C}\ }\textbf {\bibinfo {volume} {114}},\ \bibinfo {pages}
  {7341} (\bibinfo {year} {2010})},\ \Eprint
  {http://arxiv.org/abs/http://pubs.acs.org/doi/pdf/10.1021/jp909252z}
  {http://pubs.acs.org/doi/pdf/10.1021/jp909252z} \BibitemShut {NoStop}%
\bibitem [{\citenamefont {Carney}\ \emph {et~al.}(2012)\citenamefont {Carney},
  \citenamefont {Deutsch}, \citenamefont {Govyadinov},\ and\ \citenamefont
  {Hillenbrand}}]{NearFieldPhase}%
  \BibitemOpen
  \bibfield  {author} {\bibinfo {author} {\bibfnamefont {P.~S.}\ \bibnamefont
  {Carney}}, \bibinfo {author} {\bibfnamefont {B.}~\bibnamefont {Deutsch}},
  \bibinfo {author} {\bibfnamefont {A.~A.}\ \bibnamefont {Govyadinov}}, \ and\
  \bibinfo {author} {\bibfnamefont {R.}~\bibnamefont {Hillenbrand}},\ }\href
  {\doibase 10.1021/nn205008y} {\bibfield  {journal} {\bibinfo  {journal} {ACS
  Nano}\ }\textbf {\bibinfo {volume} {6}},\ \bibinfo {pages} {8} (\bibinfo
  {year} {2012})},\ \Eprint
  {http://arxiv.org/abs/http://pubs.acs.org/doi/pdf/10.1021/nn205008y}
  {http://pubs.acs.org/doi/pdf/10.1021/nn205008y} \BibitemShut {NoStop}%
\bibitem [{\citenamefont {Xu}\ \emph {et~al.}(2012)\citenamefont {Xu},
  \citenamefont {Rang}, \citenamefont {Craig},\ and\ \citenamefont
  {Raschke}}]{RaschkeVibration}%
  \BibitemOpen
  \bibfield  {author} {\bibinfo {author} {\bibfnamefont {X.~G.}\ \bibnamefont
  {Xu}}, \bibinfo {author} {\bibfnamefont {M.}~\bibnamefont {Rang}}, \bibinfo
  {author} {\bibfnamefont {I.~M.}\ \bibnamefont {Craig}}, \ and\ \bibinfo
  {author} {\bibfnamefont {M.~B.}\ \bibnamefont {Raschke}},\ }\href {\doibase
  10.1021/jz300463d} {\bibfield  {journal} {\bibinfo  {journal} {The Journal of
  Physical Chemistry Letters}\ }\textbf {\bibinfo {volume} {3}},\ \bibinfo
  {pages} {1836} (\bibinfo {year} {2012})},\ \Eprint
  {http://arxiv.org/abs/http://pubs.acs.org/doi/pdf/10.1021/jz300463d}
  {http://pubs.acs.org/doi/pdf/10.1021/jz300463d} \BibitemShut {NoStop}%
\bibitem [{\citenamefont {Amarie}\ \emph {et~al.}(2012)\citenamefont {Amarie},
  \citenamefont {Zaslansky}, \citenamefont {Kajihara}, \citenamefont
  {Griesshaber}, \citenamefont {Schmahl},\ and\ \citenamefont
  {Keilmann}}]{AmarieBones}%
  \BibitemOpen
  \bibfield  {author} {\bibinfo {author} {\bibfnamefont {S.}~\bibnamefont
  {Amarie}}, \bibinfo {author} {\bibfnamefont {P.}~\bibnamefont {Zaslansky}},
  \bibinfo {author} {\bibfnamefont {Y.}~\bibnamefont {Kajihara}}, \bibinfo
  {author} {\bibfnamefont {E.}~\bibnamefont {Griesshaber}}, \bibinfo {author}
  {\bibfnamefont {W.~W.}\ \bibnamefont {Schmahl}}, \ and\ \bibinfo {author}
  {\bibfnamefont {F.}~\bibnamefont {Keilmann}},\ }\href {\doibase
  10.3762/bjnano.3.35} {\bibfield  {journal} {\bibinfo  {journal} {Beilstein
  Journal of Nanotechnology}\ }\textbf {\bibinfo {volume} {3}},\ \bibinfo
  {pages} {312} (\bibinfo {year} {2012})}\BibitemShut {NoStop}%
\bibitem [{\citenamefont {Atkin}\ \emph {et~al.}(2012)\citenamefont {Atkin},
  \citenamefont {Berweger}, \citenamefont {Jones},\ and\ \citenamefont
  {Raschke}}]{RaschkeReview}%
  \BibitemOpen
  \bibfield  {author} {\bibinfo {author} {\bibfnamefont {J.~M.}\ \bibnamefont
  {Atkin}}, \bibinfo {author} {\bibfnamefont {S.}~\bibnamefont {Berweger}},
  \bibinfo {author} {\bibfnamefont {A.~C.}\ \bibnamefont {Jones}}, \ and\
  \bibinfo {author} {\bibfnamefont {M.~B.}\ \bibnamefont {Raschke}},\ }\href
  {\doibase 10.1080/00018732.2012.737982} {\bibfield  {journal} {\bibinfo
  {journal} {Advances in Physics}\ }\textbf {\bibinfo {volume} {61}},\ \bibinfo
  {pages} {745} (\bibinfo {year} {2012})}\BibitemShut {NoStop}%
\bibitem [{\citenamefont {Qazilbash}\ \emph {et~al.}(2007)\citenamefont
  {Qazilbash}, \citenamefont {Brehm}, \citenamefont {Chae}, \citenamefont {Ho},
  \citenamefont {Andreev}, \citenamefont {Kim}, \citenamefont {Yun},
  \citenamefont {Balatsky}, \citenamefont {Maple}, \citenamefont {Keilmann},
  \citenamefont {Kim},\ and\ \citenamefont {Basov}}]{Qazilbash}%
  \BibitemOpen
  \bibfield  {author} {\bibinfo {author} {\bibfnamefont {M.~M.}\ \bibnamefont
  {Qazilbash}}, \bibinfo {author} {\bibfnamefont {M.}~\bibnamefont {Brehm}},
  \bibinfo {author} {\bibfnamefont {B.-G.}\ \bibnamefont {Chae}}, \bibinfo
  {author} {\bibfnamefont {P.-C.}\ \bibnamefont {Ho}}, \bibinfo {author}
  {\bibfnamefont {G.~O.}\ \bibnamefont {Andreev}}, \bibinfo {author}
  {\bibfnamefont {B.-J.}\ \bibnamefont {Kim}}, \bibinfo {author} {\bibfnamefont
  {S.~J.}\ \bibnamefont {Yun}}, \bibinfo {author} {\bibfnamefont {A.~V.}\
  \bibnamefont {Balatsky}}, \bibinfo {author} {\bibfnamefont {M.~B.}\
  \bibnamefont {Maple}}, \bibinfo {author} {\bibfnamefont {F.}~\bibnamefont
  {Keilmann}}, \bibinfo {author} {\bibfnamefont {H.-T.}\ \bibnamefont {Kim}}, \
  and\ \bibinfo {author} {\bibfnamefont {D.~N.}\ \bibnamefont {Basov}},\ }\href
  {\doibase 10.1126/science.1150124} {\bibfield  {journal} {\bibinfo  {journal}
  {Science}\ }\textbf {\bibinfo {volume} {318}},\ \bibinfo {pages} {1750}
  (\bibinfo {year} {2007})},\ \Eprint
  {http://arxiv.org/abs/http://www.sciencemag.org/content/318/5857/1750.full.pdf}
  {http://www.sciencemag.org/content/318/5857/1750.full.pdf} \BibitemShut
  {NoStop}%
\bibitem [{\citenamefont {Jones}\ \emph {et~al.}(2010)\citenamefont {Jones},
  \citenamefont {Berweger}, \citenamefont {Wei}, \citenamefont {Cobden},\ and\
  \citenamefont {Raschke}}]{RaschkeVO2}%
  \BibitemOpen
  \bibfield  {author} {\bibinfo {author} {\bibfnamefont {A.~C.}\ \bibnamefont
  {Jones}}, \bibinfo {author} {\bibfnamefont {S.}~\bibnamefont {Berweger}},
  \bibinfo {author} {\bibfnamefont {J.}~\bibnamefont {Wei}}, \bibinfo {author}
  {\bibfnamefont {D.}~\bibnamefont {Cobden}}, \ and\ \bibinfo {author}
  {\bibfnamefont {M.~B.}\ \bibnamefont {Raschke}},\ }\href {\doibase
  10.1021/nl903765h} {\bibfield  {journal} {\bibinfo  {journal} {Nano Letters}\
  }\textbf {\bibinfo {volume} {10}},\ \bibinfo {pages} {1574} (\bibinfo {year}
  {2010})},\ \Eprint
  {http://arxiv.org/abs/http://pubs.acs.org/doi/pdf/10.1021/nl903765h}
  {http://pubs.acs.org/doi/pdf/10.1021/nl903765h} \BibitemShut {NoStop}%
\bibitem [{\citenamefont {Fei}\ \emph {et~al.}(2011)\citenamefont {Fei},
  \citenamefont {Andreev}, \citenamefont {Bao}, \citenamefont {Zhang},
  \citenamefont {S.~McLeod}, \citenamefont {Wang}, \citenamefont {Stewart},
  \citenamefont {Zhao}, \citenamefont {Dominguez}, \citenamefont {Thiemens},
  \citenamefont {Fogler}, \citenamefont {Tauber}, \citenamefont {Castro-Neto},
  \citenamefont {Lau}, \citenamefont {Keilmann},\ and\ \citenamefont
  {Basov}}]{PhononEnhancement}%
  \BibitemOpen
  \bibfield  {author} {\bibinfo {author} {\bibfnamefont {Z.}~\bibnamefont
  {Fei}}, \bibinfo {author} {\bibfnamefont {G.~O.}\ \bibnamefont {Andreev}},
  \bibinfo {author} {\bibfnamefont {W.}~\bibnamefont {Bao}}, \bibinfo {author}
  {\bibfnamefont {L.~M.}\ \bibnamefont {Zhang}}, \bibinfo {author}
  {\bibfnamefont {A.}~\bibnamefont {S.~McLeod}}, \bibinfo {author}
  {\bibfnamefont {C.}~\bibnamefont {Wang}}, \bibinfo {author} {\bibfnamefont
  {M.~K.}\ \bibnamefont {Stewart}}, \bibinfo {author} {\bibfnamefont
  {Z.}~\bibnamefont {Zhao}}, \bibinfo {author} {\bibfnamefont {G.}~\bibnamefont
  {Dominguez}}, \bibinfo {author} {\bibfnamefont {M.}~\bibnamefont {Thiemens}},
  \bibinfo {author} {\bibfnamefont {M.~M.}\ \bibnamefont {Fogler}}, \bibinfo
  {author} {\bibfnamefont {M.~J.}\ \bibnamefont {Tauber}}, \bibinfo {author}
  {\bibfnamefont {A.~H.}\ \bibnamefont {Castro-Neto}}, \bibinfo {author}
  {\bibfnamefont {C.~N.}\ \bibnamefont {Lau}}, \bibinfo {author} {\bibfnamefont
  {F.}~\bibnamefont {Keilmann}}, \ and\ \bibinfo {author} {\bibfnamefont
  {D.~N.}\ \bibnamefont {Basov}},\ }\href {\doibase 10.1021/nl202362d}
  {\bibfield  {journal} {\bibinfo  {journal} {Nano Letters}\ }\textbf {\bibinfo
  {volume} {11}},\ \bibinfo {pages} {4701} (\bibinfo {year} {2011})},\ \Eprint
  {http://arxiv.org/abs/http://pubs.acs.org/doi/pdf/10.1021/nl202362d}
  {http://pubs.acs.org/doi/pdf/10.1021/nl202362d} \BibitemShut {NoStop}%
\bibitem [{\citenamefont {Fei}\ \emph {et~al.}(2012)\citenamefont {Fei},
  \citenamefont {Rodin}, \citenamefont {Andreev}, \citenamefont {Bao},
  \citenamefont {McLeod}, \citenamefont {Wagner}, \citenamefont {Zhang},
  \citenamefont {Zhao}, \citenamefont {Thiemens}, \citenamefont {Dominguez},
  \citenamefont {Fogler}, \citenamefont {Castro~Neto}, \citenamefont {Lau},
  \citenamefont {Keilmann},\ and\ \citenamefont {Basov}}]{FeiGraphenePlasmons}%
  \BibitemOpen
  \bibfield  {author} {\bibinfo {author} {\bibfnamefont {Z.}~\bibnamefont
  {Fei}}, \bibinfo {author} {\bibfnamefont {A.~S.}\ \bibnamefont {Rodin}},
  \bibinfo {author} {\bibfnamefont {G.~O.}\ \bibnamefont {Andreev}}, \bibinfo
  {author} {\bibfnamefont {W.}~\bibnamefont {Bao}}, \bibinfo {author}
  {\bibfnamefont {A.~S.}\ \bibnamefont {McLeod}}, \bibinfo {author}
  {\bibfnamefont {M.}~\bibnamefont {Wagner}}, \bibinfo {author} {\bibfnamefont
  {L.~M.}\ \bibnamefont {Zhang}}, \bibinfo {author} {\bibfnamefont
  {Z.}~\bibnamefont {Zhao}}, \bibinfo {author} {\bibfnamefont {M.}~\bibnamefont
  {Thiemens}}, \bibinfo {author} {\bibfnamefont {G.}~\bibnamefont {Dominguez}},
  \bibinfo {author} {\bibfnamefont {M.~M.}\ \bibnamefont {Fogler}}, \bibinfo
  {author} {\bibfnamefont {A.~H.}\ \bibnamefont {Castro~Neto}}, \bibinfo
  {author} {\bibfnamefont {C.~N.}\ \bibnamefont {Lau}}, \bibinfo {author}
  {\bibfnamefont {F.}~\bibnamefont {Keilmann}}, \ and\ \bibinfo {author}
  {\bibfnamefont {D.~N.}\ \bibnamefont {Basov}},\ }\href@noop {} {\bibfield
  {journal} {\bibinfo  {journal} {Nature}\ }\textbf {\bibinfo {volume} {487}},\
  \bibinfo {pages} {82} (\bibinfo {year} {2012})}\BibitemShut {NoStop}%
\bibitem [{\citenamefont {Chen}\ \emph {et~al.}(2012)\citenamefont {Chen},
  \citenamefont {Badioli}, \citenamefont {Alonso-Gonzalez}, \citenamefont
  {Thongrattanasiri}, \citenamefont {Huth}, \citenamefont {Osmond},
  \citenamefont {Spasenovic}, \citenamefont {Centeno}, \citenamefont
  {Pesquera}, \citenamefont {Godignon}, \citenamefont {Zurutuza~Elorza},
  \citenamefont {Camara}, \citenamefont {Garcia~de Abajo}, \citenamefont
  {Hillenbrand},\ and\ \citenamefont {Koppens}}]{ChenGraphenePlasmons}%
  \BibitemOpen
  \bibfield  {author} {\bibinfo {author} {\bibfnamefont {J.}~\bibnamefont
  {Chen}}, \bibinfo {author} {\bibfnamefont {M.}~\bibnamefont {Badioli}},
  \bibinfo {author} {\bibfnamefont {P.}~\bibnamefont {Alonso-Gonzalez}},
  \bibinfo {author} {\bibfnamefont {S.}~\bibnamefont {Thongrattanasiri}},
  \bibinfo {author} {\bibfnamefont {F.}~\bibnamefont {Huth}}, \bibinfo {author}
  {\bibfnamefont {J.}~\bibnamefont {Osmond}}, \bibinfo {author} {\bibfnamefont
  {M.}~\bibnamefont {Spasenovic}}, \bibinfo {author} {\bibfnamefont
  {A.}~\bibnamefont {Centeno}}, \bibinfo {author} {\bibfnamefont
  {A.}~\bibnamefont {Pesquera}}, \bibinfo {author} {\bibfnamefont
  {P.}~\bibnamefont {Godignon}}, \bibinfo {author} {\bibfnamefont
  {A.}~\bibnamefont {Zurutuza~Elorza}}, \bibinfo {author} {\bibfnamefont
  {N.}~\bibnamefont {Camara}}, \bibinfo {author} {\bibfnamefont {F.~J.}\
  \bibnamefont {Garcia~de Abajo}}, \bibinfo {author} {\bibfnamefont
  {R.}~\bibnamefont {Hillenbrand}}, \ and\ \bibinfo {author} {\bibfnamefont
  {F.~H.~L.}\ \bibnamefont {Koppens}},\ }\href@noop {} {\bibfield  {journal}
  {\bibinfo  {journal} {Nature}\ }\textbf {\bibinfo {volume} {487}},\ \bibinfo
  {pages} {77} (\bibinfo {year} {2012})}\BibitemShut {NoStop}%
\bibitem [{\citenamefont {Charnukha}\ \emph {et~al.}(2012)\citenamefont
  {Charnukha}, \citenamefont {Cvitkovic}, \citenamefont {Prokscha},
  \citenamefont {Pr\"opper}, \citenamefont {Ocelic}, \citenamefont {Suter},
  \citenamefont {Salman}, \citenamefont {Morenzoni}, \citenamefont
  {Deisenhofer}, \citenamefont {Tsurkan}, \citenamefont {Loidl}, \citenamefont
  {Keimer},\ and\ \citenamefont {Boris}}]{Charnukha}%
  \BibitemOpen
  \bibfield  {author} {\bibinfo {author} {\bibfnamefont {A.}~\bibnamefont
  {Charnukha}}, \bibinfo {author} {\bibfnamefont {A.}~\bibnamefont
  {Cvitkovic}}, \bibinfo {author} {\bibfnamefont {T.}~\bibnamefont {Prokscha}},
  \bibinfo {author} {\bibfnamefont {D.}~\bibnamefont {Pr\"opper}}, \bibinfo
  {author} {\bibfnamefont {N.}~\bibnamefont {Ocelic}}, \bibinfo {author}
  {\bibfnamefont {A.}~\bibnamefont {Suter}}, \bibinfo {author} {\bibfnamefont
  {Z.}~\bibnamefont {Salman}}, \bibinfo {author} {\bibfnamefont
  {E.}~\bibnamefont {Morenzoni}}, \bibinfo {author} {\bibfnamefont
  {J.}~\bibnamefont {Deisenhofer}}, \bibinfo {author} {\bibfnamefont
  {V.}~\bibnamefont {Tsurkan}}, \bibinfo {author} {\bibfnamefont
  {A.}~\bibnamefont {Loidl}}, \bibinfo {author} {\bibfnamefont
  {B.}~\bibnamefont {Keimer}}, \ and\ \bibinfo {author} {\bibfnamefont {A.~V.}\
  \bibnamefont {Boris}},\ }\href {\doibase 10.1103/PhysRevLett.109.017003}
  {\bibfield  {journal} {\bibinfo  {journal} {Phys. Rev. Lett.}\ }\textbf
  {\bibinfo {volume} {109}},\ \bibinfo {pages} {017003} (\bibinfo {year}
  {2012})}\BibitemShut {NoStop}%
\bibitem [{\citenamefont {Wagner}\ \emph {et~al.}(2013)\citenamefont {Wagner},
  \citenamefont {Fei}, \citenamefont {McLeod}, \citenamefont {Rodin},
  \citenamefont {Bao}, \citenamefont {Iwinski}, \citenamefont {Zhao},
  \citenamefont {Goldflam}, \citenamefont {Liu}, \citenamefont {Dominguez},
  \citenamefont {Thiemens}, \citenamefont {Fogler}, \citenamefont
  {Castro-Neto}, \citenamefont {Lau}, \citenamefont {Amarie}, \citenamefont
  {Keilmann},\ and\ \citenamefont {Basov}}]{MartinPumpProbe}%
  \BibitemOpen
  \bibfield  {author} {\bibinfo {author} {\bibfnamefont {M.}~\bibnamefont
  {Wagner}}, \bibinfo {author} {\bibfnamefont {Z.}~\bibnamefont {Fei}},
  \bibinfo {author} {\bibfnamefont {A.~S.}\ \bibnamefont {McLeod}}, \bibinfo
  {author} {\bibfnamefont {A.~S.}\ \bibnamefont {Rodin}}, \bibinfo {author}
  {\bibfnamefont {W.}~\bibnamefont {Bao}}, \bibinfo {author} {\bibfnamefont
  {E.~G.}\ \bibnamefont {Iwinski}}, \bibinfo {author} {\bibfnamefont
  {Z.}~\bibnamefont {Zhao}}, \bibinfo {author} {\bibfnamefont {M.}~\bibnamefont
  {Goldflam}}, \bibinfo {author} {\bibfnamefont {M.}~\bibnamefont {Liu}},
  \bibinfo {author} {\bibfnamefont {G.}~\bibnamefont {Dominguez}}, \bibinfo
  {author} {\bibfnamefont {M.}~\bibnamefont {Thiemens}}, \bibinfo {author}
  {\bibfnamefont {M.~M.}\ \bibnamefont {Fogler}}, \bibinfo {author}
  {\bibfnamefont {A.~H.}\ \bibnamefont {Castro-Neto}}, \bibinfo {author}
  {\bibfnamefont {C.~N.}\ \bibnamefont {Lau}}, \bibinfo {author} {\bibfnamefont
  {S.}~\bibnamefont {Amarie}}, \bibinfo {author} {\bibfnamefont
  {F.}~\bibnamefont {Keilmann}}, \ and\ \bibinfo {author} {\bibfnamefont
  {D.~N.}\ \bibnamefont {Basov}},\ }\href@noop {} {\bibfield  {journal}
  {\bibinfo  {journal} {In submission}\ } (\bibinfo {year} {2013})}\BibitemShut
  {NoStop}%
\bibitem [{\citenamefont {Ford}\ and\ \citenamefont {Weber}(1984)}]{FordWeber}%
  \BibitemOpen
  \bibfield  {author} {\bibinfo {author} {\bibfnamefont {G.~W.}\ \bibnamefont
  {Ford}}\ and\ \bibinfo {author} {\bibfnamefont {W.~H.}\ \bibnamefont
  {Weber}},\ }\href@noop {} {\bibfield  {journal} {\bibinfo  {journal} {Physics
  Reports}\ }\textbf {\bibinfo {volume} {113}},\ \bibinfo {pages} {195}
  (\bibinfo {year} {1984})}\BibitemShut {NoStop}%
\bibitem [{\citenamefont {Hillenbrand}\ \emph {et~al.}(2001)\citenamefont
  {Hillenbrand}, \citenamefont {Knoll},\ and\ \citenamefont
  {Keilmann}}]{PureOpticalContrast}%
  \BibitemOpen
  \bibfield  {author} {\bibinfo {author} {\bibfnamefont {R.}~\bibnamefont
  {Hillenbrand}}, \bibinfo {author} {\bibfnamefont {B.}~\bibnamefont {Knoll}},
  \ and\ \bibinfo {author} {\bibfnamefont {F.}~\bibnamefont {Keilmann}},\
  }\href {\doibase 10.1046/j.1365-2818.2001.00794.x} {\bibfield  {journal}
  {\bibinfo  {journal} {Journal of Microscopy}\ }\textbf {\bibinfo {volume}
  {202}},\ \bibinfo {pages} {77} (\bibinfo {year} {2001})}\BibitemShut
  {NoStop}%
\bibitem [{\citenamefont {Aizpurua}\ \emph {et~al.}(2008)\citenamefont
  {Aizpurua}, \citenamefont {Taubner}, \citenamefont {de~Abajo}, \citenamefont
  {Brehm},\ and\ \citenamefont {Hillenbrand}}]{AizpuruaTaubner}%
  \BibitemOpen
  \bibfield  {author} {\bibinfo {author} {\bibfnamefont {J.}~\bibnamefont
  {Aizpurua}}, \bibinfo {author} {\bibfnamefont {T.}~\bibnamefont {Taubner}},
  \bibinfo {author} {\bibfnamefont {F.~J.~G.}\ \bibnamefont {de~Abajo}},
  \bibinfo {author} {\bibfnamefont {M.}~\bibnamefont {Brehm}}, \ and\ \bibinfo
  {author} {\bibfnamefont {R.}~\bibnamefont {Hillenbrand}},\ }\href {\doibase
  10.1364/OE.16.001529} {\bibfield  {journal} {\bibinfo  {journal} {Opt.
  Express}\ }\textbf {\bibinfo {volume} {16}},\ \bibinfo {pages} {1529}
  (\bibinfo {year} {2008})}\BibitemShut {NoStop}%
\bibitem [{\citenamefont {Porto}\ \emph {et~al.}(2003)\citenamefont {Porto},
  \citenamefont {Johansson}, \citenamefont {Apell},\ and\ \citenamefont
  {L\'opez-R\'ios}}]{Porto}%
  \BibitemOpen
  \bibfield  {author} {\bibinfo {author} {\bibfnamefont {J.~A.}\ \bibnamefont
  {Porto}}, \bibinfo {author} {\bibfnamefont {P.}~\bibnamefont {Johansson}},
  \bibinfo {author} {\bibfnamefont {S.~P.}\ \bibnamefont {Apell}}, \ and\
  \bibinfo {author} {\bibfnamefont {T.}~\bibnamefont {L\'opez-R\'ios}},\ }\href
  {\doibase 10.1103/PhysRevB.67.085409} {\bibfield  {journal} {\bibinfo
  {journal} {Phys. Rev. B}\ }\textbf {\bibinfo {volume} {67}},\ \bibinfo
  {pages} {085409} (\bibinfo {year} {2003})}\BibitemShut {NoStop}%
\bibitem [{\citenamefont {Moon}\ \emph {et~al.}(2011)\citenamefont {Moon},
  \citenamefont {Jung}, \citenamefont {Lim}, \citenamefont {Do},\ and\
  \citenamefont {Han}}]{Moon}%
  \BibitemOpen
  \bibfield  {author} {\bibinfo {author} {\bibfnamefont {K.}~\bibnamefont
  {Moon}}, \bibinfo {author} {\bibfnamefont {E.}~\bibnamefont {Jung}}, \bibinfo
  {author} {\bibfnamefont {M.}~\bibnamefont {Lim}}, \bibinfo {author}
  {\bibfnamefont {Y.}~\bibnamefont {Do}}, \ and\ \bibinfo {author}
  {\bibfnamefont {H.}~\bibnamefont {Han}},\ }\href {\doibase
  10.1364/OE.19.011539} {\bibfield  {journal} {\bibinfo  {journal} {Opt.
  Express}\ }\textbf {\bibinfo {volume} {19}},\ \bibinfo {pages} {11539}
  (\bibinfo {year} {2011})}\BibitemShut {NoStop}%
\bibitem [{\citenamefont {Cvitkovic}\ \emph {et~al.}(2007)\citenamefont
  {Cvitkovic}, \citenamefont {Ocelic},\ and\ \citenamefont
  {Hillenbrand}}]{Cvitkovic}%
  \BibitemOpen
  \bibfield  {author} {\bibinfo {author} {\bibfnamefont {A.}~\bibnamefont
  {Cvitkovic}}, \bibinfo {author} {\bibfnamefont {N.}~\bibnamefont {Ocelic}}, \
  and\ \bibinfo {author} {\bibfnamefont {R.}~\bibnamefont {Hillenbrand}},\
  }\href {\doibase 10.1364/OE.15.008550} {\bibfield  {journal} {\bibinfo
  {journal} {Opt. Express}\ }\textbf {\bibinfo {volume} {15}},\ \bibinfo
  {pages} {8550} (\bibinfo {year} {2007})}\BibitemShut {NoStop}%
\bibitem [{\citenamefont {Hauer}\ \emph {et~al.}(2012)\citenamefont {Hauer},
  \citenamefont {Engelhardt},\ and\ \citenamefont {Taubner}}]{Hauer}%
  \BibitemOpen
  \bibfield  {author} {\bibinfo {author} {\bibfnamefont {B.}~\bibnamefont
  {Hauer}}, \bibinfo {author} {\bibfnamefont {A.~P.}\ \bibnamefont
  {Engelhardt}}, \ and\ \bibinfo {author} {\bibfnamefont {T.}~\bibnamefont
  {Taubner}},\ }\href {\doibase 10.1364/OE.20.013173} {\bibfield  {journal}
  {\bibinfo  {journal} {Opt. Express}\ }\textbf {\bibinfo {volume} {20}},\
  \bibinfo {pages} {13173} (\bibinfo {year} {2012})}\BibitemShut {NoStop}%
\bibitem [{\citenamefont {Govyadinov}\ \emph {et~al.}(2013)\citenamefont
  {Govyadinov}, \citenamefont {Amenabar}, \citenamefont {Huth}, \citenamefont
  {Carney},\ and\ \citenamefont {Hillenbrand}}]{FDPInverse}%
  \BibitemOpen
  \bibfield  {author} {\bibinfo {author} {\bibfnamefont {A.~A.}\ \bibnamefont
  {Govyadinov}}, \bibinfo {author} {\bibfnamefont {I.}~\bibnamefont
  {Amenabar}}, \bibinfo {author} {\bibfnamefont {F.}~\bibnamefont {Huth}},
  \bibinfo {author} {\bibfnamefont {P.~S.}\ \bibnamefont {Carney}}, \ and\
  \bibinfo {author} {\bibfnamefont {R.}~\bibnamefont {Hillenbrand}},\ }\href
  {\doibase 10.1021/jz400453r} {\bibfield  {journal} {\bibinfo  {journal} {The
  Journal of Physical Chemistry Letters}\ }\textbf {\bibinfo {volume} {4}},\
  \bibinfo {pages} {1526} (\bibinfo {year} {2013})}\BibitemShut {NoStop}%
\bibitem [{\citenamefont {Carminati}\ and\ \citenamefont
  {S\'aenz}(2000)}]{Carminati}%
  \BibitemOpen
  \bibfield  {author} {\bibinfo {author} {\bibfnamefont {R.}~\bibnamefont
  {Carminati}}\ and\ \bibinfo {author} {\bibfnamefont {J.~J.}\ \bibnamefont
  {S\'aenz}},\ }\href {\doibase 10.1103/PhysRevLett.84.5156} {\bibfield
  {journal} {\bibinfo  {journal} {Phys. Rev. Lett.}\ }\textbf {\bibinfo
  {volume} {84}},\ \bibinfo {pages} {5156} (\bibinfo {year}
  {2000})}\BibitemShut {NoStop}%
\bibitem [{\citenamefont {Esslinger}\ and\ \citenamefont
  {Vogelgesang}(2012)}]{Esslinger}%
  \BibitemOpen
  \bibfield  {author} {\bibinfo {author} {\bibfnamefont {M.}~\bibnamefont
  {Esslinger}}\ and\ \bibinfo {author} {\bibfnamefont {R.}~\bibnamefont
  {Vogelgesang}},\ }\href {\doibase 10.1021/nn302864d} {\bibfield  {journal}
  {\bibinfo  {journal} {ACS Nano}\ }\textbf {\bibinfo {volume} {6}},\ \bibinfo
  {pages} {8173} (\bibinfo {year} {2012})},\ \Eprint
  {http://arxiv.org/abs/http://pubs.acs.org/doi/pdf/10.1021/nn302864d}
  {http://pubs.acs.org/doi/pdf/10.1021/nn302864d} \BibitemShut {NoStop}%
\bibitem [{\citenamefont {Valle}\ \emph {et~al.}(1998)\citenamefont {Valle},
  \citenamefont {Carminati},\ and\ \citenamefont {Greffet}}]{Valle}%
  \BibitemOpen
  \bibfield  {author} {\bibinfo {author} {\bibfnamefont {P.~J.}\ \bibnamefont
  {Valle}}, \bibinfo {author} {\bibfnamefont {R.}~\bibnamefont {Carminati}}, \
  and\ \bibinfo {author} {\bibfnamefont {J.-J.}\ \bibnamefont {Greffet}},\
  }\href {\doibase 10.1016/S0304-3991(97)00115-0} {\bibfield  {journal}
  {\bibinfo  {journal} {Ultramicroscopy}\ }\textbf {\bibinfo {volume} {71}},\
  \bibinfo {pages} {39 } (\bibinfo {year} {1998})}\BibitemShut {NoStop}%
\bibitem [{\citenamefont {Pike}\ and\ \citenamefont
  {Sabatier}(2002)}]{Scattering}%
  \BibitemOpen
  \bibinfo {editor} {\bibfnamefont {R.}~\bibnamefont {Pike}}\ and\ \bibinfo
  {editor} {\bibfnamefont {P.}~\bibnamefont {Sabatier}},\ eds.,\ \href@noop {}
  {\emph {\bibinfo {title} {Scattering and Inverse Scattering in Pure and
  Applied Science}}},\ \bibinfo {edition} {1st}\ ed.,\ Vol.~\bibinfo {volume}
  {1}\ (\bibinfo  {publisher} {Academic Press},\ \bibinfo {address} {525 B
  Street, Suite 1900, San Diego, California 92101-4495},\ \bibinfo {year}
  {2002})\BibitemShut {NoStop}%
\bibitem [{\citenamefont {Nystr\"{o}m}(1930)}]{Nystrom}%
  \BibitemOpen
  \bibfield  {author} {\bibinfo {author} {\bibfnamefont {E.}~\bibnamefont
  {Nystr\"{o}m}},\ }\href {http://dx.doi.org/10.1007/BF02547521} {\bibfield
  {journal} {\bibinfo  {journal} {Acta Mathematica}\ }\textbf {\bibinfo
  {volume} {54}},\ \bibinfo {pages} {185} (\bibinfo {year} {1930})},\ \bibinfo
  {note} {10.1007/BF02547521}\BibitemShut {NoStop}%
\bibitem [{\citenamefont {Zhang}\ \emph {et~al.}(2012)\citenamefont {Zhang},
  \citenamefont {Andreev}, \citenamefont {Fei}, \citenamefont {McLeod},
  \citenamefont {Dominguez}, \citenamefont {Thiemens}, \citenamefont
  {Castro-Neto}, \citenamefont {Basov},\ and\ \citenamefont {Fogler}}]{Zhang}%
  \BibitemOpen
  \bibfield  {author} {\bibinfo {author} {\bibfnamefont {L.~M.}\ \bibnamefont
  {Zhang}}, \bibinfo {author} {\bibfnamefont {G.~O.}\ \bibnamefont {Andreev}},
  \bibinfo {author} {\bibfnamefont {Z.}~\bibnamefont {Fei}}, \bibinfo {author}
  {\bibfnamefont {A.~S.}\ \bibnamefont {McLeod}}, \bibinfo {author}
  {\bibfnamefont {G.}~\bibnamefont {Dominguez}}, \bibinfo {author}
  {\bibfnamefont {M.}~\bibnamefont {Thiemens}}, \bibinfo {author}
  {\bibfnamefont {A.~H.}\ \bibnamefont {Castro-Neto}}, \bibinfo {author}
  {\bibfnamefont {D.~N.}\ \bibnamefont {Basov}}, \ and\ \bibinfo {author}
  {\bibfnamefont {M.~M.}\ \bibnamefont {Fogler}},\ }\href {\doibase
  10.1103/PhysRevB.85.075419} {\bibfield  {journal} {\bibinfo  {journal} {Phys.
  Rev. B}\ }\textbf {\bibinfo {volume} {85}},\ \bibinfo {pages} {075419}
  (\bibinfo {year} {2012})}\BibitemShut {NoStop}%
\bibitem [{\citenamefont {Behr}\ and\ \citenamefont
  {Raschke}(2008)}]{RaschkeHyperboloid}%
  \BibitemOpen
  \bibfield  {author} {\bibinfo {author} {\bibfnamefont {N.}~\bibnamefont
  {Behr}}\ and\ \bibinfo {author} {\bibfnamefont {M.~B.}\ \bibnamefont
  {Raschke}},\ }\href@noop {} {\bibfield  {journal} {\bibinfo  {journal} {J.
  Phys. Chem. C}\ }\textbf {\bibinfo {volume} {112}},\ \bibinfo {pages} {3766}
  (\bibinfo {year} {2008})}\BibitemShut {NoStop}%
\bibitem [{\citenamefont {Hartschuh}(2008)}]{HartschuhReview}%
  \BibitemOpen
  \bibfield  {author} {\bibinfo {author} {\bibfnamefont {A.}~\bibnamefont
  {Hartschuh}},\ }\href {\doibase 10.1002/anie.200801605} {\bibfield  {journal}
  {\bibinfo  {journal} {Angewandte Chemie International Edition}\ }\textbf
  {\bibinfo {volume} {47}},\ \bibinfo {pages} {8178} (\bibinfo {year}
  {2008})}\BibitemShut {NoStop}%
\bibitem [{\citenamefont {Kildemo}(1998)}]{Kildemo}%
  \BibitemOpen
  \bibfield  {author} {\bibinfo {author} {\bibfnamefont {M.}~\bibnamefont
  {Kildemo}},\ }\href {\doibase 10.1364/AO.37.000113} {\bibfield  {journal}
  {\bibinfo  {journal} {Appl. Opt.}\ }\textbf {\bibinfo {volume} {37}},\
  \bibinfo {pages} {113} (\bibinfo {year} {1998})}\BibitemShut {NoStop}%
\bibitem [{\citenamefont {Ku\v{c}\'{i}rkov\'{a}}\ and\ \citenamefont
  {Navr\'{a}til}(1994)}]{SiO2}%
  \BibitemOpen
  \bibfield  {author} {\bibinfo {author} {\bibfnamefont {A.}~\bibnamefont
  {Ku\v{c}\'{i}rkov\'{a}}}\ and\ \bibinfo {author} {\bibfnamefont
  {K.}~\bibnamefont {Navr\'{a}til}},\ }\href
  {http://as.osa.org/abstract.cfm?URI=as-48-1-113} {\bibfield  {journal}
  {\bibinfo  {journal} {Appl. Spectrosc.}\ }\textbf {\bibinfo {volume} {48}},\
  \bibinfo {pages} {113} (\bibinfo {year} {1994})}\BibitemShut {NoStop}%
\bibitem [{\citenamefont {Garc\'{i}a-Etxarri}\ \emph
  {et~al.}(2009)\citenamefont {Garc\'{i}a-Etxarri}, \citenamefont {Romero},
  \citenamefont {Garc\'{i}a~de Abajo}, \citenamefont {Hillenbrand},\ and\
  \citenamefont {Aizpurua}}]{StrongWeakCoupling}%
  \BibitemOpen
  \bibfield  {author} {\bibinfo {author} {\bibfnamefont {A.}~\bibnamefont
  {Garc\'{i}a-Etxarri}}, \bibinfo {author} {\bibfnamefont {I.}~\bibnamefont
  {Romero}}, \bibinfo {author} {\bibfnamefont {F.~J.}\ \bibnamefont
  {Garc\'{i}a~de Abajo}}, \bibinfo {author} {\bibfnamefont {R.}~\bibnamefont
  {Hillenbrand}}, \ and\ \bibinfo {author} {\bibfnamefont {J.}~\bibnamefont
  {Aizpurua}},\ }\href@noop {} {\bibfield  {journal} {\bibinfo  {journal}
  {Phys. Rev. B}\ }\textbf {\bibinfo {volume} {79}} (\bibinfo {year}
  {2009})}\BibitemShut {NoStop}%
\bibitem [{\citenamefont {Schell}\ \emph {et~al.}(2009)\citenamefont {Schell},
  \citenamefont {Garc\'{i}a-Etxarri}, \citenamefont {Huber}, \citenamefont
  {Crozier}, \citenamefont {Aizpurua},\ and\ \citenamefont
  {Hillenbrand}}]{MappingNanoantennas}%
  \BibitemOpen
  \bibfield  {author} {\bibinfo {author} {\bibfnamefont {M.}~\bibnamefont
  {Schell}}, \bibinfo {author} {\bibfnamefont {A.}~\bibnamefont
  {Garc\'{i}a-Etxarri}}, \bibinfo {author} {\bibfnamefont {A.~J.}\ \bibnamefont
  {Huber}}, \bibinfo {author} {\bibfnamefont {K.}~\bibnamefont {Crozier}},
  \bibinfo {author} {\bibfnamefont {J.}~\bibnamefont {Aizpurua}}, \ and\
  \bibinfo {author} {\bibfnamefont {R.}~\bibnamefont {Hillenbrand}},\
  }\href@noop {} {\bibfield  {journal} {\bibinfo  {journal} {Nat. Photonics}\
  }\textbf {\bibinfo {volume} {3}},\ \bibinfo {pages} {287} (\bibinfo {year}
  {2009})}\BibitemShut {NoStop}%
\bibitem [{\citenamefont {Novotny}\ and\ \citenamefont
  {Stranick}(2006)}]{NovotnyReview}%
  \BibitemOpen
  \bibfield  {author} {\bibinfo {author} {\bibfnamefont {L.}~\bibnamefont
  {Novotny}}\ and\ \bibinfo {author} {\bibfnamefont {S.~J.}\ \bibnamefont
  {Stranick}},\ }\href {\doibase 10.1146/annurev.physchem.56.092503.141236}
  {\bibfield  {journal} {\bibinfo  {journal} {Annual Review of Physical
  Chemistry}\ }\textbf {\bibinfo {volume} {57}},\ \bibinfo {pages} {303}
  (\bibinfo {year} {2006})},\ \bibinfo {note} {pMID: 16599813}\BibitemShut
  {NoStop}%
\bibitem [{\citenamefont {Brehm}\ \emph {et~al.}(2008)\citenamefont {Brehm},
  \citenamefont {Schliesser}, \citenamefont {Cajko}, \citenamefont
  {Tsukerman},\ and\ \citenamefont {Keilmann}}]{KeilmannAntenna}%
  \BibitemOpen
  \bibfield  {author} {\bibinfo {author} {\bibfnamefont {M.}~\bibnamefont
  {Brehm}}, \bibinfo {author} {\bibfnamefont {A.}~\bibnamefont {Schliesser}},
  \bibinfo {author} {\bibfnamefont {F.}~\bibnamefont {Cajko}}, \bibinfo
  {author} {\bibfnamefont {I.}~\bibnamefont {Tsukerman}}, \ and\ \bibinfo
  {author} {\bibfnamefont {F.}~\bibnamefont {Keilmann}},\ }\href {\doibase
  10.1364/OE.16.011203} {\bibfield  {journal} {\bibinfo  {journal} {Opt.
  Express}\ }\textbf {\bibinfo {volume} {16}},\ \bibinfo {pages} {11203}
  (\bibinfo {year} {2008})}\BibitemShut {NoStop}%
\bibitem [{\citenamefont {Huth}\ \emph {et~al.}(2013)\citenamefont {Huth},
  \citenamefont {Chuvilin}, \citenamefont {Schnell}, \citenamefont {Amenabar},
  \citenamefont {Krutokhvostov}, \citenamefont {Lopatin},\ and\ \citenamefont
  {Hillenbrand}}]{ResonantAntennas}%
  \BibitemOpen
  \bibfield  {author} {\bibinfo {author} {\bibfnamefont {F.}~\bibnamefont
  {Huth}}, \bibinfo {author} {\bibfnamefont {A.}~\bibnamefont {Chuvilin}},
  \bibinfo {author} {\bibfnamefont {M.}~\bibnamefont {Schnell}}, \bibinfo
  {author} {\bibfnamefont {I.}~\bibnamefont {Amenabar}}, \bibinfo {author}
  {\bibfnamefont {R.}~\bibnamefont {Krutokhvostov}}, \bibinfo {author}
  {\bibfnamefont {S.}~\bibnamefont {Lopatin}}, \ and\ \bibinfo {author}
  {\bibfnamefont {R.}~\bibnamefont {Hillenbrand}},\ }\href {\doibase
  10.1021/nl304289g} {\bibfield  {journal} {\bibinfo  {journal} {Nano Letters}\
  }\textbf {\bibinfo {volume} {13}},\ \bibinfo {pages} {1065} (\bibinfo {year}
  {2013})},\ \Eprint
  {http://arxiv.org/abs/http://pubs.acs.org/doi/pdf/10.1021/nl304289g}
  {http://pubs.acs.org/doi/pdf/10.1021/nl304289g} \BibitemShut {NoStop}%
\bibitem [{\citenamefont {Raschke}\ and\ \citenamefont
  {Lienau}(2003)}]{Raschke}%
  \BibitemOpen
  \bibfield  {author} {\bibinfo {author} {\bibfnamefont {M.~B.}\ \bibnamefont
  {Raschke}}\ and\ \bibinfo {author} {\bibfnamefont {C.}~\bibnamefont
  {Lienau}},\ }\href {\doibase 10.1063/1.1632023} {\bibfield  {journal}
  {\bibinfo  {journal} {Applied Physics Letters}\ }\textbf {\bibinfo {volume}
  {83}},\ \bibinfo {pages} {5089} (\bibinfo {year} {2003})}\BibitemShut
  {NoStop}%
\bibitem [{\citenamefont {Aigouy}\ \emph {et~al.}(1999)\citenamefont {Aigouy},
  \citenamefont {Lahrech}, \citenamefont {{A}sillon}, \citenamefont {Cory},
  \citenamefont {Boccara},\ and\ \citenamefont {Rivoal}}]{Polarization}%
  \BibitemOpen
  \bibfield  {author} {\bibinfo {author} {\bibfnamefont {L.}~\bibnamefont
  {Aigouy}}, \bibinfo {author} {\bibfnamefont {A.}~\bibnamefont {Lahrech}},
  \bibinfo {author} {\bibfnamefont {S.~G.}\ \bibnamefont {{A}sillon}}, \bibinfo
  {author} {\bibfnamefont {H.}~\bibnamefont {Cory}}, \bibinfo {author}
  {\bibfnamefont {A.~C.}\ \bibnamefont {Boccara}}, \ and\ \bibinfo {author}
  {\bibfnamefont {J.~C.}\ \bibnamefont {Rivoal}},\ }\href {\doibase
  10.1364/OL.24.000187} {\bibfield  {journal} {\bibinfo  {journal} {Opt.
  Lett.}\ }\textbf {\bibinfo {volume} {24}},\ \bibinfo {pages} {187} (\bibinfo
  {year} {1999})}\BibitemShut {NoStop}%
\bibitem [{\citenamefont {Biagioni}\ \emph {et~al.}(2012)\citenamefont
  {Biagioni}, \citenamefont {Huang},\ and\ \citenamefont {Hecht}}]{Biagioni}%
  \BibitemOpen
  \bibfield  {author} {\bibinfo {author} {\bibfnamefont {P.}~\bibnamefont
  {Biagioni}}, \bibinfo {author} {\bibfnamefont {J.-S.}\ \bibnamefont {Huang}},
  \ and\ \bibinfo {author} {\bibfnamefont {B.}~\bibnamefont {Hecht}},\ }\href
  {http://stacks.iop.org/0034-4885/75/i=2/a=024402} {\bibfield  {journal}
  {\bibinfo  {journal} {Reports on Progress in Physics}\ }\textbf {\bibinfo
  {volume} {75}},\ \bibinfo {pages} {024402} (\bibinfo {year}
  {2012})}\BibitemShut {NoStop}%
\bibitem [{\citenamefont {Bao}\ \emph {et~al.}(2013)\citenamefont {Bao},
  \citenamefont {Staffaroni}, \citenamefont {Bokor}, \citenamefont {Salmeron},
  \citenamefont {Yablonovitch}, \citenamefont {Cabrini}, \citenamefont
  {Weber-Bargioni},\ and\ \citenamefont {Schuck}}]{Campanile}%
  \BibitemOpen
  \bibfield  {author} {\bibinfo {author} {\bibfnamefont {W.}~\bibnamefont
  {Bao}}, \bibinfo {author} {\bibfnamefont {M.}~\bibnamefont {Staffaroni}},
  \bibinfo {author} {\bibfnamefont {J.}~\bibnamefont {Bokor}}, \bibinfo
  {author} {\bibfnamefont {M.~B.}\ \bibnamefont {Salmeron}}, \bibinfo {author}
  {\bibfnamefont {E.}~\bibnamefont {Yablonovitch}}, \bibinfo {author}
  {\bibfnamefont {S.}~\bibnamefont {Cabrini}}, \bibinfo {author} {\bibfnamefont
  {A.}~\bibnamefont {Weber-Bargioni}}, \ and\ \bibinfo {author} {\bibfnamefont
  {P.~J.}\ \bibnamefont {Schuck}},\ }\href {\doibase 10.1364/OE.21.008166}
  {\bibfield  {journal} {\bibinfo  {journal} {Opt. Express}\ }\textbf {\bibinfo
  {volume} {21}},\ \bibinfo {pages} {8166} (\bibinfo {year}
  {2013})}\BibitemShut {NoStop}%
\bibitem [{\citenamefont {Alaverdyan}\ \emph {et~al.}(2011)\citenamefont
  {Alaverdyan}, \citenamefont {Vamivakas}, \citenamefont {Barnes},
  \citenamefont {Lebouteiller}, \citenamefont {Hare},\ and\ \citenamefont
  {Atat\"{u}re}}]{DielectricAntennaLoading}%
  \BibitemOpen
  \bibfield  {author} {\bibinfo {author} {\bibfnamefont {Y.}~\bibnamefont
  {Alaverdyan}}, \bibinfo {author} {\bibfnamefont {N.}~\bibnamefont
  {Vamivakas}}, \bibinfo {author} {\bibfnamefont {J.}~\bibnamefont {Barnes}},
  \bibinfo {author} {\bibfnamefont {C.}~\bibnamefont {Lebouteiller}}, \bibinfo
  {author} {\bibfnamefont {J.}~\bibnamefont {Hare}}, \ and\ \bibinfo {author}
  {\bibfnamefont {M.}~\bibnamefont {Atat\"{u}re}},\ }\href {\doibase
  10.1364/OE.19.018175} {\bibfield  {journal} {\bibinfo  {journal} {Opt.
  Express}\ }\textbf {\bibinfo {volume} {19}},\ \bibinfo {pages} {18175}
  (\bibinfo {year} {2011})}\BibitemShut {NoStop}%
\bibitem [{\citenamefont {Large}\ \emph {et~al.}(2010)\citenamefont {Large},
  \citenamefont {Abb}, \citenamefont {Aizpurua},\ and\ \citenamefont
  {Muskens}}]{PhotoexcitedNanoantenna}%
  \BibitemOpen
  \bibfield  {author} {\bibinfo {author} {\bibfnamefont {N.}~\bibnamefont
  {Large}}, \bibinfo {author} {\bibfnamefont {M.}~\bibnamefont {Abb}}, \bibinfo
  {author} {\bibfnamefont {J.}~\bibnamefont {Aizpurua}}, \ and\ \bibinfo
  {author} {\bibfnamefont {O.~L.}\ \bibnamefont {Muskens}},\ }\href {\doibase
  10.1021/nl1001636} {\bibfield  {journal} {\bibinfo  {journal} {Nano Letters}\
  }\textbf {\bibinfo {volume} {10}},\ \bibinfo {pages} {1741} (\bibinfo {year}
  {2010})},\ \Eprint
  {http://arxiv.org/abs/http://pubs.acs.org/doi/pdf/10.1021/nl1001636}
  {http://pubs.acs.org/doi/pdf/10.1021/nl1001636} \BibitemShut {NoStop}%
\bibitem [{\citenamefont {Fromm}\ \emph {et~al.}(2004)\citenamefont {Fromm},
  \citenamefont {Sundaramurthy}, \citenamefont {Schuck}, \citenamefont {Kino},\
  and\ \citenamefont {Moerner}}]{SchuckBowties}%
  \BibitemOpen
  \bibfield  {author} {\bibinfo {author} {\bibfnamefont {D.~P.}\ \bibnamefont
  {Fromm}}, \bibinfo {author} {\bibfnamefont {A.}~\bibnamefont
  {Sundaramurthy}}, \bibinfo {author} {\bibfnamefont {P.~J.}\ \bibnamefont
  {Schuck}}, \bibinfo {author} {\bibfnamefont {G.}~\bibnamefont {Kino}}, \ and\
  \bibinfo {author} {\bibfnamefont {W.~E.}\ \bibnamefont {Moerner}},\ }\href
  {\doibase 10.1021/nl049951r} {\bibfield  {journal} {\bibinfo  {journal} {Nano
  Letters}\ }\textbf {\bibinfo {volume} {4}},\ \bibinfo {pages} {957} (\bibinfo
  {year} {2004})},\ \Eprint
  {http://arxiv.org/abs/http://pubs.acs.org/doi/pdf/10.1021/nl049951r}
  {http://pubs.acs.org/doi/pdf/10.1021/nl049951r} \BibitemShut {NoStop}%
\bibitem [{\citenamefont {Alonso-Gonz\'{a}lez}\ \emph
  {et~al.}(2013)\citenamefont {Alonso-Gonz\'{a}lez}, \citenamefont {Albella},
  \citenamefont {Golmar}, \citenamefont {Arzubiaga}, \citenamefont {Casanova},
  \citenamefont {Hueso}, \citenamefont {Aizpurua},\ and\ \citenamefont
  {Hillenbrand}}]{PabloAntenna}%
  \BibitemOpen
  \bibfield  {author} {\bibinfo {author} {\bibfnamefont {P.}~\bibnamefont
  {Alonso-Gonz\'{a}lez}}, \bibinfo {author} {\bibfnamefont {P.}~\bibnamefont
  {Albella}}, \bibinfo {author} {\bibfnamefont {F.}~\bibnamefont {Golmar}},
  \bibinfo {author} {\bibfnamefont {L.}~\bibnamefont {Arzubiaga}}, \bibinfo
  {author} {\bibfnamefont {F.}~\bibnamefont {Casanova}}, \bibinfo {author}
  {\bibfnamefont {L.~E.}\ \bibnamefont {Hueso}}, \bibinfo {author}
  {\bibfnamefont {J.}~\bibnamefont {Aizpurua}}, \ and\ \bibinfo {author}
  {\bibfnamefont {R.}~\bibnamefont {Hillenbrand}},\ }\href {\doibase
  10.1364/OE.21.001270} {\bibfield  {journal} {\bibinfo  {journal} {Opt.
  Express}\ }\textbf {\bibinfo {volume} {21}},\ \bibinfo {pages} {1270}
  (\bibinfo {year} {2013})}\BibitemShut {NoStop}%
\bibitem [{\citenamefont {Giustino}\ and\ \citenamefont
  {Pasquarello}(2005)}]{Giustino}%
  \BibitemOpen
  \bibfield  {author} {\bibinfo {author} {\bibfnamefont {F.}~\bibnamefont
  {Giustino}}\ and\ \bibinfo {author} {\bibfnamefont {A.}~\bibnamefont
  {Pasquarello}},\ }\href {\doibase 10.1103/PhysRevLett.95.187402} {\bibfield
  {journal} {\bibinfo  {journal} {Phys. Rev. Lett.}\ }\textbf {\bibinfo
  {volume} {95}},\ \bibinfo {pages} {187402} (\bibinfo {year}
  {2005})}\BibitemShut {NoStop}%
\bibitem [{\citenamefont {Balandin}\ \emph {et~al.}(2007)\citenamefont
  {Balandin}, \citenamefont {Pokatilov},\ and\ \citenamefont
  {Nika}}]{PhononConfinement}%
  \BibitemOpen
  \bibfield  {author} {\bibinfo {author} {\bibfnamefont {A.~A.}\ \bibnamefont
  {Balandin}}, \bibinfo {author} {\bibfnamefont {E.~P.}\ \bibnamefont
  {Pokatilov}}, \ and\ \bibinfo {author} {\bibfnamefont {D.}~\bibnamefont
  {Nika}},\ }\href {\doibase doi:10.1166/jno.2007.201} {\bibfield  {journal}
  {\bibinfo  {journal} {Journal of Nanoelectronics and Optoelectronics}\
  }\textbf {\bibinfo {volume} {2}},\ \bibinfo {pages} {140} (\bibinfo {year}
  {2007})}\BibitemShut {NoStop}%
\bibitem [{\citenamefont {Huber}\ \emph {et~al.}(2006)\citenamefont {Huber},
  \citenamefont {Ocelic}, \citenamefont {Taubner},\ and\ \citenamefont
  {Hillenbrand}}]{SiCNearField}%
  \BibitemOpen
  \bibfield  {author} {\bibinfo {author} {\bibfnamefont {A.}~\bibnamefont
  {Huber}}, \bibinfo {author} {\bibfnamefont {N.}~\bibnamefont {Ocelic}},
  \bibinfo {author} {\bibfnamefont {T.}~\bibnamefont {Taubner}}, \ and\
  \bibinfo {author} {\bibfnamefont {R.}~\bibnamefont {Hillenbrand}},\ }\href
  {\doibase 10.1021/nl060092b} {\bibfield  {journal} {\bibinfo  {journal} {Nano
  Letters}\ }\textbf {\bibinfo {volume} {6}},\ \bibinfo {pages} {774} (\bibinfo
  {year} {2006})},\ \bibinfo {note} {pMID: 16608282},\ \Eprint
  {http://arxiv.org/abs/http://pubs.acs.org/doi/pdf/10.1021/nl060092b}
  {http://pubs.acs.org/doi/pdf/10.1021/nl060092b} \BibitemShut {NoStop}%
\bibitem [{\citenamefont {Mutschke}\ \emph {et~al.}(1999)\citenamefont
  {Mutschke}, \citenamefont {Andersen}, \citenamefont {Clement}, \citenamefont
  {Henning},\ and\ \citenamefont {Peiter}}]{SiC}%
  \BibitemOpen
  \bibfield  {author} {\bibinfo {author} {\bibfnamefont {H.}~\bibnamefont
  {Mutschke}}, \bibinfo {author} {\bibfnamefont {A.}~\bibnamefont {Andersen}},
  \bibinfo {author} {\bibfnamefont {D.}~\bibnamefont {Clement}}, \bibinfo
  {author} {\bibfnamefont {T.}~\bibnamefont {Henning}}, \ and\ \bibinfo
  {author} {\bibfnamefont {G.}~\bibnamefont {Peiter}},\ }\href@noop {}
  {\bibfield  {journal} {\bibinfo  {journal} {arXiv:astro-ph/9903031}\ }
  (\bibinfo {year} {1999})}\BibitemShut {NoStop}%
\bibitem [{\citenamefont {Hillenbrand}(2004)}]{Phononics}%
  \BibitemOpen
  \bibfield  {author} {\bibinfo {author} {\bibfnamefont {R.}~\bibnamefont
  {Hillenbrand}},\ }\href {\doibase 10.1016/j.ultramic.2003.11.017} {\bibfield
  {journal} {\bibinfo  {journal} {Ultramicroscopy}\ }\textbf {\bibinfo {volume}
  {100}},\ \bibinfo {pages} {421 } (\bibinfo {year} {2004})}\BibitemShut
  {NoStop}%
\bibitem [{\citenamefont {Schmucker}\ \emph {et~al.}(2012)\citenamefont
  {Schmucker}, \citenamefont {Kumar}, \citenamefont {Abelson}, \citenamefont
  {Daly}, \citenamefont {Girolami}, \citenamefont {Bischof}, \citenamefont
  {Jaeger}, \citenamefont {Reidy}, \citenamefont {Gorman}, \citenamefont
  {J.~Alexander}, \citenamefont {Randall},\ and\ \citenamefont
  {Lyding}}]{TipSharpening}%
  \BibitemOpen
  \bibfield  {author} {\bibinfo {author} {\bibfnamefont {S.}~\bibnamefont
  {Schmucker}}, \bibinfo {author} {\bibfnamefont {N.}~\bibnamefont {Kumar}},
  \bibinfo {author} {\bibfnamefont {J.}~\bibnamefont {Abelson}}, \bibinfo
  {author} {\bibfnamefont {S.}~\bibnamefont {Daly}}, \bibinfo {author}
  {\bibfnamefont {G.}~\bibnamefont {Girolami}}, \bibinfo {author}
  {\bibfnamefont {M.}~\bibnamefont {Bischof}}, \bibinfo {author} {\bibfnamefont
  {D.}~\bibnamefont {Jaeger}}, \bibinfo {author} {\bibfnamefont
  {R.}~\bibnamefont {Reidy}}, \bibinfo {author} {\bibfnamefont
  {B.}~\bibnamefont {Gorman}}, \bibinfo {author} {\bibfnamefont {J.~B.}\
  \bibnamefont {J.~Alexander}}, \bibinfo {author} {\bibfnamefont
  {J.}~\bibnamefont {Randall}}, \ and\ \bibinfo {author} {\bibfnamefont
  {J.}~\bibnamefont {Lyding}},\ }\href {\doibase 10.1038/ncomms1907} {\bibfield
   {journal} {\bibinfo  {journal} {Nature Communications}\ }\textbf {\bibinfo
  {volume} {3}} (\bibinfo {year} {2012}),\ 10.1038/ncomms1907}\BibitemShut
  {NoStop}%
\bibitem [{\citenamefont {Kiusalaas}(2010)}]{Kiusalaas}%
  \BibitemOpen
  \bibfield  {author} {\bibinfo {author} {\bibfnamefont {J.}~\bibnamefont
  {Kiusalaas}},\ }\href@noop {} {\emph {\bibinfo {title} {Numerical Methods in
  Engineering with Python}}},\ \bibinfo {edition} {2nd}\ ed.\ (\bibinfo
  {publisher} {Cambridge University Press},\ \bibinfo {address} {New York, NY,
  USA},\ \bibinfo {year} {2010})\ pp.\ \bibinfo {pages} {246--262 \&
  216--229}\BibitemShut {NoStop}%
\bibitem [{\citenamefont {Marquardt}(1963)}]{Marquardt}%
  \BibitemOpen
  \bibfield  {author} {\bibinfo {author} {\bibfnamefont {D.~W.}\ \bibnamefont
  {Marquardt}},\ }\href {http://www.jstor.org/stable/2098941} {\bibfield
  {journal} {\bibinfo  {journal} {Journal of the Society for Industrial and
  Applied Mathematics}\ }\textbf {\bibinfo {volume} {11}},\ \bibinfo {pages}
  {pp. 431} (\bibinfo {year} {1963})}\BibitemShut {NoStop}%
\bibitem [{\citenamefont {Kuzmenko}(2005)}]{Kuzmenko}%
  \BibitemOpen
  \bibfield  {author} {\bibinfo {author} {\bibfnamefont {A.~B.}\ \bibnamefont
  {Kuzmenko}},\ }\href {\doibase 10.1063/1.1979470} {\bibfield  {journal}
  {\bibinfo  {journal} {Review of Scientific Instruments}\ }\textbf {\bibinfo
  {volume} {76}},\ \bibinfo {eid} {083108} (\bibinfo {year}
  {2005})}\BibitemShut {NoStop}%
\bibitem [{\citenamefont {Arico}\ \emph {et~al.}(2005)\citenamefont {Arico},
  \citenamefont {Bruce}, \citenamefont {Scrosati}, \citenamefont {Tarascon},\
  and\ \citenamefont {van Schalkwijk}}]{EnergyNanostructures}%
  \BibitemOpen
  \bibfield  {author} {\bibinfo {author} {\bibfnamefont {A.~S.}\ \bibnamefont
  {Arico}}, \bibinfo {author} {\bibfnamefont {P.}~\bibnamefont {Bruce}},
  \bibinfo {author} {\bibfnamefont {B.}~\bibnamefont {Scrosati}}, \bibinfo
  {author} {\bibfnamefont {J.-M.}\ \bibnamefont {Tarascon}}, \ and\ \bibinfo
  {author} {\bibfnamefont {W.}~\bibnamefont {van Schalkwijk}},\ }\href@noop {}
  {\bibfield  {journal} {\bibinfo  {journal} {Nat. Mater.}\ ,\ \bibinfo {pages}
  {366}} (\bibinfo {year} {2005})}\BibitemShut {NoStop}%
\bibitem [{\citenamefont {Hwang}\ \emph {et~al.}(2012)\citenamefont {Hwang},
  \citenamefont {Iwasa}, \citenamefont {Kawasaki}, \citenamefont {Keimer},
  \citenamefont {Nagaosa},\ and\ \citenamefont
  {Tokura}}]{OxideHeterostructures}%
  \BibitemOpen
  \bibfield  {author} {\bibinfo {author} {\bibfnamefont {H.~Y.}\ \bibnamefont
  {Hwang}}, \bibinfo {author} {\bibfnamefont {Y.}~\bibnamefont {Iwasa}},
  \bibinfo {author} {\bibfnamefont {M.}~\bibnamefont {Kawasaki}}, \bibinfo
  {author} {\bibfnamefont {B.}~\bibnamefont {Keimer}}, \bibinfo {author}
  {\bibfnamefont {N.}~\bibnamefont {Nagaosa}}, \ and\ \bibinfo {author}
  {\bibfnamefont {Y.}~\bibnamefont {Tokura}},\ }\href@noop {} {\bibfield
  {journal} {\bibinfo  {journal} {Nat. Mater.}\ ,\ \bibinfo {pages} {103}}
  (\bibinfo {year} {2012})}\BibitemShut {NoStop}%
\bibitem [{\citenamefont {Garc\'ia~de Abajo}\ and\ \citenamefont
  {Aizpurua}(1997)}]{AizpuruaDielectrics}%
  \BibitemOpen
  \bibfield  {author} {\bibinfo {author} {\bibfnamefont {F.~J.}\ \bibnamefont
  {Garc\'ia~de Abajo}}\ and\ \bibinfo {author} {\bibfnamefont {J.}~\bibnamefont
  {Aizpurua}},\ }\href {\doibase 10.1103/PhysRevB.56.15873} {\bibfield
  {journal} {\bibinfo  {journal} {Phys. Rev. B}\ }\textbf {\bibinfo {volume}
  {56}},\ \bibinfo {pages} {15873} (\bibinfo {year} {1997})}\BibitemShut
  {NoStop}%
\bibitem [{\citenamefont {Agudin}\ and\ \citenamefont
  {Platzeck}(1980)}]{FieldDecomposition}%
  \BibitemOpen
  \bibfield  {author} {\bibinfo {author} {\bibfnamefont {J.~L.}\ \bibnamefont
  {Agudin}}\ and\ \bibinfo {author} {\bibfnamefont {A.~M.}\ \bibnamefont
  {Platzeck}},\ }\href {\doibase 10.1364/JOSA.70.001329} {\bibfield  {journal}
  {\bibinfo  {journal} {J. Opt. Soc. Am.}\ }\textbf {\bibinfo {volume} {70}},\
  \bibinfo {pages} {1329} (\bibinfo {year} {1980})}\BibitemShut {NoStop}%
\bibitem [{\citenamefont {Twomey}(1963)}]{Twomey}%
  \BibitemOpen
  \bibfield  {author} {\bibinfo {author} {\bibfnamefont {S.}~\bibnamefont
  {Twomey}},\ }\href {\doibase http://doi.acm.org/10.1145/321150.321157}
  {\bibfield  {journal} {\bibinfo  {journal} {J. ACM}\ }\textbf {\bibinfo
  {volume} {10}},\ \bibinfo {pages} {97} (\bibinfo {year} {1963})}\BibitemShut
  {NoStop}%
\bibitem [{\citenamefont {Phillips}(1962)}]{Phillips}%
  \BibitemOpen
  \bibfield  {author} {\bibinfo {author} {\bibfnamefont {D.~L.}\ \bibnamefont
  {Phillips}},\ }\href {\doibase http://doi.acm.org/10.1145/321105.321114}
  {\bibfield  {journal} {\bibinfo  {journal} {J. ACM}\ }\textbf {\bibinfo
  {volume} {9}},\ \bibinfo {pages} {84} (\bibinfo {year} {1962})}\BibitemShut
  {NoStop}%
\bibitem [{\citenamefont {Novotny}\ and\ \citenamefont
  {Hecht}(2006)}]{NovotnyOptics}%
  \BibitemOpen
  \bibfield  {author} {\bibinfo {author} {\bibfnamefont {L.}~\bibnamefont
  {Novotny}}\ and\ \bibinfo {author} {\bibfnamefont {B.}~\bibnamefont
  {Hecht}},\ }\href@noop {} {\emph {\bibinfo {title} {{Principles of
  Nano-Optics}}}}\ (\bibinfo  {publisher} {Cambridge University Press},\
  \bibinfo {year} {2006})\BibitemShut {NoStop}%
\bibitem [{\citenamefont {Novotny}\ \emph {et~al.}(1998)\citenamefont
  {Novotny}, \citenamefont {Sanchez},\ and\ \citenamefont
  {Xie}}]{NovotnyCharge}%
  \BibitemOpen
  \bibfield  {author} {\bibinfo {author} {\bibfnamefont {L.}~\bibnamefont
  {Novotny}}, \bibinfo {author} {\bibfnamefont {E.~J.}\ \bibnamefont
  {Sanchez}}, \ and\ \bibinfo {author} {\bibfnamefont {X.~S.}\ \bibnamefont
  {Xie}},\ }\href {\doibase DOI: 10.1016/S0304-3991(97)00077-6} {\bibfield
  {journal} {\bibinfo  {journal} {Ultramicroscopy}\ }\textbf {\bibinfo {volume}
  {71}},\ \bibinfo {pages} {21 } (\bibinfo {year} {1998})}\BibitemShut
  {NoStop}%
\bibitem [{\citenamefont {Taubner}\ \emph {et~al.}(2005)\citenamefont
  {Taubner}, \citenamefont {Keilmann},\ and\ \citenamefont
  {Hillenbrand}}]{TaubnerSubsurface}%
  \BibitemOpen
  \bibfield  {author} {\bibinfo {author} {\bibfnamefont {T.}~\bibnamefont
  {Taubner}}, \bibinfo {author} {\bibfnamefont {F.}~\bibnamefont {Keilmann}}, \
  and\ \bibinfo {author} {\bibfnamefont {R.}~\bibnamefont {Hillenbrand}},\
  }\href {\doibase 10.1364/OPEX.13.008893} {\bibfield  {journal} {\bibinfo
  {journal} {Opt. Express}\ }\textbf {\bibinfo {volume} {13}},\ \bibinfo
  {pages} {8893} (\bibinfo {year} {2005})}\BibitemShut {NoStop}%
\bibitem [{\citenamefont {Zhang}\ \emph
  {et~al.}(2013{\natexlab{a}})\citenamefont {Zhang}, \citenamefont {Jiang},\
  and\ \citenamefont {Fogler}}]{Spheroid}%
  \BibitemOpen
  \bibfield  {author} {\bibinfo {author} {\bibfnamefont {L.~M.}\ \bibnamefont
  {Zhang}}, \bibinfo {author} {\bibfnamefont {B.~Y.}\ \bibnamefont {Jiang}}, \
  and\ \bibinfo {author} {\bibfnamefont {M.~M.}\ \bibnamefont {Fogler}},\
  }\href@noop {} {\bibfield  {journal} {\bibinfo  {journal} {\textit{in
  preparation}}\ } (\bibinfo {year} {2013}{\natexlab{a}})}\BibitemShut
  {NoStop}%
\bibitem [{\citenamefont {Zhang}\ \emph
  {et~al.}(2013{\natexlab{b}})\citenamefont {Zhang}, \citenamefont {Zhang},
  \citenamefont {Dong}, \citenamefont {Jian}, \citenamefont {Zhang},
  \citenamefont {Chen}, \citenamefont {Zhang}, \citenamefont {Liao},
  \citenamefont {Aizpurua}, \citenamefont {Luo}, \citenamefont {Yang},\ and\
  \citenamefont {Hou}}]{SingleMoleculeRaman}%
  \BibitemOpen
  \bibfield  {author} {\bibinfo {author} {\bibfnamefont {R.}~\bibnamefont
  {Zhang}}, \bibinfo {author} {\bibfnamefont {Y.}~\bibnamefont {Zhang}},
  \bibinfo {author} {\bibfnamefont {Z.~C.}\ \bibnamefont {Dong}}, \bibinfo
  {author} {\bibfnamefont {S.}~\bibnamefont {Jian}}, \bibinfo {author}
  {\bibfnamefont {C.}~\bibnamefont {Zhang}}, \bibinfo {author} {\bibfnamefont
  {L.~G.}\ \bibnamefont {Chen}}, \bibinfo {author} {\bibfnamefont
  {L.}~\bibnamefont {Zhang}}, \bibinfo {author} {\bibfnamefont
  {Y.}~\bibnamefont {Liao}}, \bibinfo {author} {\bibfnamefont {J.}~\bibnamefont
  {Aizpurua}}, \bibinfo {author} {\bibfnamefont {Y.}~\bibnamefont {Luo}},
  \bibinfo {author} {\bibfnamefont {J.~L.}\ \bibnamefont {Yang}}, \ and\
  \bibinfo {author} {\bibfnamefont {J.~G.}\ \bibnamefont {Hou}},\ }\href@noop
  {} {\bibfield  {journal} {\bibinfo  {journal} {Nature}\ ,\ \bibinfo {pages}
  {82}} (\bibinfo {year} {2013}{\natexlab{b}})}\BibitemShut {NoStop}%
\bibitem [{\citenamefont {Qiu}\ \emph {et~al.}(2003)\citenamefont {Qiu},
  \citenamefont {Nazin},\ and\ \citenamefont {Ho}}]{QiuFluorescence}%
  \BibitemOpen
  \bibfield  {author} {\bibinfo {author} {\bibfnamefont {X.~H.}\ \bibnamefont
  {Qiu}}, \bibinfo {author} {\bibfnamefont {G.~V.}\ \bibnamefont {Nazin}}, \
  and\ \bibinfo {author} {\bibfnamefont {W.}~\bibnamefont {Ho}},\ }\href
  {\doibase 10.1126/science.1078675} {\bibfield  {journal} {\bibinfo  {journal}
  {Science}\ }\textbf {\bibinfo {volume} {299}},\ \bibinfo {pages} {542}
  (\bibinfo {year} {2003})},\ \Eprint
  {http://arxiv.org/abs/http://www.sciencemag.org/content/299/5606/542.full.pdf}
  {http://www.sciencemag.org/content/299/5606/542.full.pdf} \BibitemShut
  {NoStop}%
\bibitem [{\citenamefont {Pechenezhskiy}\ \emph {et~al.}(2013)\citenamefont
  {Pechenezhskiy}, \citenamefont {Hong}, \citenamefont {Nguyen}, \citenamefont
  {Dahl}, \citenamefont {Carlson}, \citenamefont {Wang},\ and\ \citenamefont
  {Crommie}}]{CrommieInfrared}%
  \BibitemOpen
  \bibfield  {author} {\bibinfo {author} {\bibfnamefont {I.~V.}\ \bibnamefont
  {Pechenezhskiy}}, \bibinfo {author} {\bibfnamefont {X.}~\bibnamefont {Hong}},
  \bibinfo {author} {\bibfnamefont {G.~D.}\ \bibnamefont {Nguyen}}, \bibinfo
  {author} {\bibfnamefont {J.~E.~P.}\ \bibnamefont {Dahl}}, \bibinfo {author}
  {\bibfnamefont {R.~M.~K.}\ \bibnamefont {Carlson}}, \bibinfo {author}
  {\bibfnamefont {F.}~\bibnamefont {Wang}}, \ and\ \bibinfo {author}
  {\bibfnamefont {M.~F.}\ \bibnamefont {Crommie}},\ }\href {\doibase
  10.1103/PhysRevLett.111.126101} {\bibfield  {journal} {\bibinfo  {journal}
  {Phys. Rev. Lett.}\ }\textbf {\bibinfo {volume} {111}},\ \bibinfo {pages}
  {126101} (\bibinfo {year} {2013})}\BibitemShut {NoStop}%
\bibitem [{\citenamefont {Jackson}(1998)}]{Jackson}%
  \BibitemOpen
  \bibfield  {author} {\bibinfo {author} {\bibfnamefont {J.~D.}\ \bibnamefont
  {Jackson}},\ }\href
  {http://www.amazon.com/exec/obidos/redirect?tag=citeulike07-20\&path=ASIN/047130932X}
  {\emph {\bibinfo {title} {{Classical Electrodynamics Third Edition}}}},\
  \bibinfo {edition} {3rd}\ ed.\ (\bibinfo  {publisher} {Wiley},\ \bibinfo
  {year} {1998})\BibitemShut {NoStop}%
\bibitem [{\citenamefont {Wyld}(1999)}]{Wyld}%
  \BibitemOpen
  \bibfield  {author} {\bibinfo {author} {\bibfnamefont {H.~W.}\ \bibnamefont
  {Wyld}},\ }\href@noop {} {\emph {\bibinfo {title} {{Mathematical Methods for
  Physics}}}}\ (\bibinfo  {publisher} {Perseus Books Publishing},\ \bibinfo
  {address} {Reading, Massachussetts},\ \bibinfo {year} {1999})\BibitemShut
  {NoStop}%
\end{thebibliography}%

\end{document}